\documentclass[final,3p,times]{elsarticle}

\usepackage{amsmath,amssymb,amsthm, mathrsfs}
\usepackage{mathtools}
\usepackage{graphicx}
\usepackage{stmaryrd}
\usepackage{multirow}
\usepackage{wrapfig}
\usepackage[ruled,vlined]{algorithm2e}
\usepackage{hyperref}
\usepackage[dvipsnames]{xcolor}
\hypersetup{
    colorlinks=true,       
}

\newcommand{\mynabla}{\widetilde{\nabla}} 
\newcommand{\jump}[1]{[\![#1]\!]} 

\makeatletter
\def\ps@pprintTitle{%
 \let\@oddhead\@empty
 \let\@evenhead\@empty
 \def\@oddfoot{}%
 \let\@evenfoot\@oddfoot}
\makeatother

\begin{document}

\begin{frontmatter}




\title{A Hybridizable Discontinuous Galerkin solver \\ for the Grad-Shafranov equation}

\author[tsv]{Tonatiuh S\'anchez-Vizuet}
\ead{tonatiuh@cims.nyu.edu}
\address[tsv]{New York University, Courant Institute of Mathematical Sciences}
\author[ms]{Manuel Solano}
\ead{msolano@ing-mat.udec.cl}
\address[ms]{Universidad de Concepci\'on, Department of Mathematical Engineering and Center for Research in Mathematical Engineering (CI$^2$MA)}

\begin{abstract}
In axisymmetric fusion reactors, the equilibrium magnetic configuration can be expressed in terms of the solution to a semi-linear elliptic equation known as the Grad-Shafranov equation, the solution of which determines the poloidal component of the magnetic field. When the geometry of the confinement region is known, the problem becomes an interior Dirichlet boundary value problem. We propose a high order solver based on the Hybridizable Discontinuous Galerkin method. The resulting algorithm (1) provides high order of convergence for the flux function and its gradient, (2) incorporates a novel method for handling piecewise smooth geometries by extension from polygonal meshes, (3) can handle geometries with non-smooth boundaries and x-points, and (4) deals with the semi-linearity through an accelerated two-grid fixed-point iteration. The effectiveness of the algorithm is verified with computations for cases where analytic solutions are known on configurations similar to those of actual devices (ITER with single null and double null divertor, NSTX, ASDEX upgrade, and Field Reversed Configurations).
\end{abstract}


\end{frontmatter}



\section{Introduction}
In the current work we are interested in the computational simulation of the static magnetic equilibrium configuration of a plasma in an axially symmetric magnetic confinement device. In an axisymmetric geometry with the standard cylindrical coordinates denoted by $(r,\phi,z)$ along the unit direction vectors $\hat{\mathbf r}$, $\hat{\boldsymbol \phi}$, and $\hat{\mathbf z}$, the magnetic field $\mathbf B$ can be written as
\[
\mathbf B = \left(\frac{\widehat{\boldsymbol\phi}}{r}\right) \times grad\,\psi(r,z) + g(\psi)\left(\frac{\widehat{\boldsymbol\phi}}{r}\right),
\]
where the function $\psi$ is known as the \textit{poloidal flux function} and $g(\psi)/r$ is the \textit{toroidal field function} \cite{Jardin:2010,TaTo:1991}. The latter is also related to the net current flowing in the plasma and the external coils in the poloidal direction, $I_p$, through the relation
\[
I_p = \frac{2\pi}{\mu_0}g(\psi).
\]
In every cross section $\phi=constant$, the plasma will be confined to the region where the level sets of $\psi$ are closed curves. Since the magnetic field depends on $\psi$ only through its gradient, with the introduction of an appropriate shift, the boundary between the confinement and free regions can be made to correspond to the level set $\psi=0$ and we will assume so without loss of generality.

Under axisymmetry requirements, it can be shown that the equilibrium condition 
\[
grad\, p = \mathbf J \times \mathbf B,
\] 
between the force due to the kinetic pressure $p$ in the plasma and that one produced by the effect of the magnetic field $\mathbf B$ on the current density $\mathbf J$ can be expressed entirely in terms of $\psi$, $p$, and $g$. The resulting equivalent expression
\begin{equation}\label{eq:GS}
-\Delta^*\psi = \mu_0r^2\frac{dp}{d\psi} + g\frac{dg}{d\psi}
\end{equation}
was derived independently by Grad and Rubin, \cite{GrRu:1958}, Shafranov \cite{Shafranov:1958}, and Lust and Schl\"uter \cite{LuSc:1957} and is known as the \textit{Grad-Shafranov} equation. In the above expression, the magnetic permeability of vacuum $\mu_0$ is constant and the toroidal operator $\Delta^*$ is defined by
\[
\Delta^*\psi := r^2 div\left(\frac{1}{r^2} grad\,\psi\right) = r\partial_r\left(\frac{1}{r}\partial_r\psi\right) + \partial_z^2\psi= r\partial_r\left(\frac{1}{r}\partial_r\psi\right) + r\partial_z\left(\frac{1}{r}\partial_z\psi\right)=r\widetilde\nabla\cdot\left(\frac{1}{r}\widetilde\nabla\psi\right),
\] 
where the operator $\widetilde\nabla:=(\partial_r,\partial_z)$ acts formally like a vector of partial derivatives independent of the coordinate system. The function $g(\psi)$ and the pressure $p(\psi)$ depend purely on $\psi$ and the specific functional form of the dependence can be considered to be user provided.  In view of this, it is convenient to write \eqref{eq:GS} in the form
\begin{equation}\label{eq:WeakGS}
-\widetilde\nabla\cdot\left(\frac{1}{r}\widetilde\nabla\psi\right) = \frac{F(r,z,\psi)}{r}, \qquad F(r,z,\psi):= \mu_0r^2\frac{dp}{d\psi} + g\frac{dg}{d\psi},
\end{equation}
which highlights the semi-linear nature of the equation and leads naturally to a weak formulation. If the right hand side is truly nonlinear as a function of $\psi$, the equation can be solved iteratively, as will be detailed in Section \ref{sec:iterative}. 

When the precise location of the plasma-vacuum region is known a priori, the problem of determining the flux function in the plasma region becomes an interior Dirichlet boundary-value problem. In this case, the $\psi=0$ level set is of particular interest, for it becomes the boundary of the plasma domain, henceforth denoted $\Omega$, where homogeneous Dirichlet boundary conditions are imposed. This problem is often referred to as a fixed boundary problem and its numerical solution will be the focus of the present work. 

\begin{figure}[tb]\label{fig:plasmaconfig}
\center{ \begin{tabular}{cc}
\includegraphics[height=0.25\textwidth]{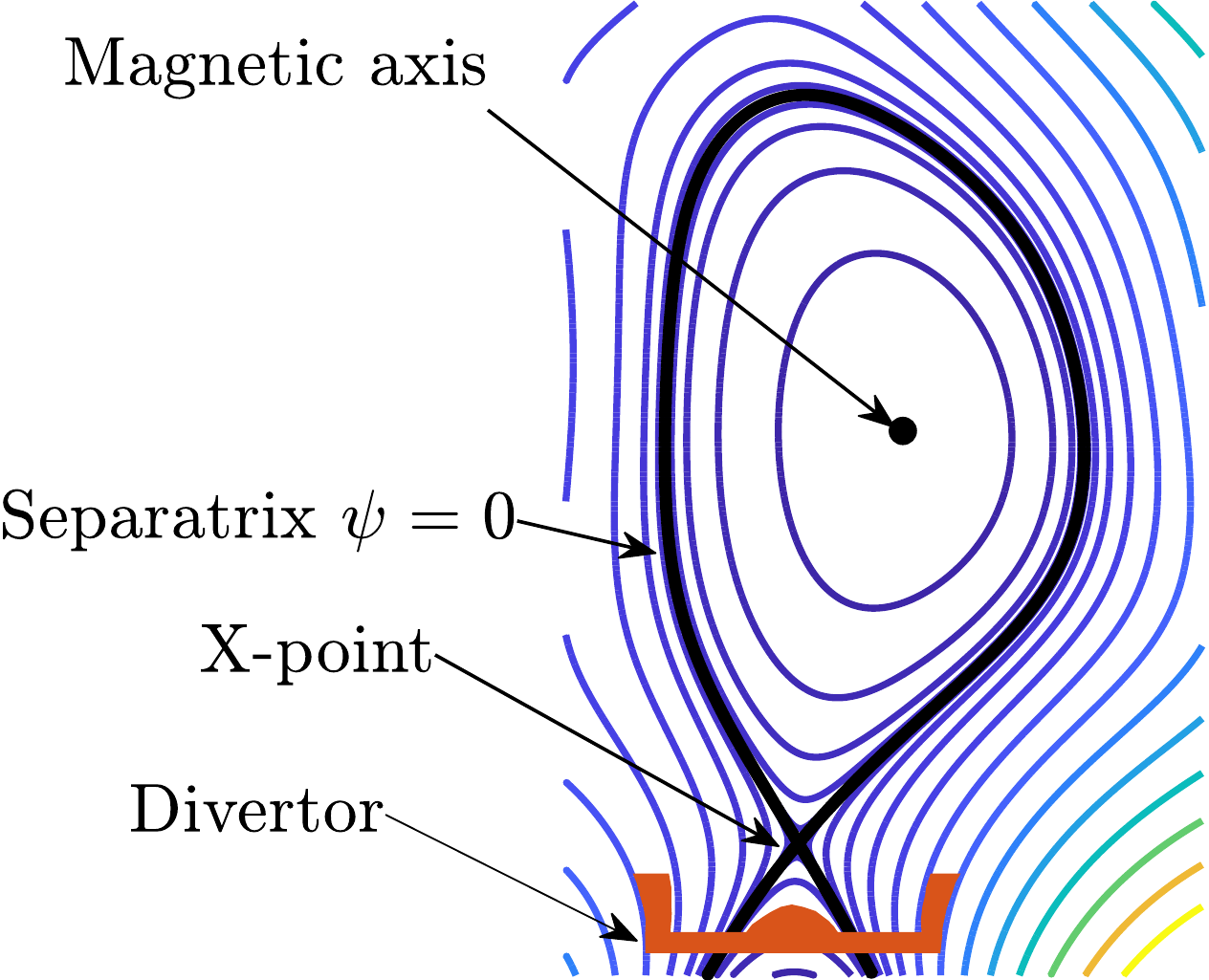} \qquad \quad & \quad \qquad
\includegraphics[height=0.25\textwidth]{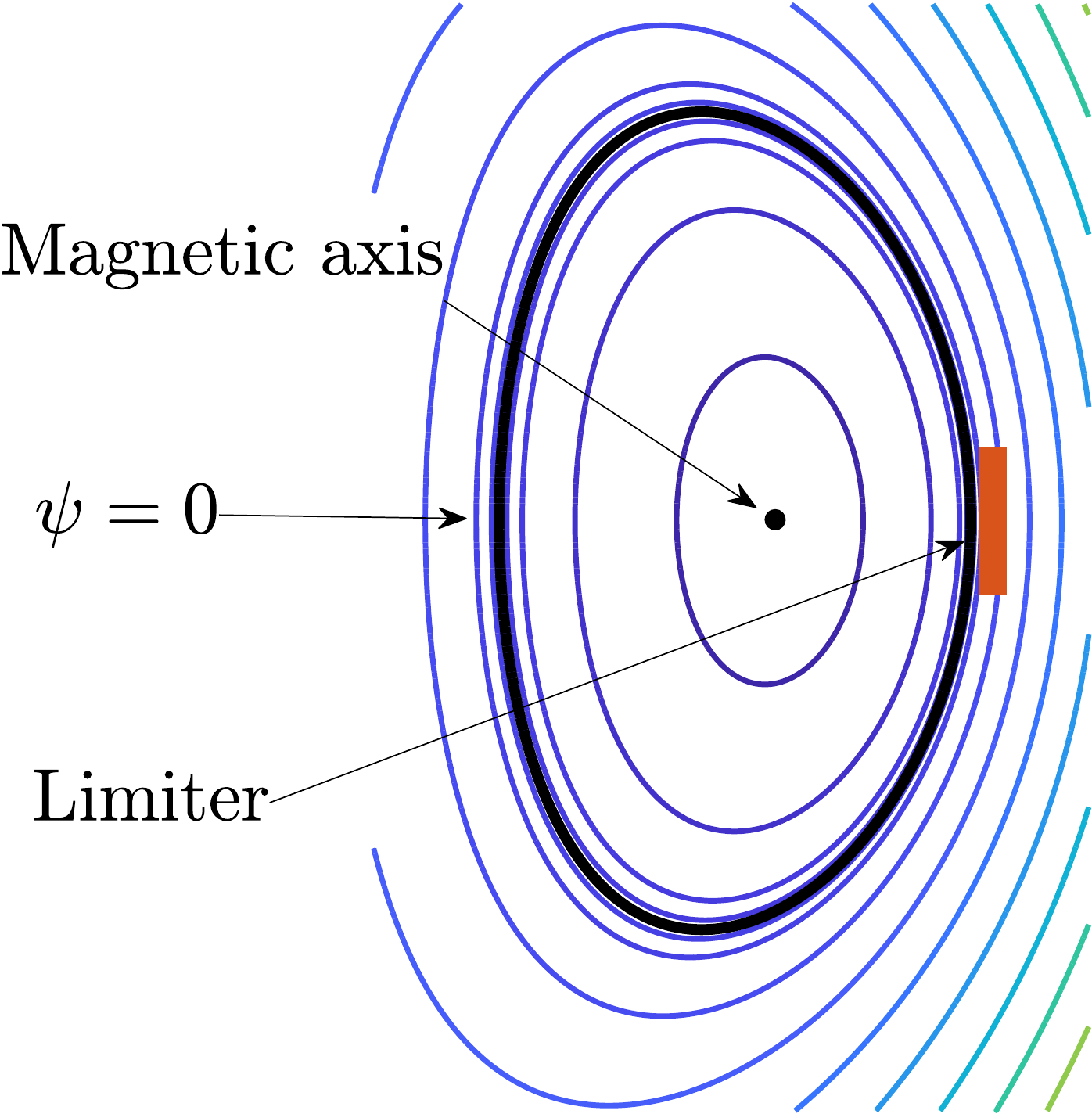}
\end{tabular}}
  \caption{{\scriptsize The plasma remains confined inside the region where the contour lines of the poloidal flux $\psi$ are closed. For engineering purposes, reactors are often designed so that the last closed flux surface has an x-point, below which a divertor is placed (left). When the contour lines do not present a separatrix, a limiter is required in order to forcefully stop the plasma from getting in contact with the reactor walls (right). Fixed-boundary computations solve the interior boundary value problem in the confinement region delimited by the level set $\psi=0$.} }
\end{figure}

Depending on the location, number, and current intensity of the external coils, the level set $\psi=0$ may become a separatrix, presenting what is known as an \textit{x-point}, as  depicted in Figure \ref{fig:plasmaconfig}. This kind of non-smooth geometry often poses additional challenges for some numerical solvers but the presence of such a point is often a desirable engineering feature and fusion reactors are frequently designed to produce such a configuration. One of the reasons for this is that in the absence of the clear division between the vacuum and plasma regions provided by the separatrix,  a physical \textit{limiter} must be introduced to prevent the plasma from touching the walls of the reactor. Since the limiter is in direct contact with the plasma, this severely limits the range of temperatures attainable in experiments and may introduce additional impurities in the plasma. Moreover, as the particles drift along the magnetic flux lines, the region between the open ends of the separatrix provides the optimal placement location for the divertor, a device that removes plasma impurities, extracts excess heat and protects the walls of the reactor \cite{JaBoFeIgKuPaPaSu:1995,DiChAnFeJaMaPaTi:1995}.  Therefore, methods that can handle these kinds of geometries can prove to be advantageous.

Many different approaches for the solution of the Grad-Shafranov equation have been employed over the years a very detailed --if dated-- description of the different approaches can be found in the review by Takeda and Tokuda \cite{TaTo:1991}, which is also a very complete reference on general aspects of plasma equilibrium. A concise discussion on plasma equilibrium and some related numerical techniques can also be found in \cite{Jardin:2010}. In what follows we briefly discuss only some relevant recent efforts without attempting a comprehensive review.

The weak form associated to \eqref{eq:WeakGS} is well suited for numerical computations and has been exploited in many computational efforts, such as the widely used Finite Element codes CHEASE \cite{LuBoRo:1992} and HELENA \cite{HuGoKe:1991,KoZi:2007} both of which use bi-cubic Hermite elements, and in more recent works involving the use of high-order spectral elements on rectangular geometries \cite{HoSo:2014} or mimetic elements \cite{PaKoFe:2016}. In recent works the free boundary problem has been addressed by employing a combination of boundary elements and finite elements \cite{Cedres:2015}, and mortar elements in overlapping grids \cite{HeRa:2017}. Nevertheless, weak/variational treatments are by no means the only way to approach the solution and many different alternatives have been employed.

The use of Multi-Grid methods can be traced back at least to the work of Braams \cite{Braams:1986} and have continued to attract attention over time both for the fixed boundary problem \cite{HaIkKa:2004} and the free boundary problem \cite{GoLeNe:2006}. In recent years, integral equation techniques combined with conformal mapping have been successfully employed to achieve fast and high order algorithms that can provide accurate approximations for both the flux function and its derivatives \cite{PaCeFrGrOn:2013}, this approach has been adopted in the flexible code ECOM \cite{LeCe:2015}.

Some other recent alternatives that have attracted attention include the hybrid approach EEC-ESC that couples Hermite elements near the plasma edge with Fourier decomposition methods in the plasma core \cite{LiZaDr:2014}, the use of meshless methods \cite{ImKaIt:2015,GhAm:2016}, the use of approximate particular solutions \cite{NaKaMu:2015}, and the method of fundamental solutions \cite{NaKa:2014}.

The references above do not focus on high order approximation of the derivatives of the flux function. However, the quantity of physical relevance in the problem is the magnetic field $\mathbf B$, therefore an important requirement for a Grad-Shafranov solver is to be able to provide accurate approximations to the partial derivatives of the flux as well, and some advanced post-processing techniques towards that goal have been developed lately \cite{RiCeRaFr:2016}. 

The present work represents an intermediate stage in a wider effort that aims to build a robust and flexible solver, in the spirit of ECOM, capable of dealing with direct equilibria in a wider variety of geometries and eventually handling situations in which other physical quantities related to the current profile are provided as input instead of $g(\psi)$. The novelty, and strength, of our approach resides on the use of a high order Hybridizable Discontinuous Galerkin (HDG) method combined with a technique for handling geometries with curved boundaries that preserves the order of accuracy of the method. The result is a robust algorithm that is able to provide high order of accuracy for the approximation of both the flux function and the magnetic field, offers good potential for parallel computations \cite{Cockburn:2010, KiShCo:2012}, and provides the flexibility of handling curved geometries (with or without x-points) relying only on polygonal meshes \cite{CoSo:2012}.

This work will focus only on the algorithmic aspects of the method and the analytical parts will be dealt with in a separate communication. The rest of the paper is structured as follows: In Section \ref{sec:continuous} we present the mixed formulation of the Grad-Shafranov equation at the continuous level; Section \ref{sec:method} describes the solution algorithm starting with the handling of curved boundaries by extension from a polygonal mesh and the HDG formulation is described afterwards. The treatment of the non-linearity through an accelerated fixed-point iteration closes the description of the algorithm. We then move on to the validation of the method in Section \ref{sec:validation} where the analytic solutions used as benchmarks are described and we present the results of the numerical experiments; concluding remarks and directions for future and ongoing work are given in the final Section \ref{sec:conclusions}.
%
\section{The continuous mixed formulation}\label{sec:continuous}
In order to apply an HDG discretization, the interior problem for the Grad-Shafranov equation must be first recast as a first order system. This is required since HDG methods include the gradient (or flux) as an additional unknown, but the reformulation is also physically motivated, since the quantity of interest is the magnetic field as opposed to the scalar poloidal flux.

Consider the fixed-boundary problem for the Grad-Shafranov equation
\begin{subequations}\label{eq:BVP2D}
\begin{alignat}{6}
-\mynabla\cdot\left(\frac{1}{r}\mynabla\psi\right) = \, & \frac{F}{r} &\qquad& \text{in }\, \Omega\subset \mathbb R^2, \label{eq:BVPa2D}\\
\psi =\, & 0 &\qquad& \text{on }\, \partial\Omega, \label{eq:BVPb2D}
\end{alignat}
\end{subequations}
where $\Omega$ is a cross section of the plasma region at constant toroidal angle. We consider a particular HDG method, which is called the \textit{hybridizable local discontinuous Galerkin (LDG-H) method in \cite{CoGoLa:2009}, that requires} us to formulate the problem in mixed form through the introduction of the \textit{flux} $\boldsymbol q := \frac{1}{r}\mynabla\psi$ as an additional unknown. This addition transforms \eqref{eq:BVP2D} into the equivalent system
\begin{subequations}\label{eq:mixedBVP2D}
\begin{alignat}{6}
\boldsymbol q - \frac{1}{r}\mynabla\psi = \,& \boldsymbol 0 &\qquad&  \text{in }\, \Omega\subset \mathbb R^2,  \label{eq:mixedBVPa2D} \\
-\mynabla\cdot\boldsymbol q =\,& \frac{F}{r} &\qquad& \text{in }\, \Omega\subset \mathbb R^2, \label{eq:mixedBVPb2D}\\
\psi =\, & 0 &\qquad& \text{on }\, \Gamma:=\partial\Omega. \label{eq:mixedBVPc2D}
\end{alignat}
\end{subequations}
If we had a standard continuous Galerkin discretization in mind, at this point the above system would be transformed into a weak formulation by the usual process of testing with arbitrary functions from appropriately chosen spaces. However, the HDG method does not rely on a continuous weak formulation. Instead, the domain is tessellated first and then a piecewise polynomial approximation is built by solving local weak problems defined over a polygonal approximation of $\Omega$. The strategy will be described in detail in Section \ref{sec:method}.
%
\section{The numerical method}\label{sec:method}
The solution strategy that we propose consists of three main components: (a) An HDG solver for a linear elliptic operator on polygonal domains, (b) a ``transfer" algorithm that allows for the handling of curved geometries through computation on polygonal sub-domains, and (c) an accelerated fixed point strategy that takes care of the semi-linearity of the problem. 

Since the HDG discretization is tied to the actual tessellation of the domain used for the computations, we start with the construction of the simplified polygonal domain used to approximate the plasma region, we then describe the method used to transfer Dirichlet data between the physical (curved) boundary and the computational (polygonal) boundary, next we introduce the HDG method applied to \eqref{eq:mixedBVP2D} and we conclude the section describing the iterative strategy used to treat the non-linearity and the eigenvalue problem.
%
\subsection{The treatment of curved boundaries}\label{sec:curved}
In general, standard numerical methods are defined over polygonal or polyhedral domains that can be triangulated. When dealing with domains having a curved boundary, \textit{fitted} \cite{Le:1986,BrKi:1996} or \textit{unfitted} \cite{Pe:1972,BrDuTh:1972} methods can be considered. In the former case, the boundary $\Gamma$ is matched or ``fitted'' by the computational boundary $\Gamma_h$. For instance, $\Gamma_h$ can be constructed by interpolating $\Gamma$. On the other hand, unfitted methods approximate the domain by a polygonal domain whose boundary $\Gamma_h$ does not necessarily  ``fit'' $\Gamma$. For example, this can be achieved by immersing $\Omega$ in a background mesh and setting the computational domain to be the union of all the elements of the mesh that lie inside $\Omega$. In both approaches the boundary condition on $\Gamma_h$ must be properly defined.

The boundary data can be easily imposed in the computational domain using fitted methods, which  is one of their main advantages. For a high order fitted method,  $\Gamma_h$ must approximate $\Gamma$ with enough accuracy to be able to recover a high order approximation of the solution. For instance, isoparametric finite elements can be used \cite{Le:1986}, where the elements near the boundary have a curved side that locally interpolates $\Gamma$. This construction might be unpractical in complicated geometries or evolving domains and that is why unfitted methods are preferred in these cases since the computational mesh is not adjusted to $\Omega$. However, the main drawback of standard unfitted methods is that only low order approximations can be obtained due to the fact that the boundary data on the computational domain is imposed ``away'' from the true boundary. 

Recently, an approach that combines the flexibility in the generation of the mesh characteristic of unfitted methods with a technique to transfer the boundary data from $\Gamma$ to $\Gamma_h$ has been developed \cite{CoSo:2012,CoQiSo:2014}. This method proposes an approximation of the boundary data by performing line integration along segments, called {\it transferring paths}, connecting $\Gamma_h$ to $\Gamma$. This strategy preserves the order of accuracy used for the approximation and has been our choice for handling the curved boundary. We describe it in detail in what follows.

\begin{figure}[tb]\center{
\begin{tabular}{ccc}
\includegraphics[width=0.2\linewidth]{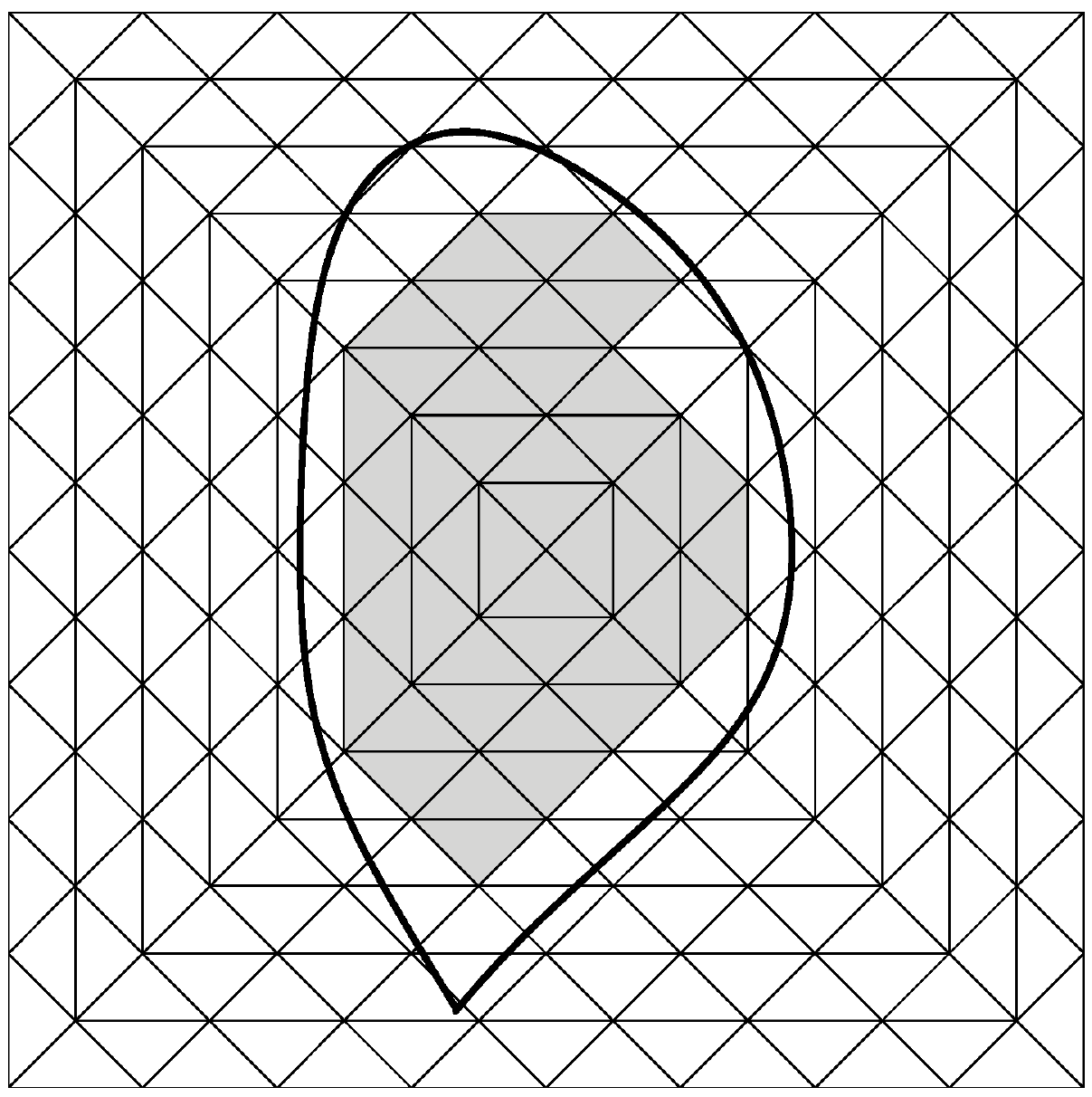} \qquad \qquad &
\includegraphics[width=0.2\linewidth]{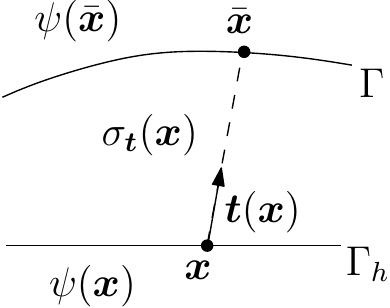} \qquad \qquad &
\includegraphics[width=0.12\linewidth]{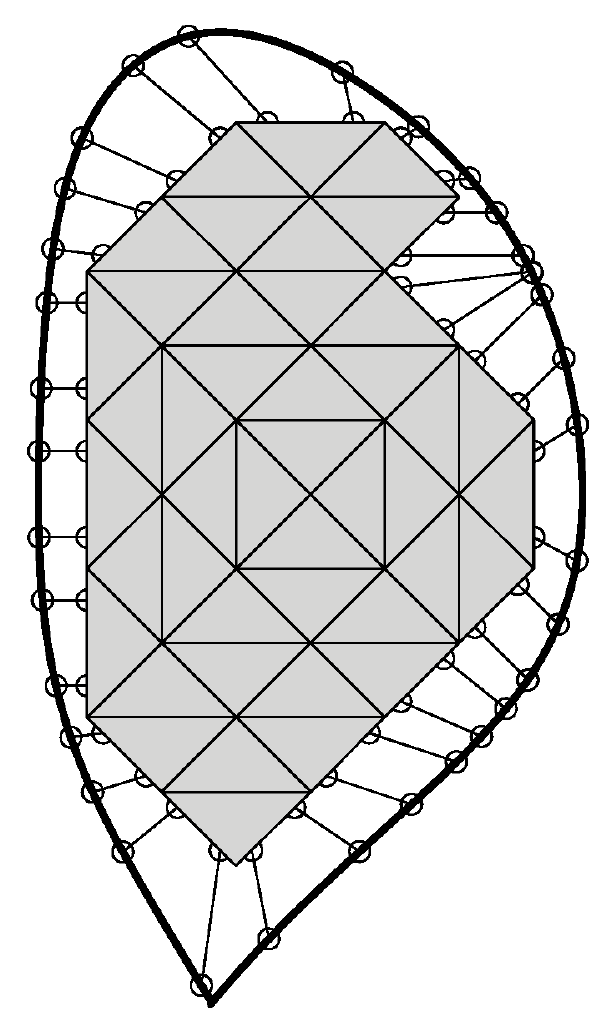}
\end{tabular}}
\caption{Left: Domain $\Omega$, background domain (square)  and polygonal subdomain (shaded). Middle: Transferring path  $\sigma_{\boldsymbol{t}}$ connecting $\boldsymbol{x}\in\Gamma_h$ to $\bar{\boldsymbol{x}} \in \Gamma$. Right: Transferring paths (segments with starting and ending points marked with $\circ$) associated to two points on each boundary edge. }\label{figure:Omegah}
\end{figure}

\paragraph{Computational domain} Let $\mathcal{B}$ be a background polygonal domain such that $\Omega \subset \mathcal{B}$ and $\mathcal{T}_h$ a triangulation of   $\mathcal{B}$ consisting of triangles $K$ that are uniformly shape-regular as Figure \ref{figure:Omegah} (left) shows. For a triangle  $K$, we denote its diameter by $h_K$ and its outward unit normal by $\boldsymbol{n}_K$, writing $\boldsymbol{n}$ instead of $\boldsymbol{n}_K$ when there is no confusion. We assume the triangulation does not have hanging nodes. Let $\mathsf{T}_h$ be the set of triangles of $\mathcal{T}_h$ that lie completely inside of $\Omega$ and $\partial \mathsf T_h : = \{\partial K: K \in \mathsf T_h \}$. Then, we define the computational domain $\Omega^h:=(\bigcup_{K\in\mathsf{T}_h} \overline{K}\big)^\circ$  (shaded region in Figure \ref{figure:Omegah}). The mesh size $h$ is defined as $\max_{K\in \mathsf{T}_h} h_K$. The set of  edges of $\mathsf{T}_h$ is denoted by  $\mathcal{E}_h$ and $\mathcal{E}_h=\mathcal{E}_h^\circ \cup \mathcal{E}_h^\partial$, where $\mathcal{E}_h^\circ$ and $\mathcal{E}_h^\partial$ are the sets of interior and boundary edges, respectively. Finally, we denote by $\Gamma_h$ the boundary of the computational domain.
%
%
\paragraph{Transferring paths} Let $\boldsymbol{x}=(r,z)$  be a point of $\Gamma_h$ to which we associate a point $\bar{\boldsymbol{x}}=(\bar{r},\bar{z})$ in $\Gamma$. We denote by $\sigma_{\boldsymbol{t}}(\boldsymbol{x})$ the segment joining $\boldsymbol{x}$ and $\bar{\boldsymbol{x}}$ with unit tangent vector $\boldsymbol{t}(\boldsymbol{x})$ and length $l(\boldsymbol{x})$ as depicted in Figure \ref{figure:Omegah} (middle). We will refer to $\sigma_{\boldsymbol{t}}(\boldsymbol{x})$ as the {\it transferring path} associated to $\boldsymbol{x}$.

\begin{figure}[tb]\center{
\begin{tabular}{cc}
\includegraphics[width=0.33\linewidth]{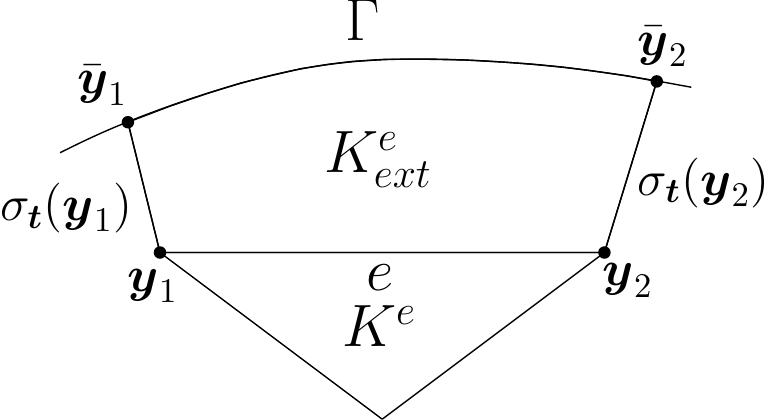}\qquad \qquad & \qquad \qquad
\includegraphics[width =0.12\linewidth]{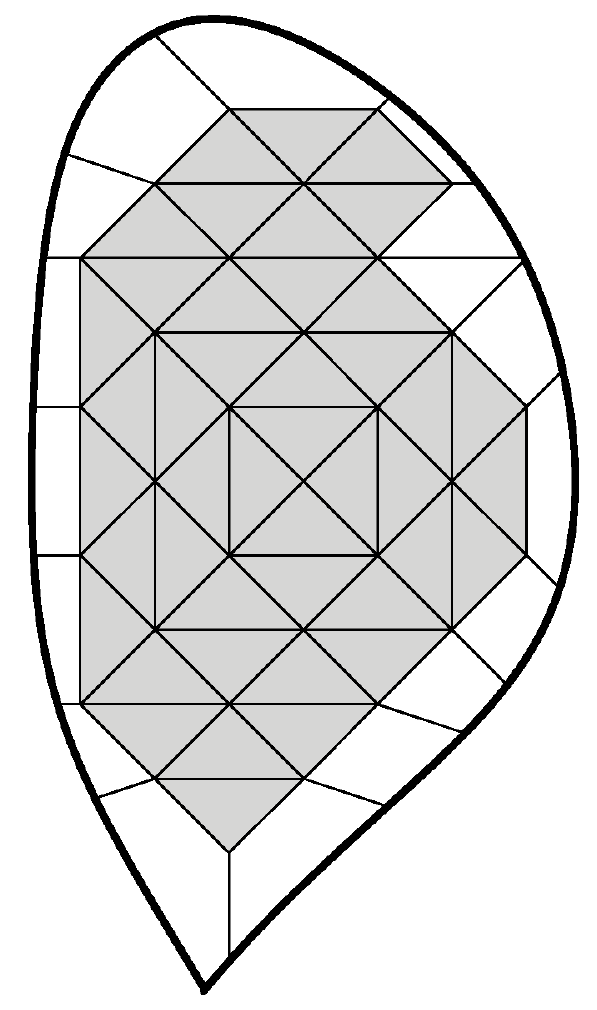}
\end{tabular}}
\caption{Left: A boundary edge $e$ with corresponding element $K^e$ and region $K^e_{ext}$. Right: $\Omega^h_{ext}:=\cup_{e\in \mathcal{E}_h^\partial} K^e_{ext}$ (white region).}\label{figure:Kext}
\end{figure}

 In principle, $\bar{\boldsymbol{x}}$ can be \textit{any} point in $\Gamma$ as long its distance to $\boldsymbol{x}$ is of order $h$. Given $\boldsymbol{x}\in \Gamma_h$, we will use the algorithm proposed in \cite{CoSo:2012} to find $\bar{\boldsymbol{x}}\in \Gamma$ such that the following three conditions are satisfied:  (1) $\sigma_{\boldsymbol{t}}(\boldsymbol{x})$ does not intersect another transferring path before terminating at $\Gamma$, (2) it does not intersect the interior of the computational domain  $\Omega^h$ and (3) the distance form $\bar{\boldsymbol{x}}$ to $\boldsymbol{x}$ is of order $h$. More details can be found in Section 2.4.1 of \cite{CoSo:2012}. Figure \ref{figure:Omegah} (right) shows the transferring paths associated to two quadrature points on each boundary edge.  For the computations, only transferring paths associated to the quadrature points and vertices of a boundary edge are needed.  
%
%
\paragraph{Extension from subdomains} Given a boundary edge $e$ with vertices at $\boldsymbol{y}_1$ and $\boldsymbol{y}_2$, we define $K^e_{ext}$ as the interior of the region determined by $e$, the segments $\sigma_{\boldsymbol{t}}(\boldsymbol{y}_1)$
 and $\sigma_{\boldsymbol{t}}(\boldsymbol{y}_2)$, and the arc of $\Gamma$ connecting $\bar{\boldsymbol{y}}_2$ and $\bar{\boldsymbol{y}}_1$ as shown in Figure \ref{figure:Kext}. By construction,  $K^e_{ext}$ does not intersect the corresponding region associated to a neighboring boundary edge. In addition, we observe that the union of all the elements $K^e_{ext}$ coincides with the complementary region $\Omega \setminus \Omega^h$ as shown in Figure \ref{figure:Kext}. Hence, we define  $\Omega^h_{ext}:= \displaystyle  \cup_{e\in \mathcal{E}_h^\partial} K^e_{ext}$ and note that $\Omega = \Omega^h \cup \Omega^{h}_{ext}$. 

Finally, it is convenient at this point to define a local polynomial extrapolation that will be used in order to build the approximation of the boundary data at $\Gamma_h$. More precisely, let $e$ be the  boundary edge that belongs to the triangle  $K^e \in \mathsf{T}_h$ and is associated to the exterior region $K_{ext}^e$. For a polynomial function $p$ defined on $K^e$, we denote by $E(p)$ the extrapolation from ${K^e}$ to $K_{ext}^e$ obtained by extending the domain of definition of $p$ to $K^e \cup K^e_{ext}$ while keeping the same  polynomial form. Hence,
\begin{equation}\label{eq:extension}
E(p): \,{K^e \cup K^e_{ext}}\, \longrightarrow \; \mathbb R, \qquad\qquad E(p)(\boldsymbol y) := p(\boldsymbol y).
\end{equation}
This definition provides a systematic way to extend a polynomial function defined only in the computational domain into the region $\Omega^h_{ext}$ enclosed by the physical boundary $\Gamma$ and the computational boundary $\Gamma_h$.
%
%
\paragraph{Approximation of the boundary data} Let $\sigma_{\boldsymbol{t}}(\boldsymbol{x})$ be the transferring path from $\boldsymbol{x}=(r,z) \in \Gamma_h$ to $\bar{\boldsymbol{x}}=(\bar{r},\bar{z}) \in \Gamma$. 
Multiplying  \eqref{eq:mixedBVPa2D} by $r$, integrating along  $\sigma_{\boldsymbol{t}}(\boldsymbol{x})$ and recalling that $\psi(\bar{\boldsymbol{x}})=0$, we obtain
\begin{equation}
\psi(\boldsymbol{x})=  - \int_{0}^ {l(\boldsymbol{x})} \! r\, \boldsymbol{q} (\boldsymbol{x}+\boldsymbol{t}(\boldsymbol{x})s)\cdot\boldsymbol{t}(\boldsymbol{x})\, ds \, =:\, \varphi(\boldsymbol{x}).\label{eq:bc}
\end{equation}
In other words, $\varphi$ is an \textit{exact} representation of the boundary data at $\Gamma_h$ that depends on the unknown $\boldsymbol{q}$. Let us emphasize that the numerical method will approximate the solution $(\boldsymbol{q},\psi)$ in $\Omega^h$, however the integral in the definition of $\varphi$ is over a segment lying outside $\Omega^h$ where $\boldsymbol{q}$ is still unknown. That is why, instead of considering the exact boundary condition $\varphi$, we will use an approximation denoted by $\varphi_h$. To be more precise, if $\boldsymbol{q}_h$ is the approximation of $\boldsymbol{q}$ computed by the HDG method on $K^e$, then motivated by \eqref{eq:bc} we define
\begin{equation}\label{eq:bch}
\varphi_h(\boldsymbol{x}):=- \int_{0}^ {l(\boldsymbol{x})} r E(\boldsymbol{q}_h) (\boldsymbol{x}+\boldsymbol{t}(\boldsymbol{x})s)\cdot\boldsymbol{t}(\boldsymbol{x})\, ds,
\end{equation}
where $E(\boldsymbol{q}_h)$ is the extension of the polynomial $\boldsymbol{q}_h\vert_{K^e}$ to the neighboring exterior element $K_{ext}^e$ defined in \eqref{eq:extension}.     
\subsection{The Hybridizable Discontinuous Galerkin Method}
The problem to solve in the computational domain is
%
\begin{alignat*}{6}
\boldsymbol q - \frac{1}{r}\mynabla\psi = \,& \boldsymbol 0 &\qquad&  \text{in }\, \Omega^h, \\
-\mynabla\cdot\boldsymbol q =\,& \frac{F}{r} &\qquad& \text{in }\, \Omega^h, \\
\psi =\, & \varphi_h &\qquad& \text{on }\, \Gamma_h,
\end{alignat*}
%
where $\varphi_h$ is given by \eqref{eq:bch}. The above problem can be thought of as a set of local boundary value problems defined on every element $K\in \mathsf T_h$. The local solutions are coupled through (a) the introduction of the \textit{hybrid} unknown $\widehat \psi:=\psi|_{\mathcal E_h}$ that corresponds to the restriction of $\psi$ to the skeleton of the mesh $\mathcal E_h$, and (b) the requirement that for every pair of elements $K^+$ and $K^-$ sharing an edge $e$ and with exterior normal vectors given by $\boldsymbol n^+$, $\boldsymbol n^-$, the normal component of the flux be continuous across their interface $\jump{\boldsymbol q}:=\boldsymbol q^+\cdot \boldsymbol n^+ + \boldsymbol q^-\cdot \boldsymbol n^-= 0$. This equivalent form can be stated as
\begin{alignat*}{6}
\boldsymbol q - \frac{1}{r}\mynabla\psi = \,& \boldsymbol 0 &\qquad& \text{ in } K \;\forall\, K\in\mathsf T_h,  \\
-\mynabla\cdot\boldsymbol q =\,& \frac{F}{r} &\qquad& \text{ in } K \; \forall\, K\in\mathsf T_h, \\
\psi =\, & \widehat{\psi} &\qquad& \text{ on }\, \partial K \; \forall\, K\in\mathsf T_h,  \\
 \jump{\boldsymbol q} =\,& 0 &\qquad& \text{ on } e\; \forall e\in \mathcal E^\circ_h,  \\
  \psi =\, & \varphi_h &\qquad& \text{ on }\, \Gamma_h. 
\end{alignat*}
The first three equations above define the local problems, while the last two define the global system that determines the hybrid unknown $\widehat{\psi}$.  The HDG method \cite{Cockburn:2010} seeks approximations $(\boldsymbol{q}_h,\psi_h,\widehat{\psi}_h)$ to the solutions $(\boldsymbol{q},\psi, \psi|_{\mathcal{E}_h})$ in the finite dimensional space $\boldsymbol{V}_h\times W_h\times M_h$ given by
\begin{eqnarray*}
\boldsymbol{V}_h & = & \{ \boldsymbol{v}\in\boldsymbol{L}^2(\mathsf{T}_h): \ \ \boldsymbol{v}|_K\in\mathbf{P}_k(K) \ \ \forall K\in\mathsf{T}_h \}, \\
W_h & = & \{ w \in L^2(\mathsf{T}_h): \ \ \ w|_K\in \mathbb{P}_k(K) \ \ \forall K\in\mathsf{T}_h \}, \\
M_h & = & \{ \mu\in L^2(\mathcal{E}_h): \ \ \mu|_e\in \mathbb{P}_k(e) \ \ \forall e\in\mathcal{E}_h \},
\end{eqnarray*}
where $\mathbb{P}_k(K)$ is the space of polynomials of degree $k$ defined on the triangle $K$, $\mathbf{P}(K):=[\mathbb{P}_k(K)]^2$, and $\mathbb{P}_k(e)$ is the space of polynomials of degree $k$ defined on the edge $e$. Following the standard notation, we will denote by $(\cdot,\cdot)_K$ and $\langle\cdot, \cdot\rangle_{\partial K}$ the $L^2$ inner products on an element $K$ and on its boundary $\partial K$ respectively, and we will define
\[
(\cdot,\cdot)_{\mathsf T_h}:= \sum_{K\in \mathsf T_h} (\cdot,\cdot)_{K}\, \qquad \qquad \langle\cdot,\cdot\rangle_{\partial \mathsf T_h} :=  \sum_{K\in \mathsf T_h} \langle\cdot,\cdot\rangle_{\partial K}. 
\]
The approximation $(\boldsymbol{q}_h,\psi_h,\widehat{\psi}_h) \in \boldsymbol{V}_h\times W_h\times M_h $ is the unique solution of 
\begin{subequations}\label{eq:hdg}
\begin{eqnarray}
\label{eq:hdg1}(r\boldsymbol{q}_h,\boldsymbol{v})_{\mathsf{T}_h} + (\psi_h,\mynabla\cdot\boldsymbol{v})_{\mathsf{T}_h} - \langle\widehat{\psi}_h,\boldsymbol{v}\cdot\boldsymbol{n} \rangle_{\partial \mathsf T_h} & = & 0, \\
\label{eq:hdg2}(\boldsymbol{q}_h,\mynabla w)_{\mathsf{T}_h}   - \langle\widehat{\boldsymbol{q}}_h\cdot \boldsymbol{n},w\rangle_{\partial \mathsf T_h} & = & (F/r,w)_{\mathsf{T}_h},\\
\label{eq:hdg3}\langle\widehat{\psi}_h,\mu\rangle_{\Gamma_h} & = & \langle \varphi_h,\mu\rangle_{\Gamma_h}, \\ 
\label{eq:hdg4}\langle\widehat{\boldsymbol{q}}_h \cdot\boldsymbol{n},\mu\rangle_{\partial \mathsf T_h\setminus\Gamma_h} & = & 0,
\end{eqnarray}
for all $(\boldsymbol{v},w,\mu)\in \boldsymbol{V}_h\times W_h\times M_h$, where the \textit{numerical flux} $\widehat{\boldsymbol q}_h$ is defined as
\begin{equation}\label{eq:hdg5}
\widehat{\boldsymbol{q}}_h\cdot\boldsymbol{n} := \boldsymbol{q}_h\cdot \boldsymbol{n} + \tau (\psi_h-\widehat{\psi}_h)\qquad\text{on }\partial \mathsf T_h,
\end{equation}
\end{subequations}
and $\tau$ is a non-negative piecewise constant stabilization parameter defined on $\partial \mathsf{T}_h$. This simple choice for the numerical flux has become somewhat standard, but it is by no means unique. One of the advantages of the definition \eqref{eq:hdg5} is that it keeps equations \eqref{eq:hdg1} and \eqref{eq:hdg2} completely in terms of local quantities (once the hybrid unknown has been determined). It has been shown in the linear case (in \cite{CoGoSa:2010} for polyhedral domains and in \cite{CoQiSo:2014} for curved domains) that the method achieves optimal convergence order when $\tau$ is kept of order one. In our computations the value of the stabilization parameter was set to $\tau=1$.

Once the solution $(\boldsymbol{q}_h,\psi_h$) in $\Omega^h$ is computed by solving \eqref{eq:hdg}, it can be extrapolated to the entire domain $\Omega$ using \eqref{eq:extension}. However, we can define a better approximation for $\psi$ on the exterior domain by means of the transferring technique \eqref{eq:bc}. More precisely,  consider $\boldsymbol{y}\in K_{ext}^e$ and let $\sigma_{\boldsymbol{t}}(\boldsymbol{y})$ be the transferring path from $\boldsymbol{y}=(r,z) \in \Gamma_h$ to $\bar{\boldsymbol{y}}=(\bar{r},\bar{z}) \in \Gamma$. Then, recalling that $\psi(\bar{\boldsymbol{y}})=0$, the approximation of $\psi(\boldsymbol y)$ is given by
\begin{equation}\label{eq:trasnfer_psih}
\psi_h(\boldsymbol{y}):=  - \int_{0}^ {l(\boldsymbol{y})} \! r\, \boldsymbol{q}_h (\boldsymbol{y}+\boldsymbol{t}(\boldsymbol{x})s)\cdot\boldsymbol{t}(\boldsymbol{y})\, ds. 
\end{equation}
This approximation converges to $\psi$ in $\Omega^h_{ext}$ with an additional order as shown in \cite{CoSo:2012,CoQiSo:2014}. 

It is evident that HDG methods are related to the family of \textit{mixed methods}, where the gradient of the scalar potential is introduced as an additional unknown of the problem. This reformulation results in an increased number of problem unknowns, but in return ensures that the order of accuracy provided by the discretization is the same for the original unknown and its partial derivatives.

As a matter of fact, the mixed structure of the problem, together with the hybridization, also has positive consequences from the computational point of view. For instance, the hybridization technique helps to overcome the increment on the number of degrees of freedom usually associated to discontinuous Galerkin methods and the resulting matrices have the same sparsity and size as the ones resulting from the hybrid \textit{continuous} mixed methods \cite{CoGo:2004}. This contrasts with the tendency of Discontinuous Galerkin methods to ``use too many unknowns". Moreover, since the only global unknown is the numerical trace that lies on the edges of the triangulation, once it is determined, the other unknowns are locally computed in each element, allowing for parallelization. 

The reader familiar with mixed methods may object that this family imposes very tough restrictions on the possible choices for the approximation spaces, but one can rest assured that yet another positive feature of HDG is providing with ample flexibility for the spaces used.  We refer the reader interested in the programming aspects of HDG methods to \cite{FuGaSa:2015}, where a very detailed explanation and coding strategies and tools are given. Our own implementation is based on the tools provided in that reference. 
%
\paragraph{Computational complexity} One of the points that distinguish HDG from other variants of discontinuous Galerkin methods, is its efficient use of degrees of freedom  and the advantageous sparsity pattern of the matrices arising from it. In order to see why this is the case, we will first briefly point out some of the main differences between traditional continuous (CG) and discontinuous Galerkin (DG) methods.

As a byproduct of enforcing continuity of the discrete solution across element interfaces, CG methods use fewer degrees of freedom when compared to DG methods and therefore result in comparatively smaller matrices. This is due to the fact that those degrees of freedom located at the interfaces are shared by the neighboring elements in the continuous setting, whereas in the discontinuous case they are independent of each other and thus have to be accounted roughly once per every element that shares a face or a vertex. On the other hand, the sharing of information between neighboring continuous elements gives rise to matrices with stronger coupling (i.e. bigger bandwidth) than those stemming from discontinuous elements, where the interaction between adjacent elements is much weaker.

In the CG framework an effective technique for reducing the size of the global system has been known since at least 1965 in the context of finite element methods \cite{Guyan:1965} and mixed methods \cite{Fraeijs:1965}. The technique, that has come to be known as \textit{static condensation}, described in the first reference amounts to a reorganization of the degrees of freedom separating those defined on edges from those on the interior of the elements, which can then be eliminated from the global system. In the mixed methods a similar result is attained by introducing a \textit{hybrid} variable defined at the inter element boundaries that approximates the trace (restriction) of the scalar unknown to the faces of the triangulation and by requiring the continuity of the normal component of the flux across elements. In both cases, the size of the global system that determines the unknowns on the edges of the triangulation is reduced, and the interior unknowns are then recovered by a local post-processing.

From the above, and the description of HDG in the previous section, we can see that HDG strives to attain the flexibility and reduced bandwidth of DG by solving local problems, and addresses the issue of the larger number of degrees of freedom by static condensation through the introduction of a hybrid variable. The method thus combines positive features of both continuous and discontinuous Galerkin. If the method is parallelized, the smaller bandwidth of the global system reduces communication costs between different processors while the local solves are naturally parallel.

Detailed comparative studies between HDG and CG (spectral elements) applied to linear elliptic equations have been performed in \cite{KiShCo:2012} for 2D and in \cite{YaMoKiSh:2015} for 3D. The authors there conclude that, whereas an iterative solver would in general favor CG, for polynomial degrees beyond 3 the bandwidth of the global HDG system falls below the one from the statically condensed CG. As a consequence, a parallel iterative solve would render both computations comparable and a direct solve of the HDG system would be faster. This is particularly attractive for cases when the source term is a non-linear function of the scalar unknown and the solution would potentially require multiple linear solves as will be discussed in Section \ref{sec:iterative}. All these features make HDG a very promising option with great potential for parallelization.
\paragraph{A remark on post-processing and superconvergence} One of the advantageous properties of HDG is the possibility of devising local post-processing strategies for the scalar unknown, $\psi$ in our case, that yield an approximation converging with an additional order. The choice of discrete spaces and post-processing strategies that enable superconvergence has been thoroughly analyzed in \cite{CoQiSh:2012} within the context of HDG methods for \textit{linear} elliptic problems. 

In the current application however, our main focus is to obtain accurate approximations for the partial derivatives of $\psi$ (i.e. the components of $r\boldsymbol q$), since the physical quantities of interest in magnetic confinement are related to the magnetic field and not to the scalar potential. In view of this, we have not included a post-processing stage in our implementation. Nevertheless, we present one such strategy here for the sake of completeness. 

Once the approximations $\psi_h$ and $\boldsymbol q_h$ have been determined from the solution of \eqref{eq:hdg}, one now looks for a piecewise polynomial function $\psi_h^*$ such that
\begin{alignat*}{6}
\psi_h^*\in\;& \mathbb P_{k+1}(K) &\quad& \forall\, K\in \mathsf T_h,\\
(\nabla\psi_h^*,\nabla w_h)_K =\,& (r\boldsymbol q_h,\nabla w_h)_K &\quad& \forall\,w_h\in \mathbb P_{k+1}(K), \\
(\psi_h^*,1)_K =\,& (\psi_h,1)_K. &
\end{alignat*}
The solution to this auxiliary problem was shown in \cite{CoGoSa:2010} to converge towards $\psi$ with an additional order when $k\geq 1$ .
%
\subsection{Accelerated fixed-point iterations}\label{sec:iterative}
In order to deal with the semi-linearity of the problem we will resort to an iterative strategy. Due to their simplicity and effectiveness, straightforward fixed-point iterations (also known as a Picard iterations) of the style
\[
-\Delta^* \psi^{n} = F(r,z,\psi^{n-1})
\]
have been preferred in many applications \cite{HoSo:2014,PaKoFe:2016,GoLeNe:2006,PaCeFrGrOn:2013,LeCe:2015,LiZaDr:2014,GhAm:2016}. We choose to follow a similar strategy, but enhance it with two simple yet effective acceleration methods. 

The first method consists of a simple two-grid strategy that was already included in CHEASE \cite{LuBoRo:1992} where the fixed-point iteration is carried out in a coarse grid $\mathsf T_H$ until convergence is achieved to a prescribed tolerance. The resulting coarse-grid solution is then prolonged onto a finer grid $\mathsf T_h$ where it is used as initial guess for a second round of fixed-point iterations. The computations on the coarse grid are considerably less taxing and the resulting improved initial guess decreases the number of iterations required on the fine grid. The convergence properties of this two-grid fixed-point strategy have been analyzed in \cite{Xu:1994, Xu:1996}, where rigorous conditions for optimal rates of convergence are discussed. 

The second strategy pertains to the fixed-point iteration itself and is an optimized variation of the popular back-averaging method known as \textit{Anderson acceleration} \cite{Anderson:1965}. The idea is to use an optimized convex linear combination of a predetermined number of previous iterates as input for the next update. If we denote by $M(\cdot)$ the mapping whose fixed-point is being sought for and $u_0$ the initial input then, in its simplest form, the acceleration algorithm {\tt anderson}$\left(m, u^0,M(\cdot),\epsilon\right)$ using $m$ previous iterates can be described as follows:
\begin{center}
\begin{algorithm}[H]
\SetAlgoLined
 \KwData{\\ $m$:\, Depth.\, $u^{0}$:\, Initial guess.\,$M:$\ Mapping.\, $\epsilon$:\, Stopping tolerance.}
 \KwResult{\\$u^{*}$:\, Approximate fixed-point of $M$. }
 \Begin{
 $n=0$\, , \quad $Res = 1$\;
 $\widetilde u^1 = M(u^0)$\;
 $G^1 = \widetilde u^1-u^0$\;
 $u^1 = \widetilde u^1$\;
 \While{Res $\geq \epsilon$}{
 $ n= n + 1$\;
 $k = \min\{m,n\}$\;
 $\widetilde{u}^{n+1} = M(u^{n})$\;
 $G^{n+1} = \widetilde{u}^{n+1}-u^{n}$ \; 
 Find: \, $(\alpha_1,\ldots,\alpha_{k+1})\in \mathbb R^{k+1}$ such that
 \begin{enumerate}
 \item $\sum_{j=1}^{k+1}{\alpha_j}=1$
 \item $(\alpha_1,\ldots,\alpha_{k+1}) = \text{argmin } \|\sum_{j=1}^{k+1}{\alpha_j G^{n+j-k}}\| $
 \end{enumerate}
 $u^{n} = \sum_{j=1}^{k+1}\alpha_j\widetilde{u}^{n+j-k}$\;
 $Res = \|u^{n}-u^{n-1}\|/\|u^n\|$; 
 }
  $u^* = u^n$\;
 \caption{{\tt anderson}$\left(m,u^0,M(\cdot),\epsilon\right)$}
 }
\end{algorithm}
\end{center}
It is clear that the algorithm of depth $m=0$ coincides with the simple Picard iterative scheme. By including information from more than one of the previous updates in this way, the convergence can be dramatically improved. Furthermore, in terms of the number of iterations needed to achieve a prescribed tolerance, the method can not do worse than Picard and it often provides considerable improvement, as shown by Toth and Kelley \cite{ToKe:2015}. In the same work, the authors report that, although no results are available regarding the optimal depth $m$,  empirically there is no gain from choosing $m\geq 3$. This was consistent with our own experiments and therefore we settled for an acceleration of depth $m=2$ in our implementation.  A more elaborate version of the procedure allows for the mixing of previous instances of both $\tilde{u}$ and $u$ \cite{ToKe:2015,WaNi:2011}, but for our implementation we follow the simple procedure described above.
%
\subsection{Summary of the solution method}
%
Let $S_{H,F}$ be the operator that, for a given right hand side $F$, maps the initial guess $\psi^0$ to the HDG approximation $(\boldsymbol q_H,\psi_H)$ computed on a mesh $\mathsf T_H$ with parameter $H$, and $\Pi_h$ be a prolongation operator onto the finer grid $\mathsf T_h$. With this notation we can summarize the solution strategy algorithmically as follows:
\begin{center}
\begin{algorithm}[H]
\SetAlgoLined
 \KwData{ \\ $F$:\, Right hand side.\, $\mathsf T_H$:\, Coarse triangulation.\, $\mathsf T_h$:\, Fine triangulation.\, \\ $\epsilon$:\, Stopping tolerance.}
 \KwResult{\\$(\boldsymbol q_h,\psi_h)$:\, Approximate HDG solutions. }
 \Begin{
 $\psi^0$ \tcp*{Non-trivial initial guess}
 $(\boldsymbol q_H,\psi_H)=\text{{\tt anderson}}\left(2,\psi^0,S_{H,F},\epsilon\right)$  \tcp*{Coarse grid}
 $\psi^0 = \Pi_h \psi_H$   \tcp*{Prolongation onto a fine grid}
$(\boldsymbol q_h,\psi_h)=\text{{\tt anderson}}\left(2,\psi^0,S_{h,F},\epsilon\right)$  \tcp*{Fine grid}   
}
 \caption{{\tt solveGS}$\left(F,\mathsf T_H,\mathsf T_h,\epsilon\right)$}
\end{algorithm}
\end{center}
%
\section{Numerical Experiments}\label{sec:validation}
The accuracy and convergence properties of the scheme are evaluated by comparing the approximations to the flux function and its gradient to some existing analytical solutions briefly presented below. The geometries used for the simulations are closely related to physically relevant configurations: the International Thermonuclear Experimental Reactor (ITER), the National Spherical Toroidal Experiment (NSTX), the Axially Symmetric Divertor Experiment (ASDEX upgrade), and Field Reversed Configurations (FRC). Additional experiments with non-linear source terms are performed on ITER-like geometries: one with a double null divertor and a smooth D-shaped Miller parametrization. We measure error in the standard $L^2$ norm, but also verify the performance  ``off the grid" in an $L^\infty$-related norm by sampling on random non-grid points in the computational domain and considering the maximum discrepancy as the error measure. Convergence as a function of the grid size and the polynomial degree are tested independently and the results of the numerical experiments on the different geometries are presented at the end of the section.
%
\subsection{Analytical solutions in free space}
For particular pressure and poloidal current profiles some exact solutions have been derived. These profiles determine the right hand side of equation \eqref{eq:BVP2D} that we recall here for convenience:
\[
-\widetilde\nabla\cdot\left(\frac{1}{r}\widetilde\nabla\psi\right) = \frac{F(r,z,\psi)}{r}, \qquad F(r,z,\psi):= \mu_0r^2\frac{dp}{d\psi} + g\frac{dg}{d\psi}.
\]
The concrete expressions of the solutions we use as benchmarks (and the right hand sides they give rise to) are described briefly in what follows.
\paragraph{Solov'ev profiles} These profiles for $p(\psi)$ and $g(\psi)$ arise when the terms on the right hand side of \eqref{eq:GS} are such that 
\[
\mu_0\frac{dp}{d\psi} = -C \, , \qquad g\frac{dg}{d\psi} = -A,
\]
for some constant values $A$ and $C$. If the flux is normalized so that $A+C=1$ the right hand side function becomes
\begin{equation}\label{eq:solovevF}
F(r,z,\psi) = -\left((1-A)r^2 + A\right).
\end{equation}
The solution of this equation can be split into a homogeneous part, $\psi_H$, and particular part $\psi_P$, so that $\psi = \psi_H + \psi_P$. A particular solution is
\begin{equation}\label{eq:psir}
\psi_P(r,z) = \frac{r^4}{8} + A\left(\frac{1}{2}r^2\ln r - \frac{r^4}{8}\right),
\end{equation}
while the homogeneous solution can be a linear combination the terms:
\begin{alignat}{6}
\nonumber
\psi_1 =\,& 1, & \qquad & \psi_7 = \,&& 8z^6-140z^4r^2+75z^2r^4-15r^6\ln{r} \\
\nonumber
\psi_2 =\,& r^2, & \qquad &  &&+180r^4z^2\ln{r}-120r^2z^4\ln{r},\\ 
\nonumber
\psi_3 =\,& z^2-r^2\ln{r}, &\qquad &  \psi_8 =\,&& z,\\ 
\nonumber
\psi_4 =\,& r^4-4r^2z^2, &\qquad &  \psi_9 =\,&& zr^2,\\ 
\nonumber
\psi_5 =\,& 2z^4-9z^2r^2+3r^4\ln{r} &\qquad & \psi_{10} =\,&& z^3-3zr^2\ln{r},\\ 
\nonumber
 &  -12r^2z^2\ln{r}, & \qquad &  \psi_{11} =\,&& 3zr^4-4z^3r^2,\\ 
\label{eq:solovevSol}
\psi_6 =\,& r^6-12r^4z^2+8r^2z^4, &\qquad& \psi_{12} =\,&& 8z^5-45zr^4 -80z^3r^2\ln{r} +60zr^4\ln{r}. 
\end{alignat}
Following the procedure carefully derived  in \cite{CeFr:2010} to choose the coefficients $(c_1,\ldots,c_{12})$ corresponding to each of these functions, the solution $\psi$ can be written in terms of three free parameters $(\epsilon,\delta,\kappa)$ that describe the cross section of the configuration. By adjusting the values of $\epsilon, \delta$, and $\kappa$ it is possible to obtain profiles that correspond to relevant physical configurations. The interested reader is referred to the above reference for more details on the parametrization of the plasma boundary.
%
\paragraph{Configurations with dissimilar source functions} If, as proposed in \cite{McCarthy:1999}, the pressure and toroidal flux are set to
\[
p = \frac{S}{\mu_0}\psi\,,\qquad g^2 = T\psi^2 + 2U\psi + g_0^2  
\]
for some constants $S, T, U,$ and $g_0$, the right hand side function of \eqref{eq:GS} becomes
\begin{equation}\label{eq:dissimilarF}
F(r,z,\psi) = T\psi+ Sr^ 2 + U.
\end{equation}
The solution to the Grad-Shafranov equation for these sources can be obtained by a similar procedure as before; by splitting the solution into homogeneous and particular parts $\psi = \psi_h + \psi_p$ where
\[
\psi_h = -\frac{1}{T}\left(U+Sr^2\right) \quad\text{ and } \quad -\Delta^*\psi_p = T\psi_p.
\]
Eight different families of solutions $\psi_p$ to the above eigenvalue problem are found in \cite{McCarthy:1999}. Here we will focus on one particular linear combination of them that gives rise to the solution
\begin{align}
\nonumber
\psi =\,& c_1 + c_2r^2 + rJ_1(pr)\left(c_3+c_4z\right) + c_5\cos{\left(pz\right)} + c_6\sin{\left(pz\right)} \\
\nonumber
 & +  r^2\left(c_7\cos{\left(pz\right)} + c_8\sin{\left(pz\right)}\right)  + c_9\cos{\left(p\sqrt{r^2+z^2}\right)} + c_{10}\sin{\left(p\sqrt{r^2+z^2}\right)} \\
 \nonumber
 & + rJ_1(\nu r)\left(c_{11}\cos{\left(qz\right)} + c_{12}\sin{\left(qz\right)}\right)+  rJ_1(q r)\left(c_{13}\cos{\left(\nu z\right)} + c_{14}\sin{\left(\nu z\right)}\right)\\
\label{eq:dissimilarSol}
 & + rY_1(\nu r)\left(c_{15}\cos{\left(qz\right)} + c_{16}\sin{\left(qz\right)}\right) + rY_1(q r)\left(c_{17}\cos{\left(\nu z\right)} + c_{18}\sin{\left(\nu z\right)}\right),
\end{align}
where $p = \sqrt{T}$, $q=p/2$, $\nu = \sqrt{3/4}p$, and $J_1$ (resp. $Y_1$) is the Bessel function of the first (resp. second) kind. The first two terms of the above expression correspond to the homogeneous solution $\psi_h$ where we have set $U = -c_1T$ and $S=-c_2T$.
%
\subsection{Convergence studies}
%
\paragraph{Error measures} We consider two different error measures for both the poloidal flux $\psi$ and its gradient $\nabla\psi$. First, for a function $f$ and its HDG approximation $f_h$ we define the usual mean square error as
\[
E_2(f) := \left(\|f-f_h\|^2_{L^2(\Omega^h)} + \|f-f_h\|^2_{L^2(\Omega^h_{ext})}\right)^{1/2}.
\]
This expression includes the error contributions from the computational domain $\Omega_h$ and from the corresponding exterior region  $\Omega^h_{ext}$-- where the value of $\boldsymbol q_h$ is defined by the extrapolation in \eqref{eq:extension} and the value of $\psi_h$ by the transference defined in \eqref{eq:trasnfer_psih}. 

In many applications, the output of a plasma  equilibrium code is not the final result of a simulation, but only one part of a much bigger calculation involving different physical processes simulated by several computational codes coupled together. In these cases the approximation of the magnetic field obtained is sampled and used as input for the computation of other physical processes of interest. The points where the approximation is evaluated may or may not coincide with nodes of the original mesh and therefore it is important to have a measure of the point-wise convergence of the approximation. As an estimate of the point-wise accuracy of the algorithm we sample the solution at random points in the interior of $\Omega$ and consider the maximum discrepancy with the analytic solution as a measure of accuracy. This leads to the definition
\[
E_\infty(f) := \max_{\boldsymbol x \in S}{|f(\boldsymbol x)-f_h(\boldsymbol x)|},
\]
where $S$ is comprised of five random points from every element of the discretization of both $\Omega^h$ and $\Omega^h_{ext}$. 
%
\paragraph{Stopping criterion} In all the following examples the stopping criterion for the iterative algorithm was set to a relative difference of $10^{-12}$ or below between two consecutive iterations. This effectively sets the maximum possible accuracy of the algorithm.
\paragraph{Refinement strategies} In the numerical experiments both mesh and polynomial refinement (commonly referred to as $h$ and $p$ refinements respectively) tests are considered. The computational grids are built following the steps given in Section \ref{sec:curved}. Starting from a  background shape-regular triangulation with diameter $h$, the associated computational domain $\Omega^h$ and the region $\Omega^h_{ext}$ are determined for every geometry. Mesh refinement is performed by uniform subdivision of the background mesh, which results in a finer mesh with diameter $h/2$; the computational domain and the extension are updated accordingly. For the cases requiring two-grid iterations, the coarse and fine meshes correspond to adjacent levels of refinements with diameters $h/2^n$ and $h/2^{n+1}$. In the case of $p$-refinements, the grids mentioned above were held fixed and the refinement was carried out by increasing the polynomial degree of the approximation space.

\begin{figure}
\centering
\begin{tabular}{cccc}
\includegraphics[width=0.26\linewidth]{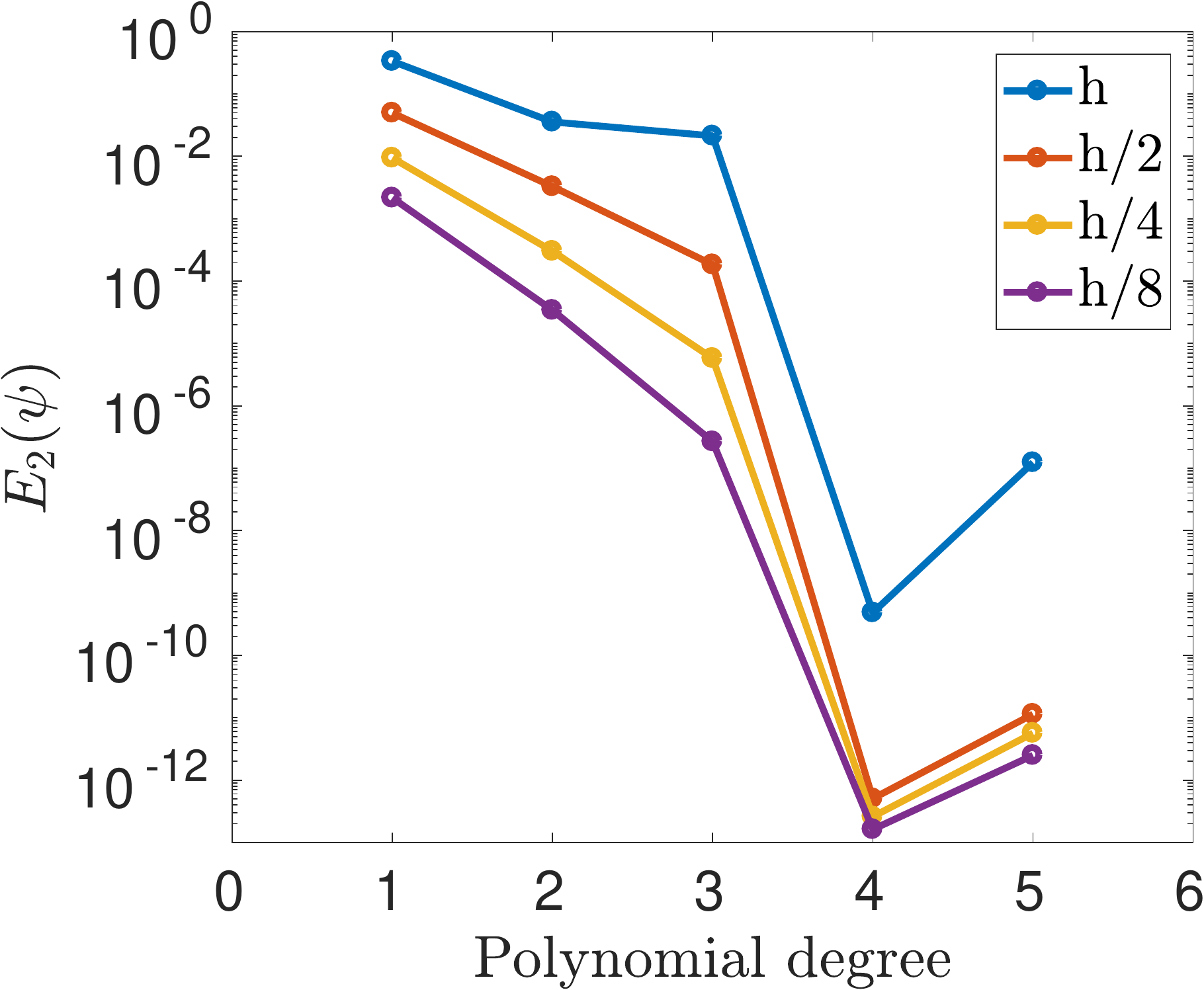} \quad & \includegraphics[width=0.26\linewidth]{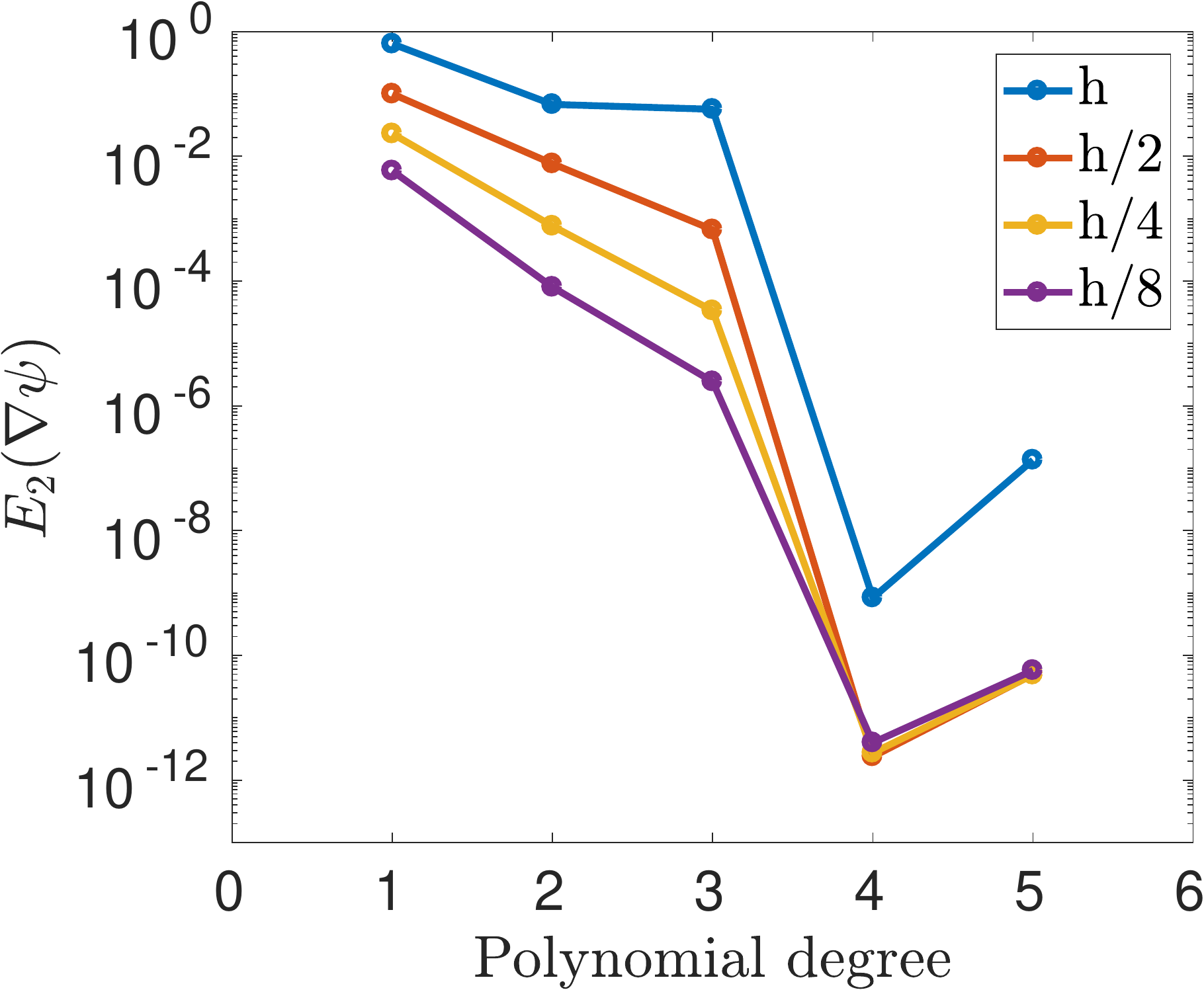} \quad & 
\multirow{5}{*}[\dimexpr1.25in]{\includegraphics[width=0.215\linewidth]{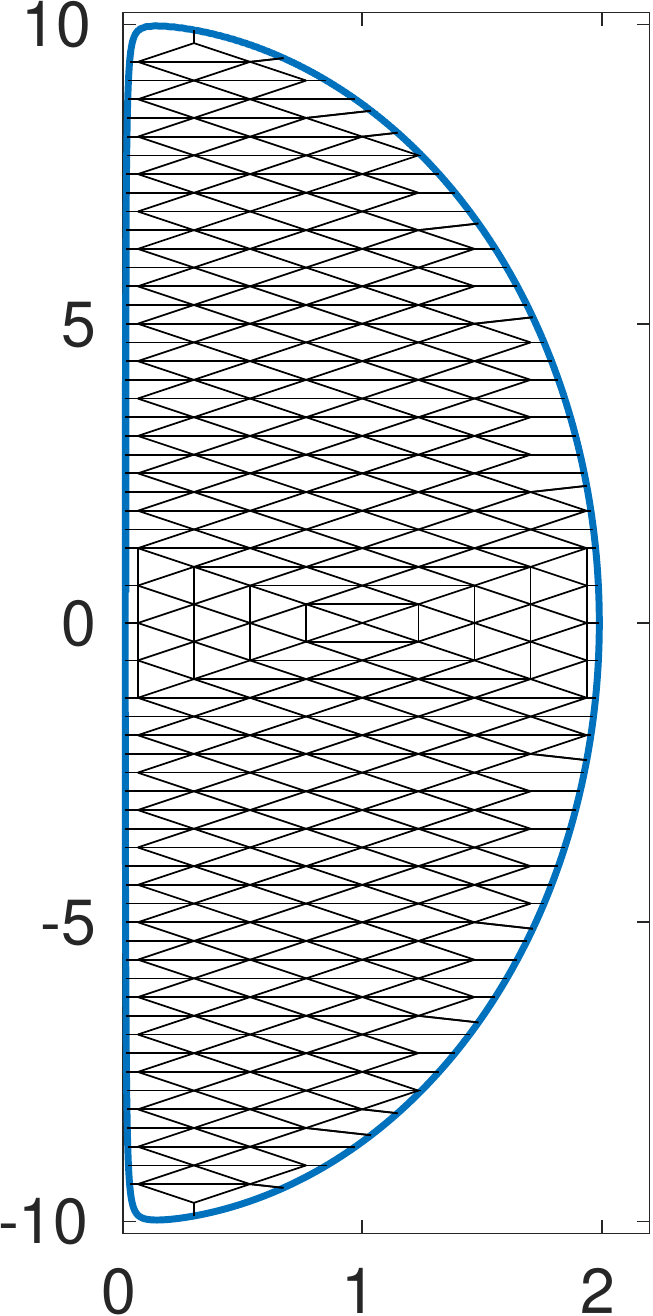}} &
\multirow{5}{*}[\dimexpr1.28in]{\includegraphics[width=0.085\linewidth]{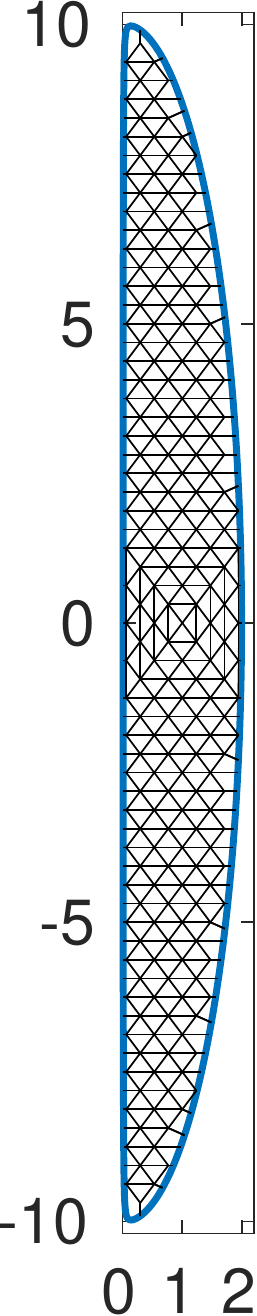}}\\ 
\includegraphics[width=0.26\linewidth]{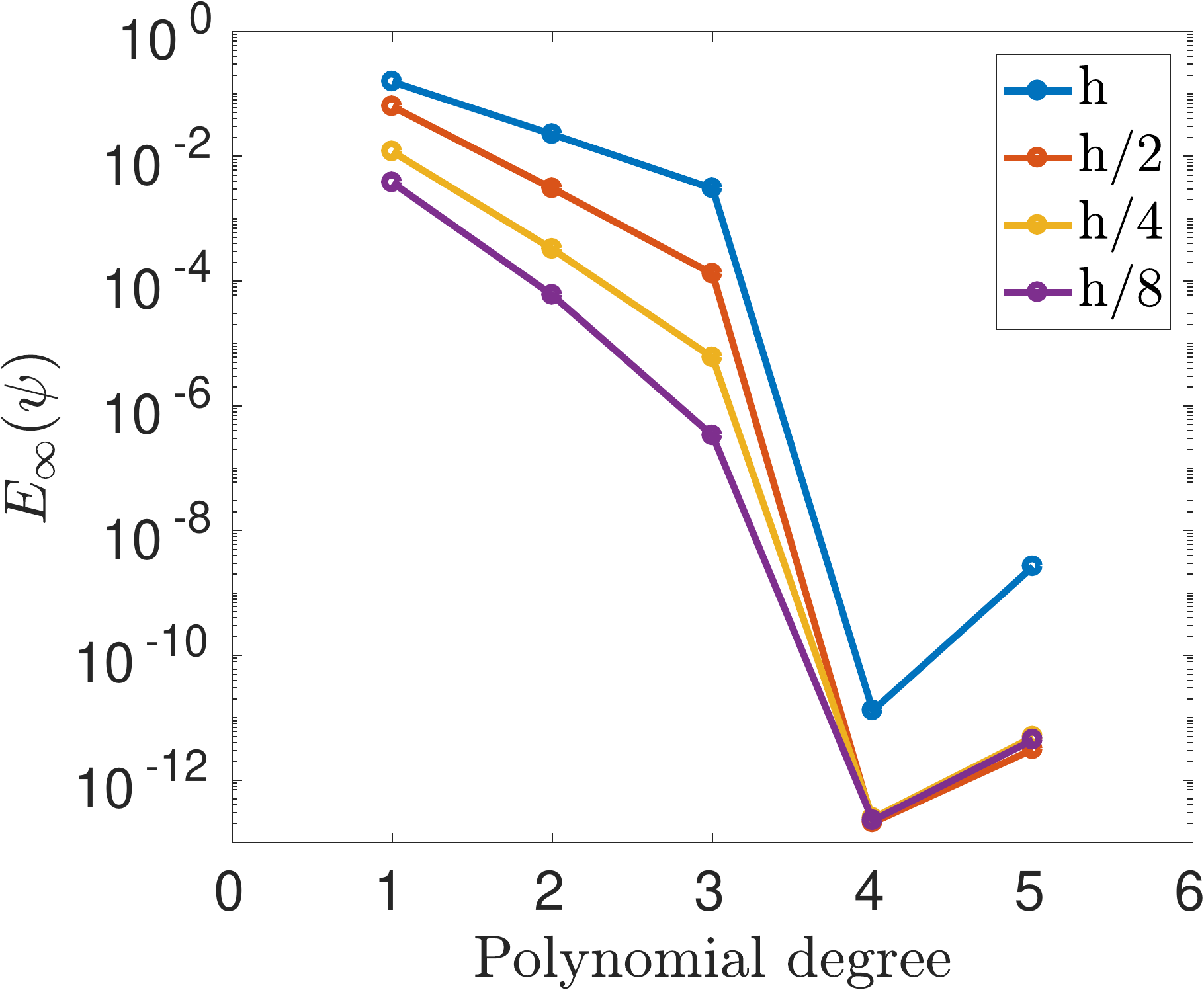} \quad & \includegraphics[width=0.26\linewidth]{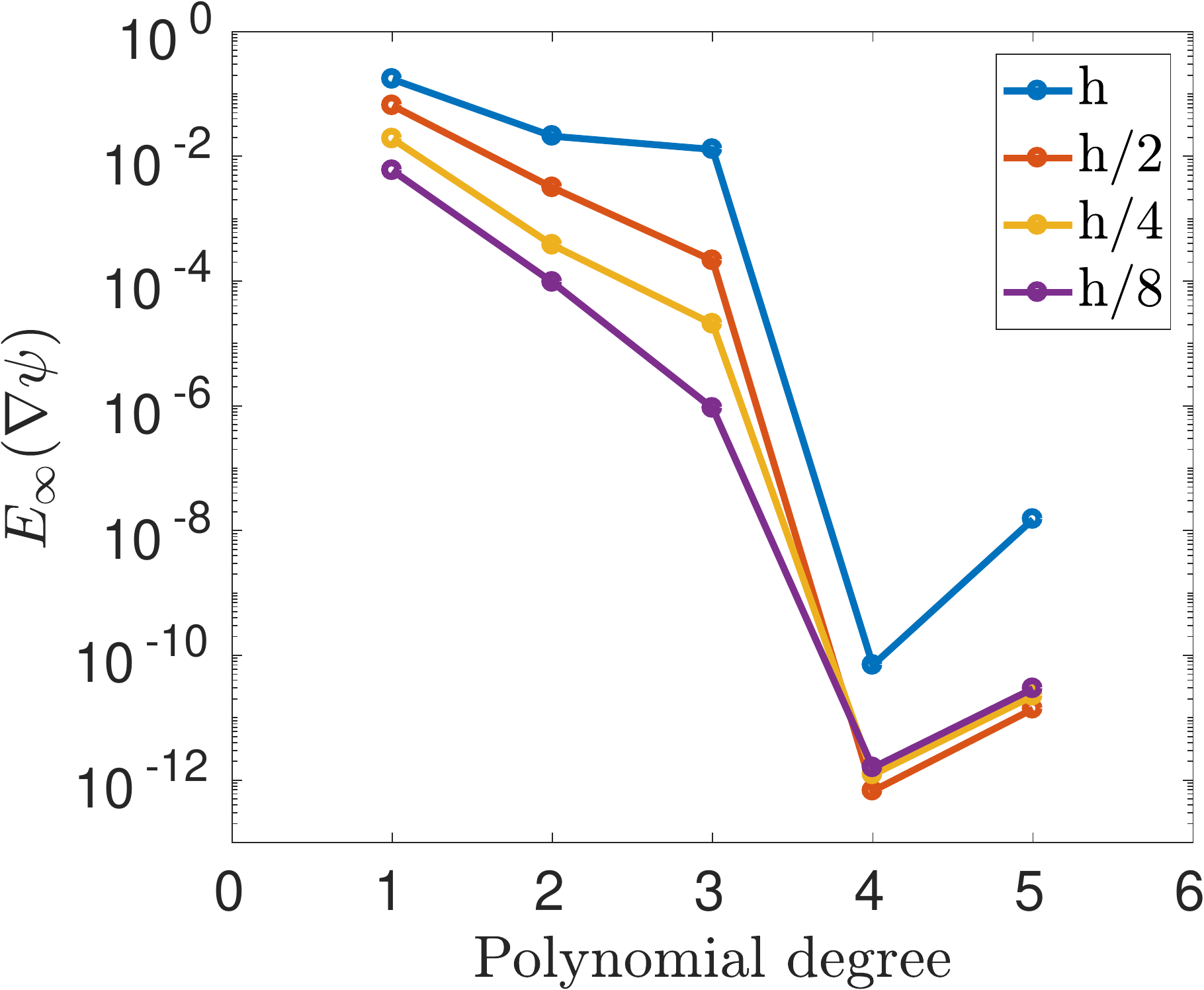} \quad & 
\end{tabular}
\caption{{\footnotesize Convergence plots for Example 1 (FRC) for successive refinements of the computational grid and increasingly higher polynomial degrees. Left column: $E_2$ (top) and $E_\infty$ (bottom) errors for the poloidal flux $\psi$. Center column: $E_2$ (top) and $E_\infty$ (bottom) errors for $\nabla\psi$. Right column: Confinement region and sample grid corresponding to the second level of refinement (red curve in the convergence plot). The geometry is shown on 1:1 scale on the right end to stress the elongation of the domain. The reader is referred to the on-line version of the manuscript for the color scheme.}}\label{fig:frc}
\end{figure}

\begin{figure}
\centering
\begin{tabular}{ccc}
\includegraphics[width=0.26\linewidth]{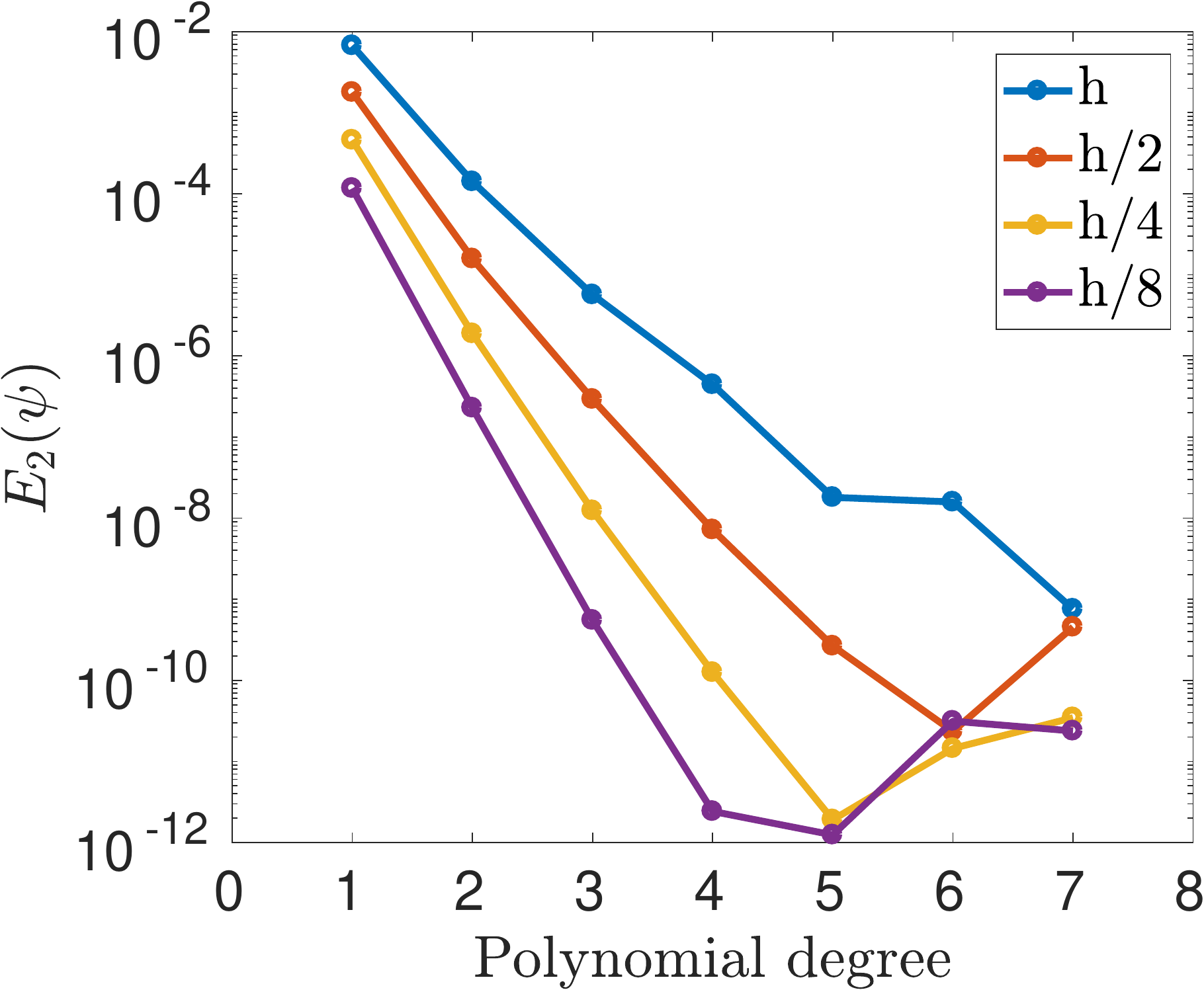} \quad & \includegraphics[width=0.26\linewidth]{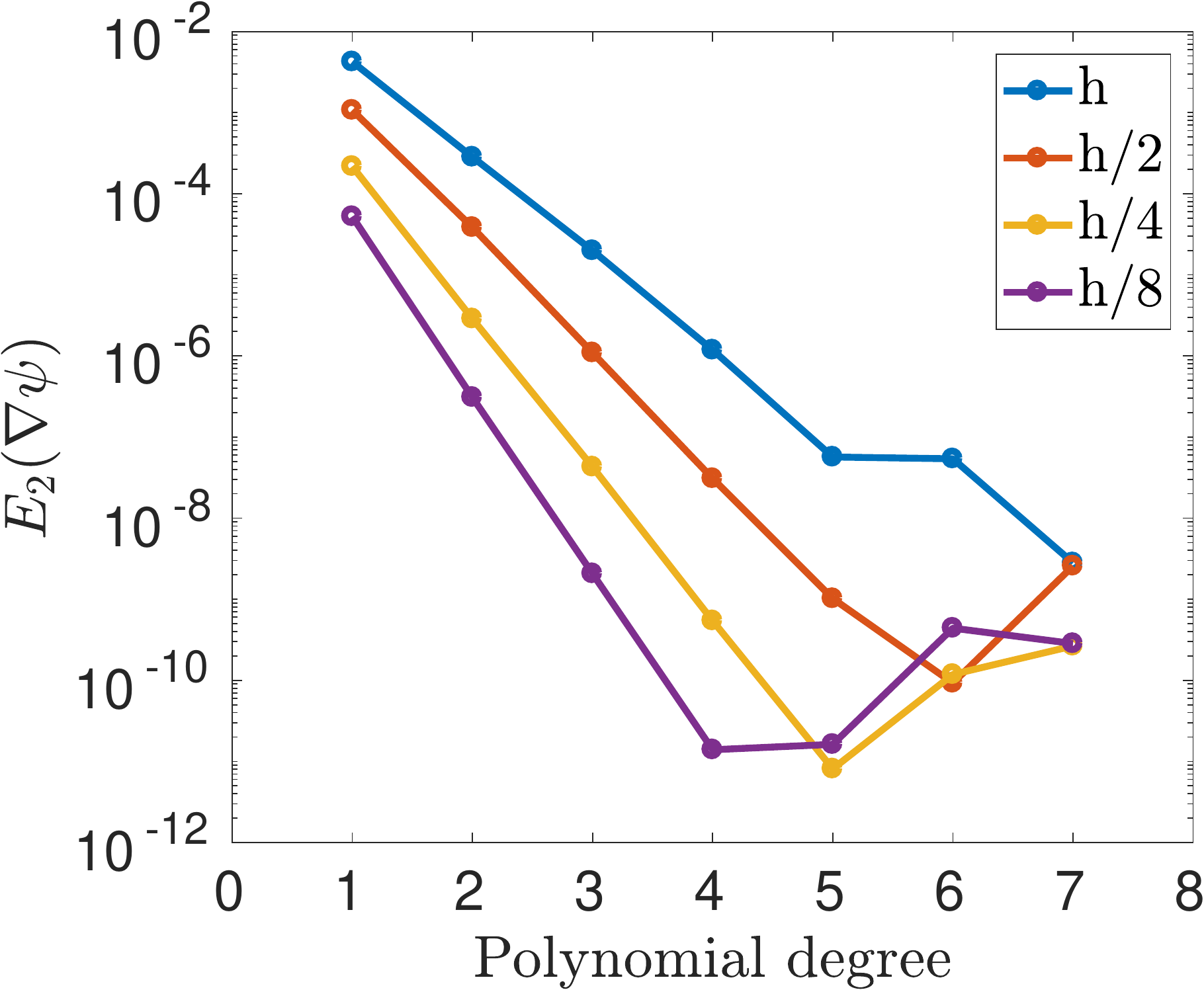} \quad & 
\multirow{5}{*}[\dimexpr1.225in]{\includegraphics[width=0.3\linewidth]{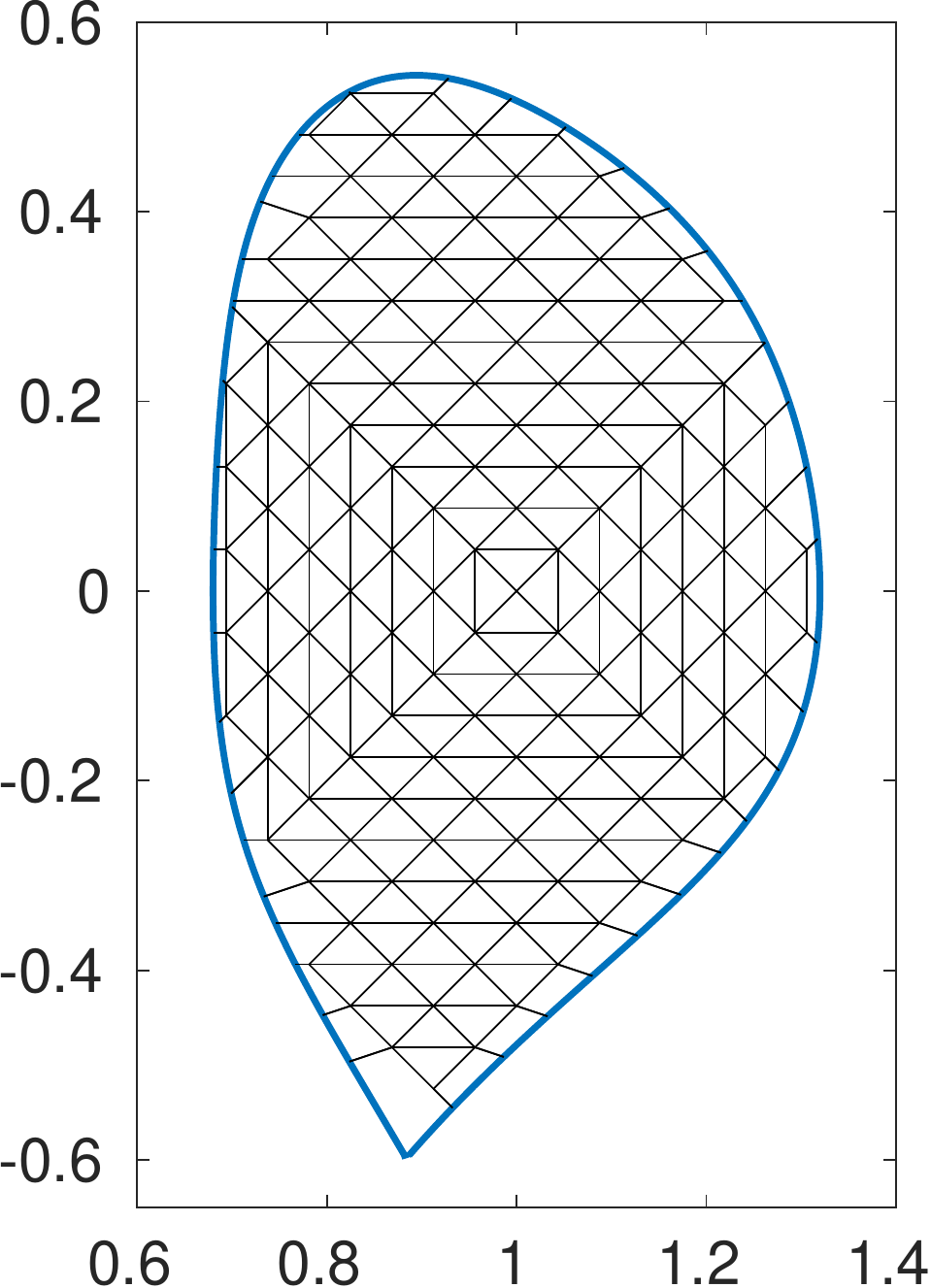}} \\ 
\includegraphics[width=0.26\linewidth]{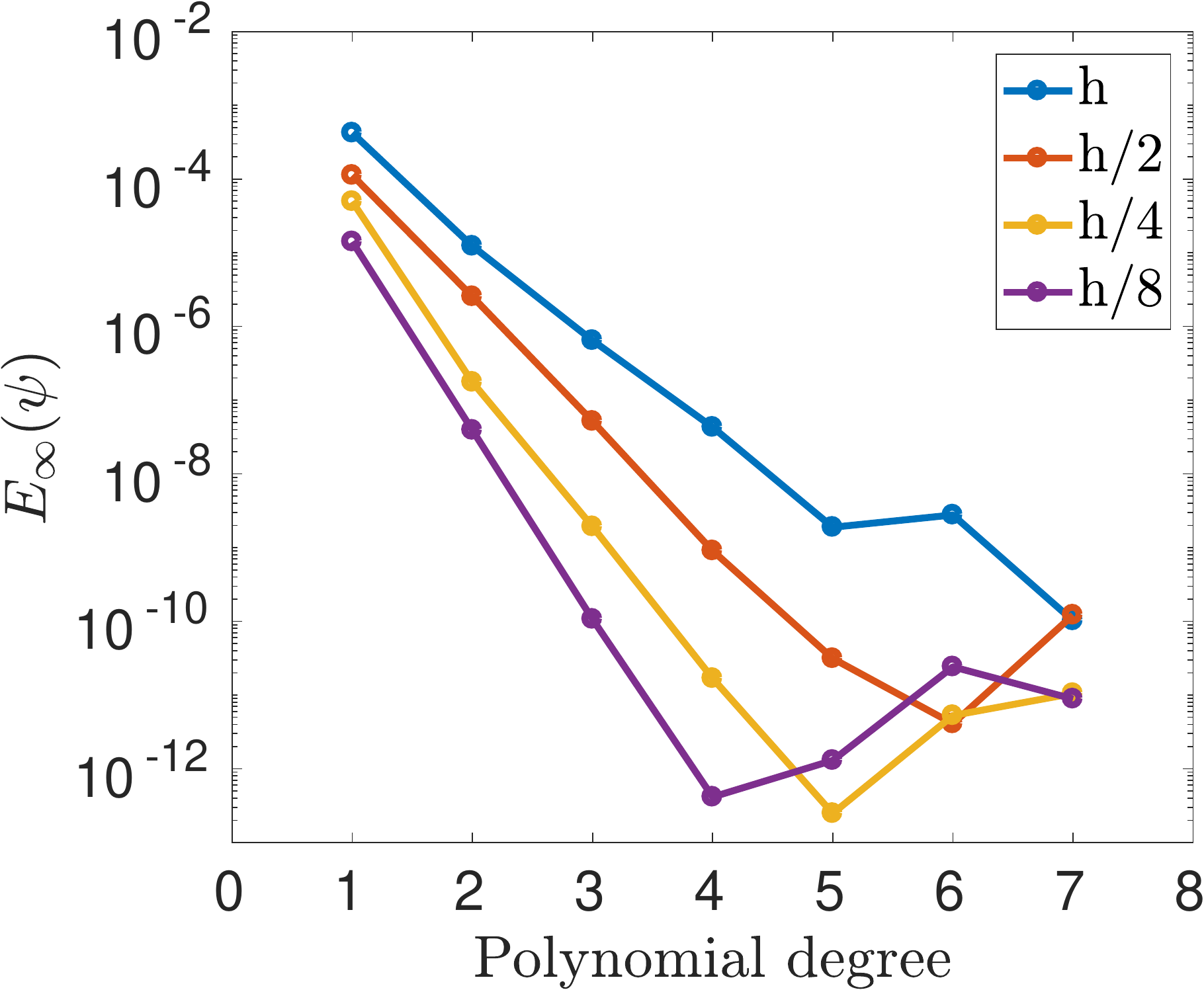} \quad & \includegraphics[width=0.26\linewidth]{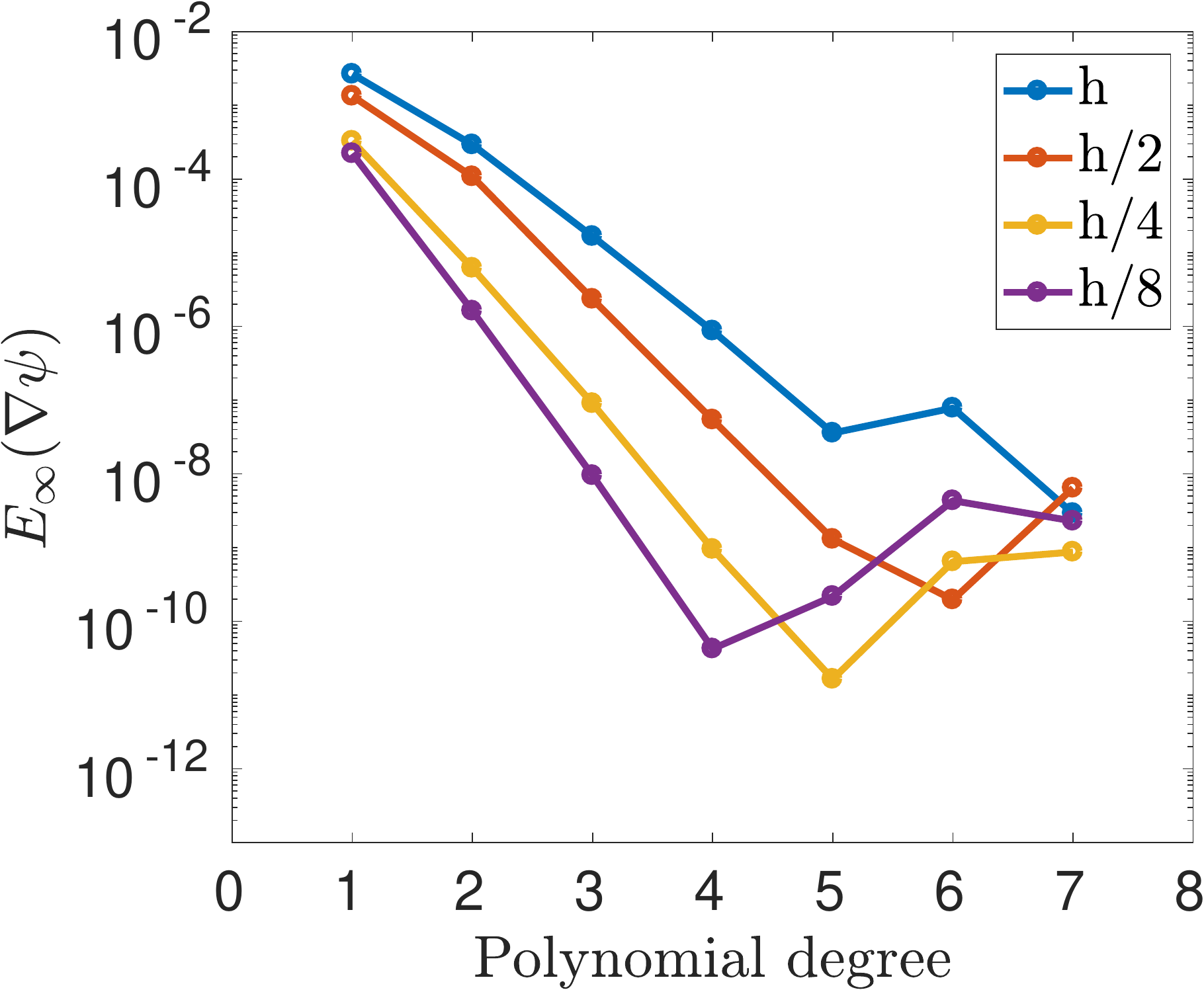} \quad & 
\end{tabular}
\caption{{\footnotesize Convergence plots for Example 2 (ITER) for successive refinements of the computational grid and increasingly higher polynomial degrees. Leftt column: $E_2$ (top) and $E_\infty$ (bottom) errors for the poloidal flux $\psi$. Center column: $E_2$ (top) and $E_\infty$ (bottom) errors for $\nabla\psi$. Right column: Confinement region and sample grid corresponding to the second level of refinement (red curve in the convergence plot). The reader is referred to the on-line version of the manuscript for the color scheme.}}\label{fig:iter}
\end{figure}

\begin{figure}
\centering
\begin{tabular}{ccc}
\includegraphics[width=0.26\linewidth]{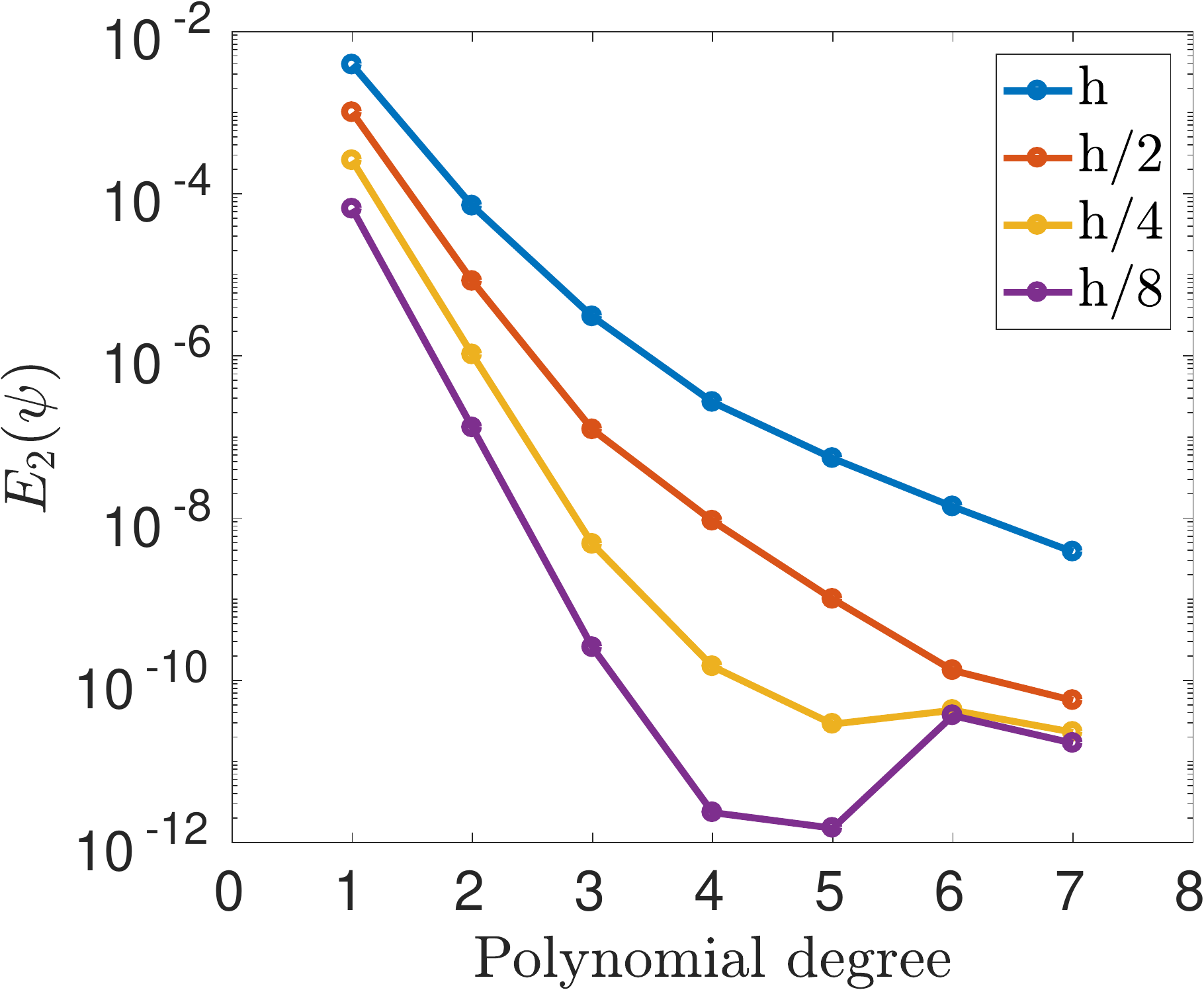} \quad & \includegraphics[width=0.26\linewidth]{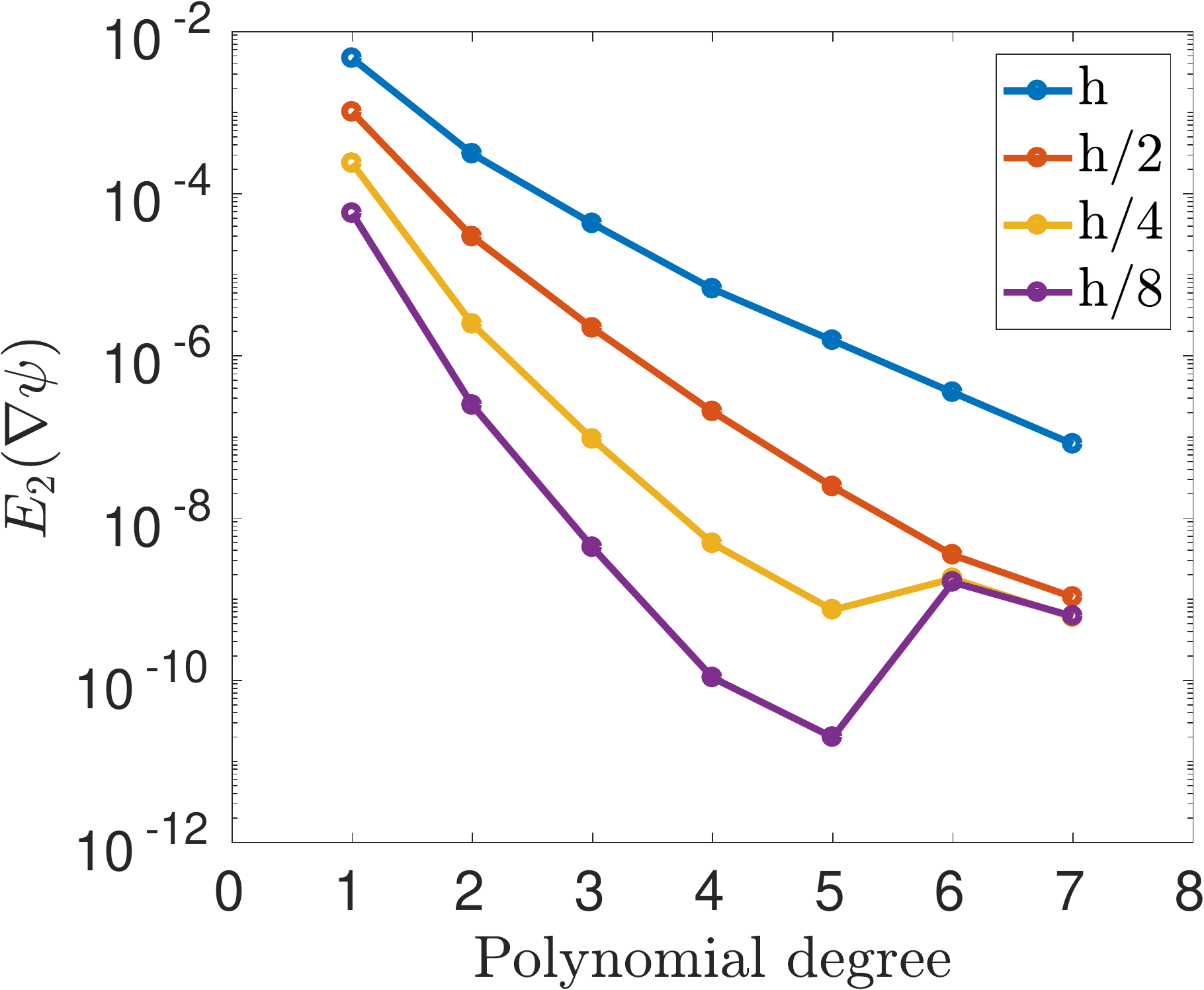} \quad & 
\multirow{5}{*}[\dimexpr 1.2in]{\includegraphics[width=0.28\linewidth]{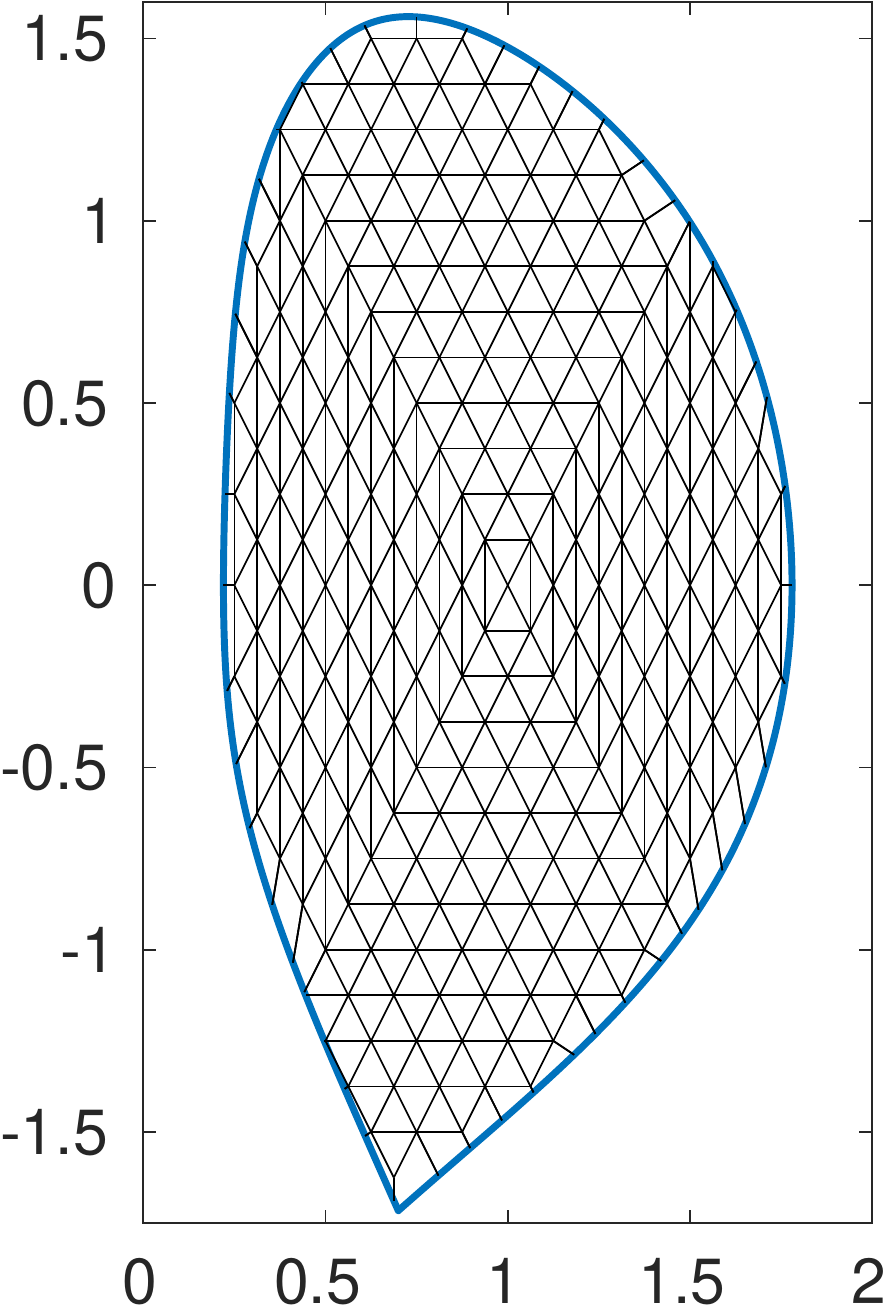}} \\ 
\includegraphics[width=0.26\linewidth]{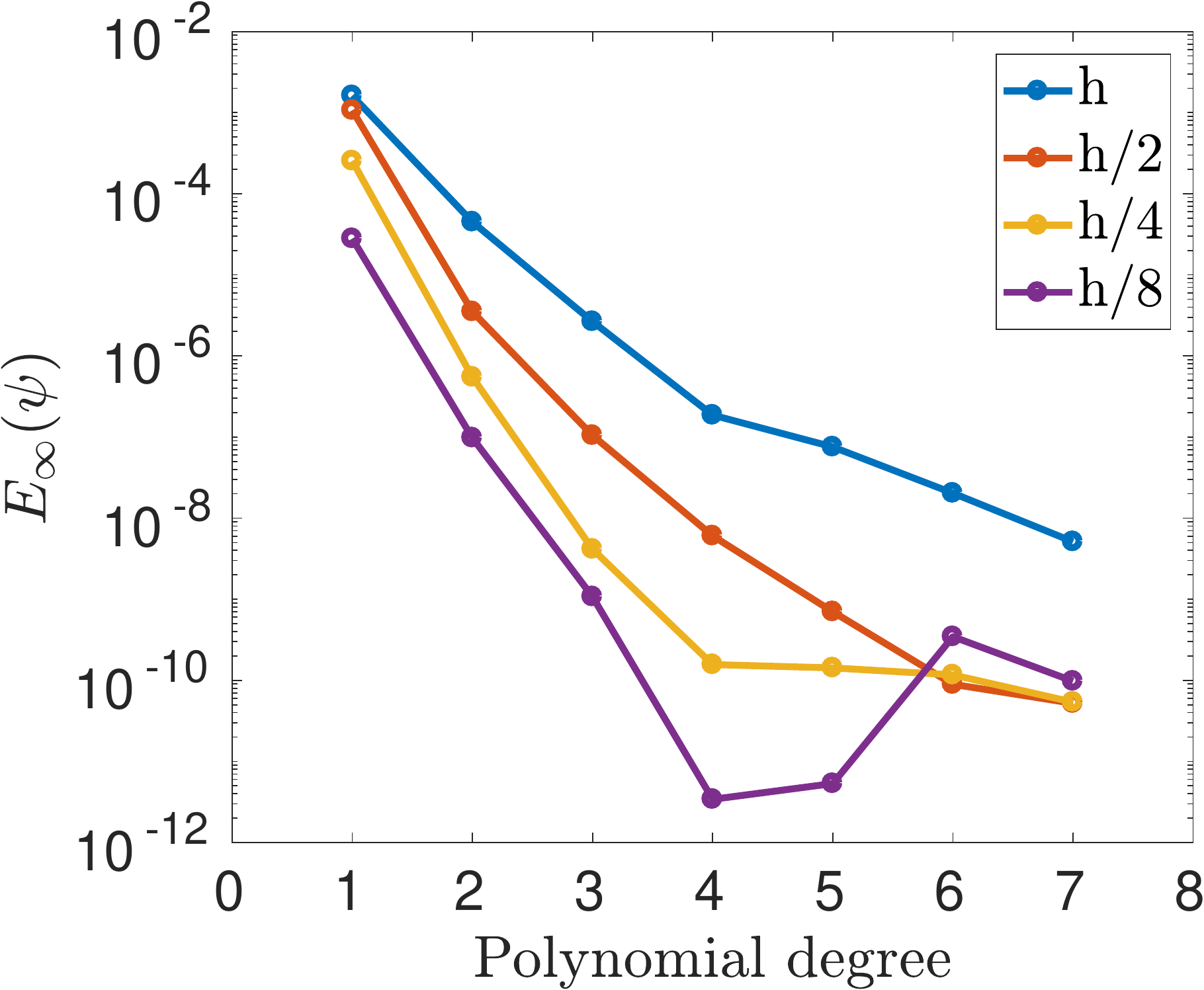} \quad & \includegraphics[width=0.26\linewidth]{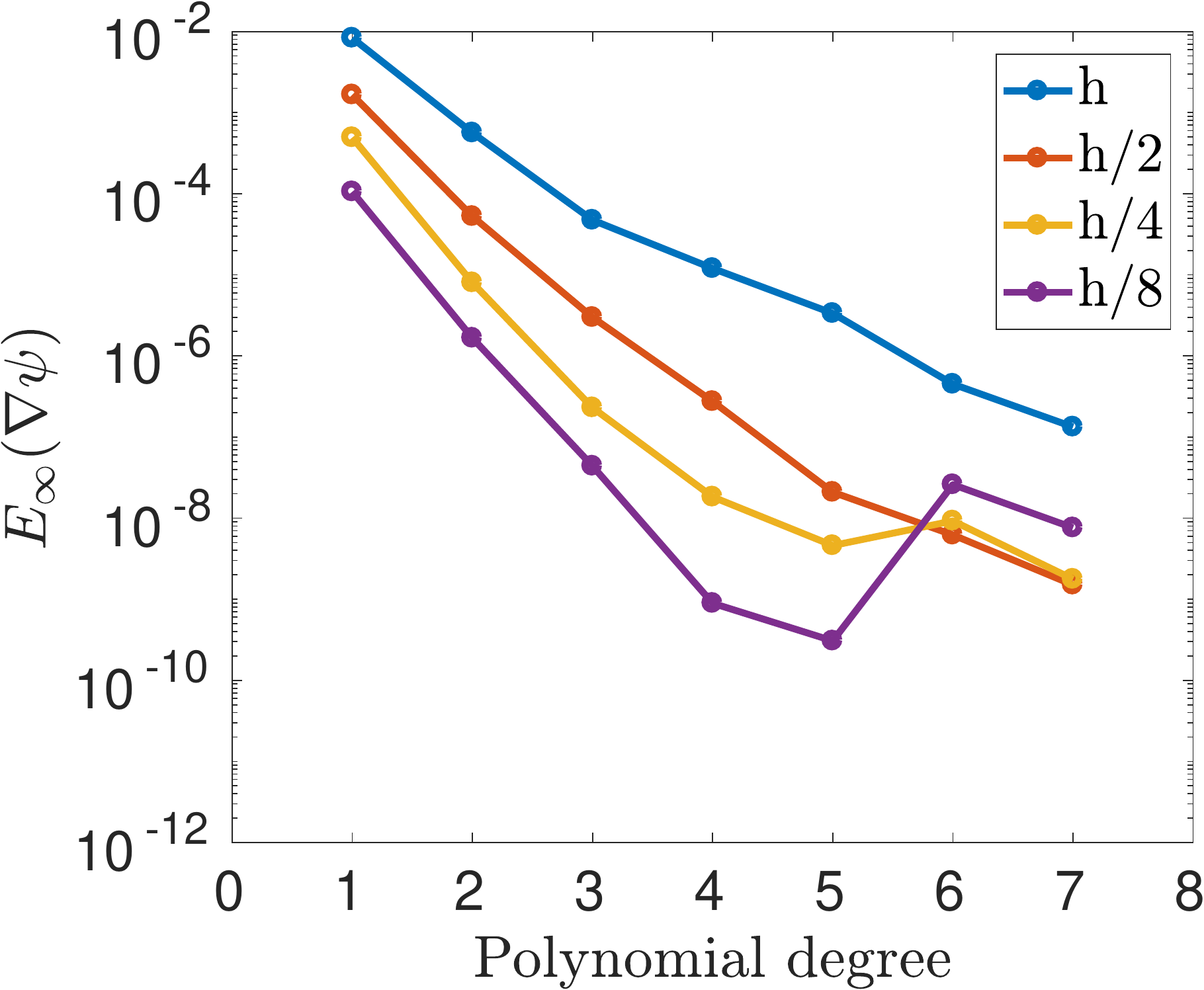} \quad & 
\end{tabular}
\caption{{\footnotesize Convergence plots for Example 3 (NSTX) for successive refinements of the computational grid and increasingly higher polynomialonclusion degrees. Left column: $E_2$ (top) and $E_\infty$ (bottom) errors for the poloidal flux $\psi$. Center column: $E_2$ (top) and $E_\infty$ (bottom) errors for $\nabla\psi$. Right column: Confinement region and sample grid corresponding to the second level of refinement (red curve in the convergence plot). The reader is referred to the on-line version of the manuscript for the color scheme.}}\label{fig:nstx}
\end{figure}

\begin{figure}
\centering
\begin{tabular}{ccc}
\includegraphics[width=0.26\linewidth]{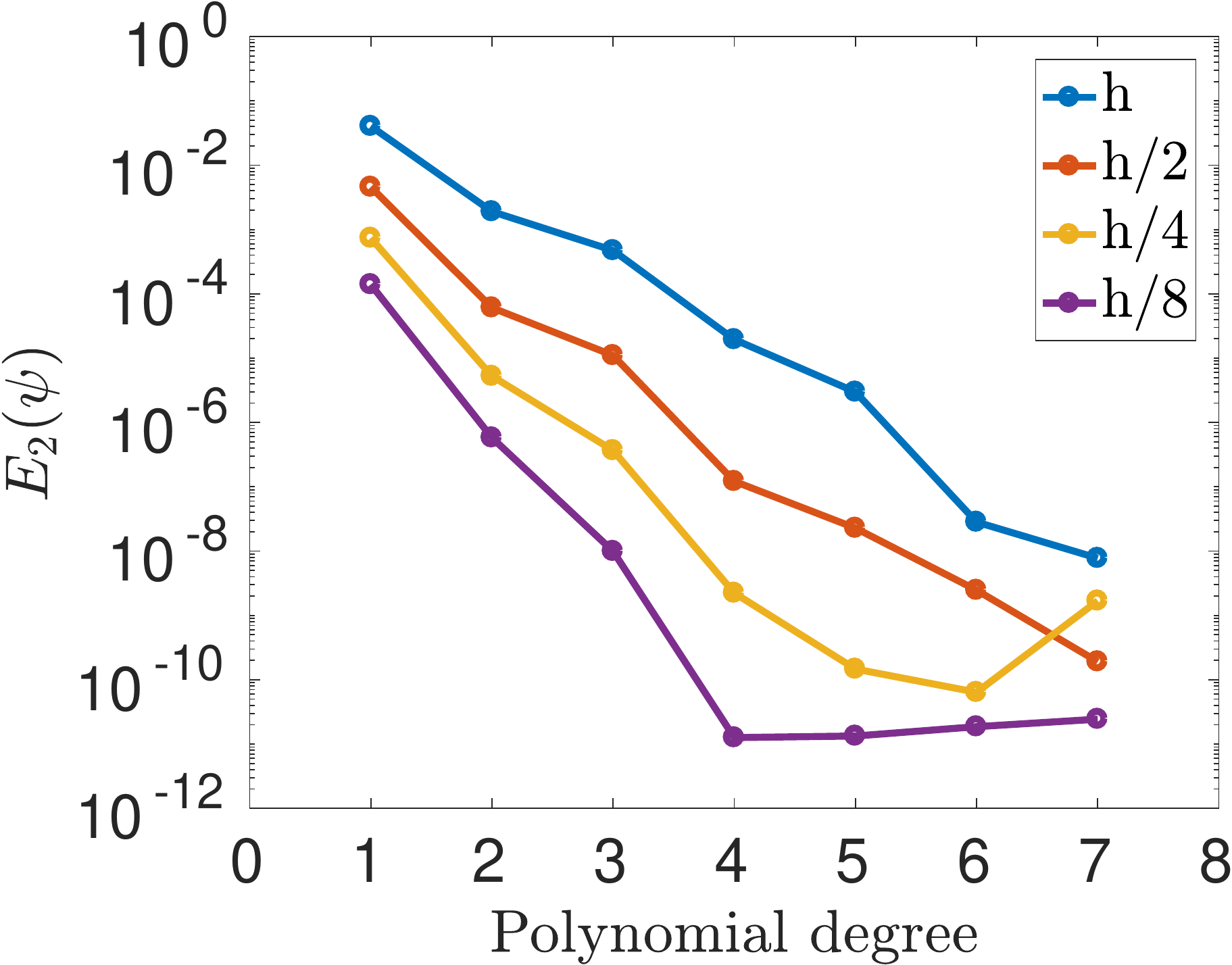} \quad & \includegraphics[width=0.26\linewidth]{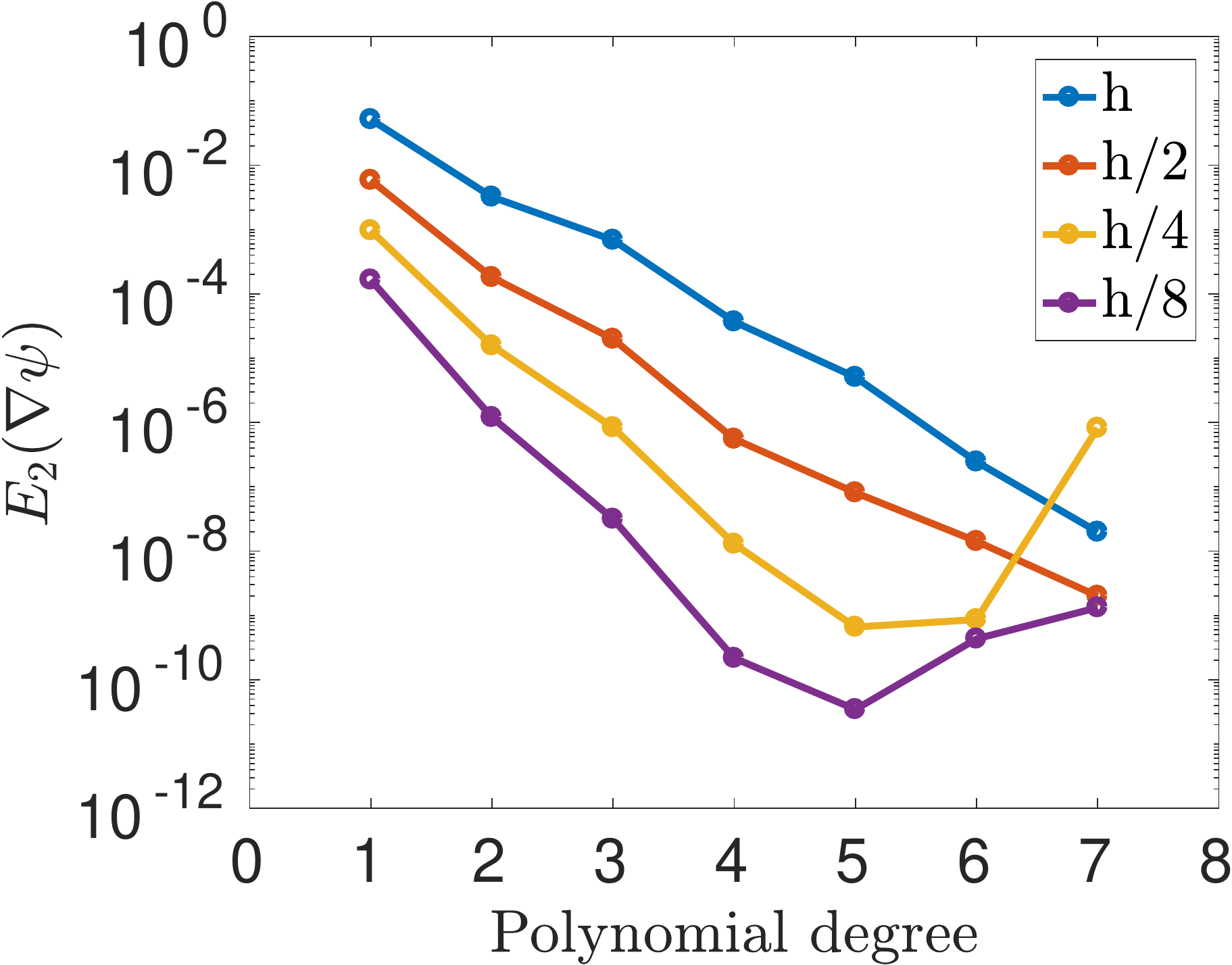} \quad & 
\multirow{5}{*}[\dimexpr 1.2in]{\includegraphics[width=0.32\linewidth]{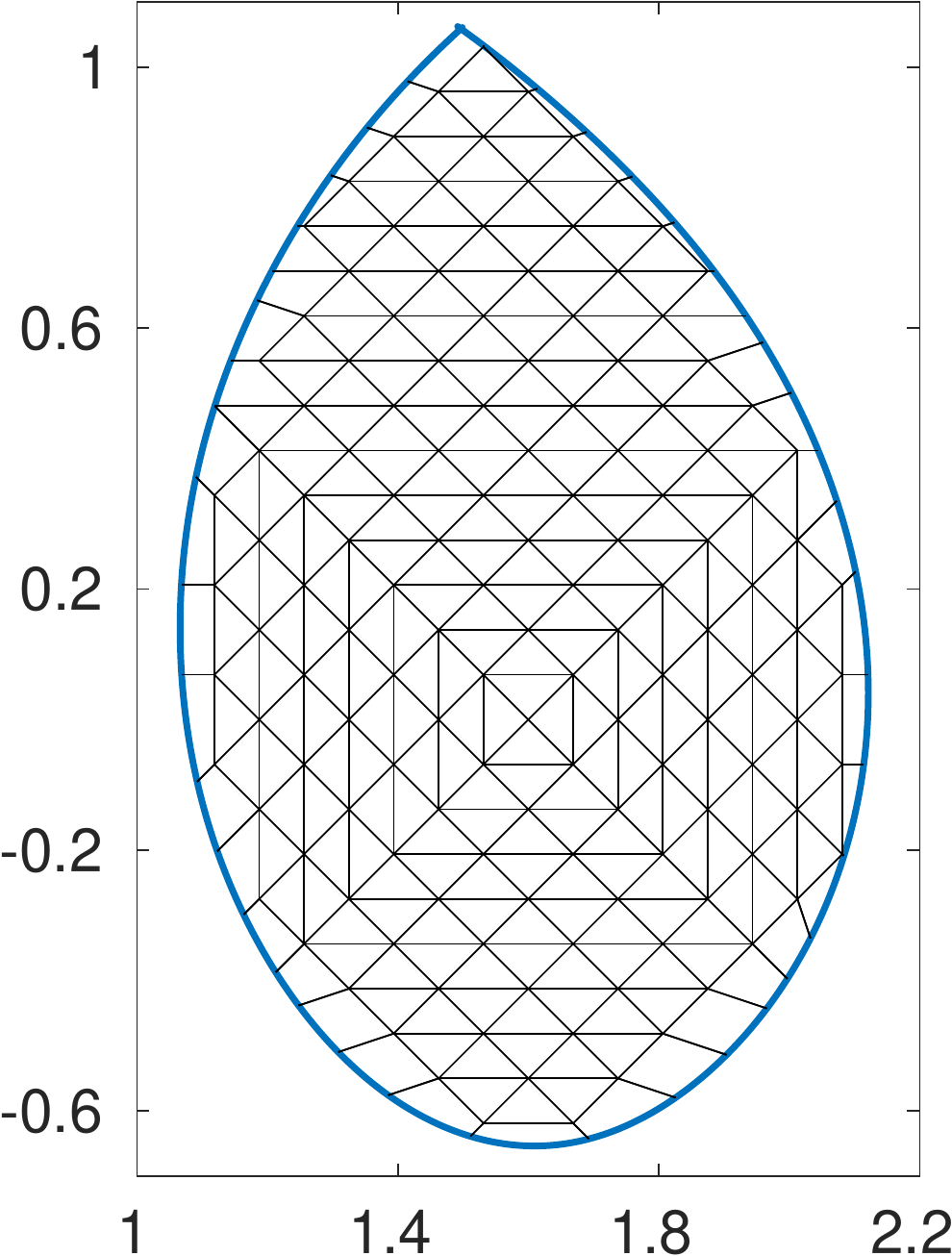}} \\ 
\includegraphics[width=0.26\linewidth]{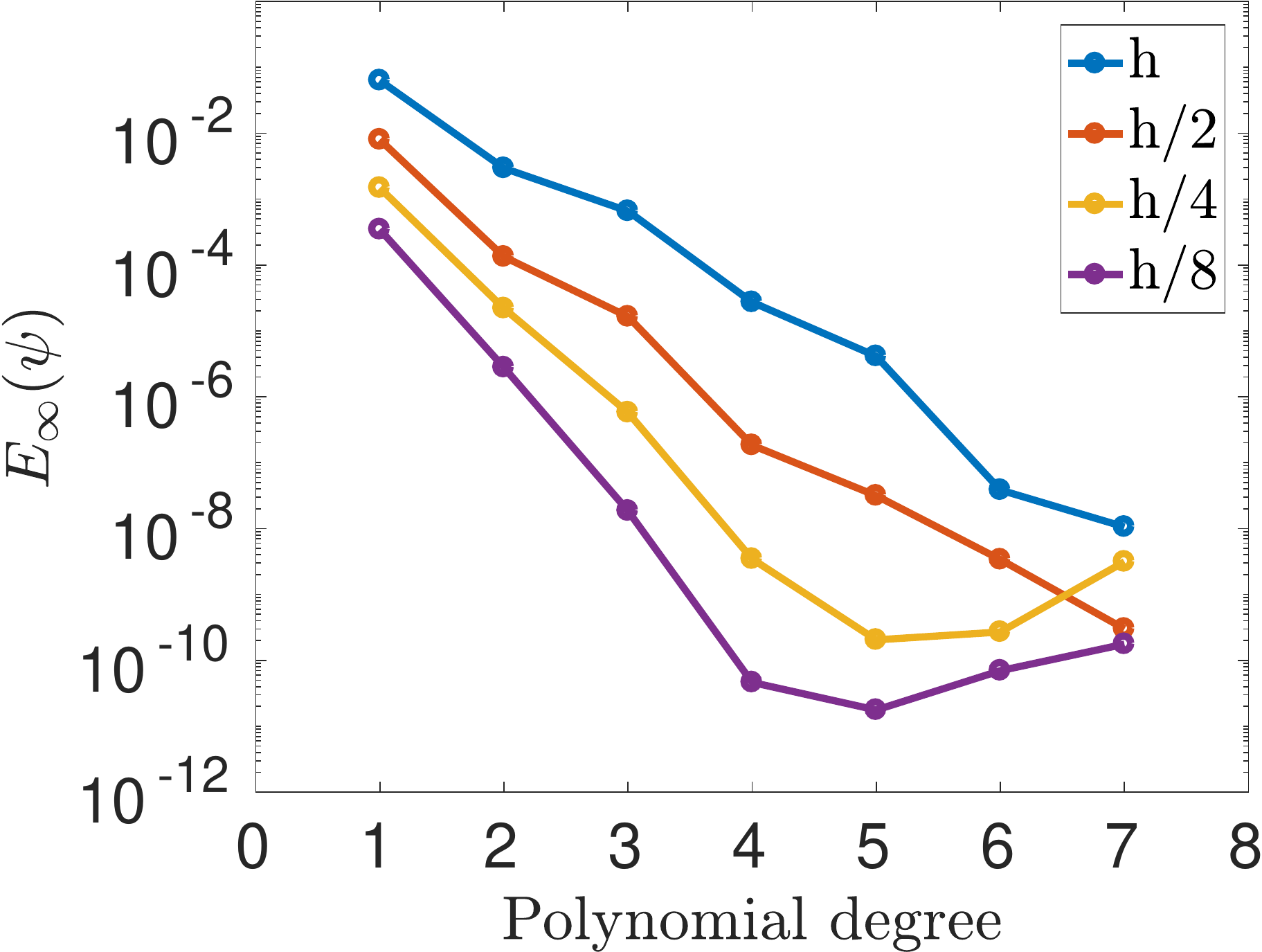} \quad & \includegraphics[width=0.26\linewidth]{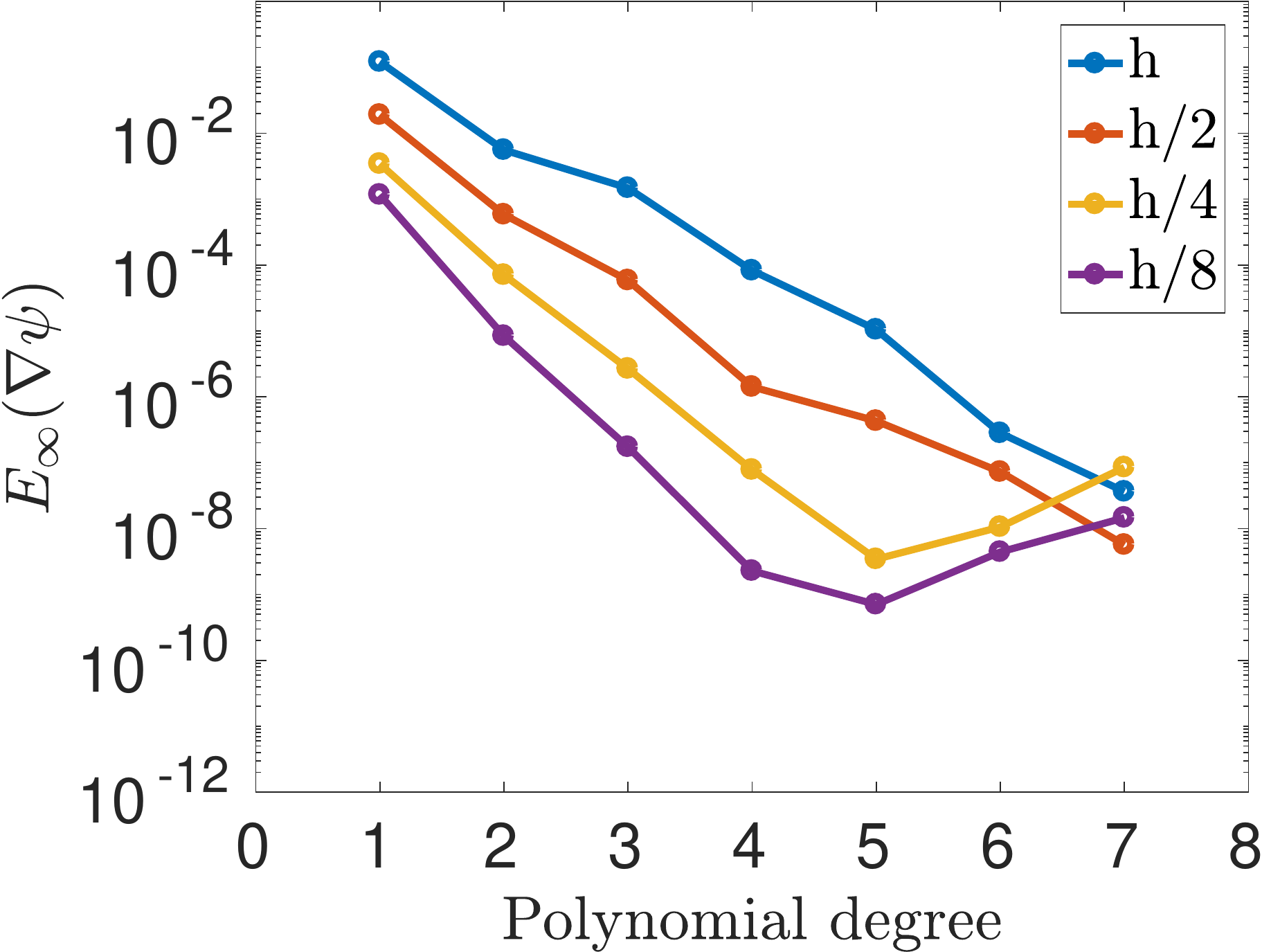} \quad & 
\end{tabular}
\caption{{\footnotesize Convergence plots for Example 4 (ASDEX upgrade with an upwards oriented x-point) for successive refinements of the computational grid and increasingly higher polynomial degrees. Left column: $E_2$ (top) and $E_\infty$ (bottom) errors for the poloidal flux $\psi$. Center column: $E_2$ (top) and $E_\infty$ (bottom) errors for $\nabla\psi$. Right column: Confinement region and sample grid corresponding to the second level of refinement (red curve in the convergence plot). The reader is referred to the on-line version of the manuscript for the color scheme.}}\label{fig:asdexup}
\end{figure}

\begin{figure}
\centering
\begin{tabular}{ccc}
\includegraphics[width=0.26\linewidth]{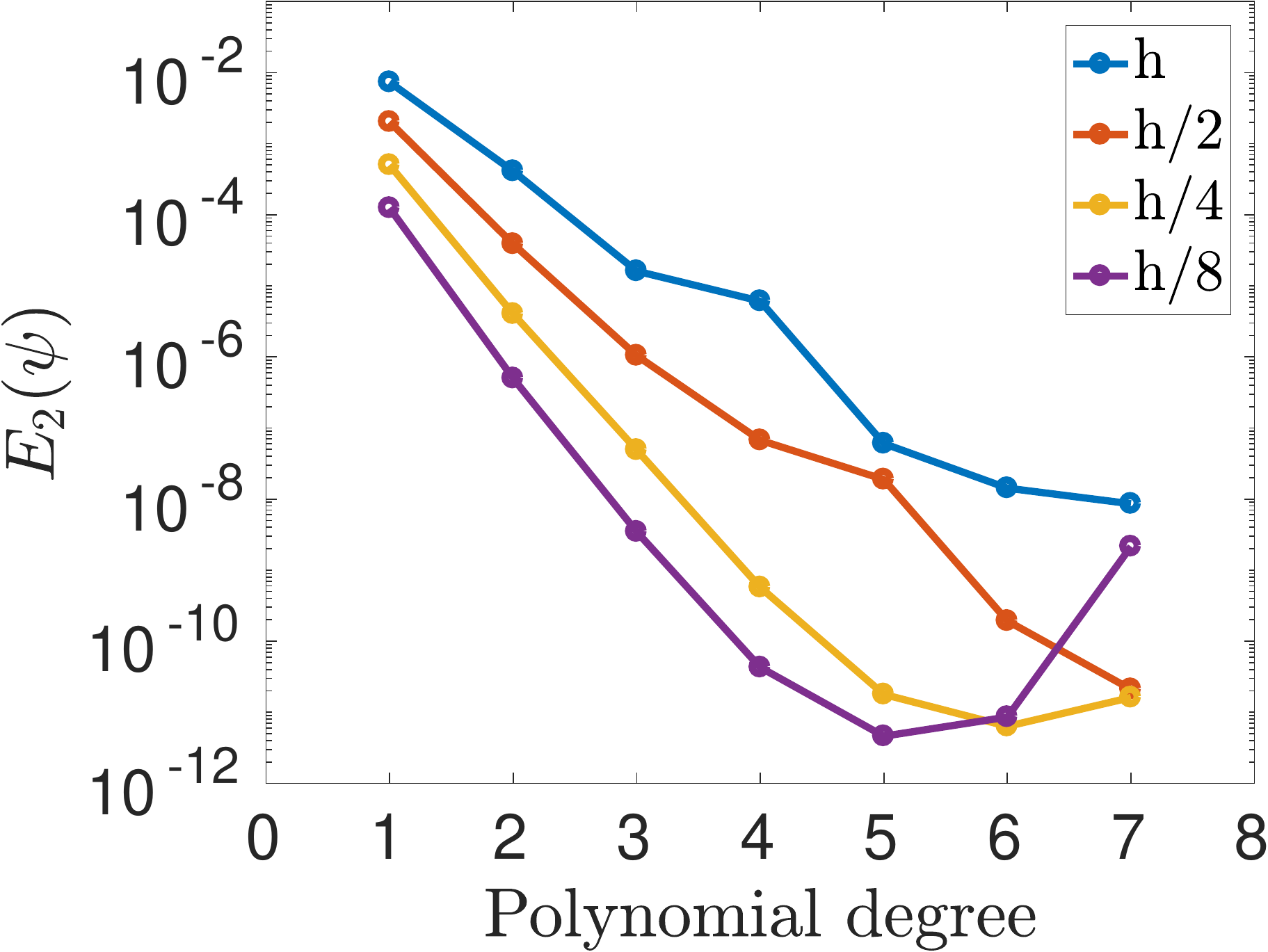} \quad & \includegraphics[width=0.26\linewidth]{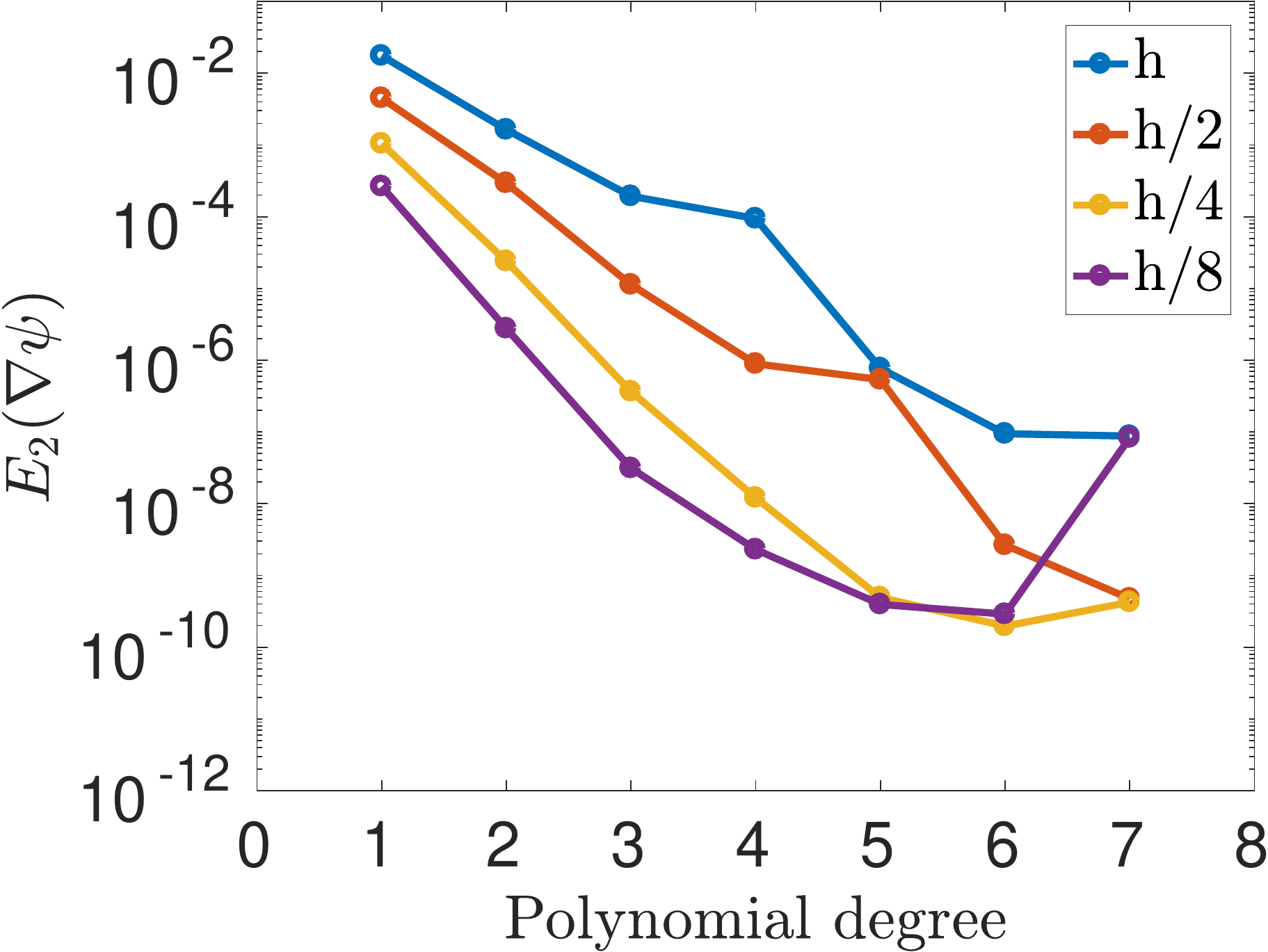} \quad & 
\multirow{5}{*}[\dimexpr 1.2in]{\includegraphics[width=0.27\linewidth]{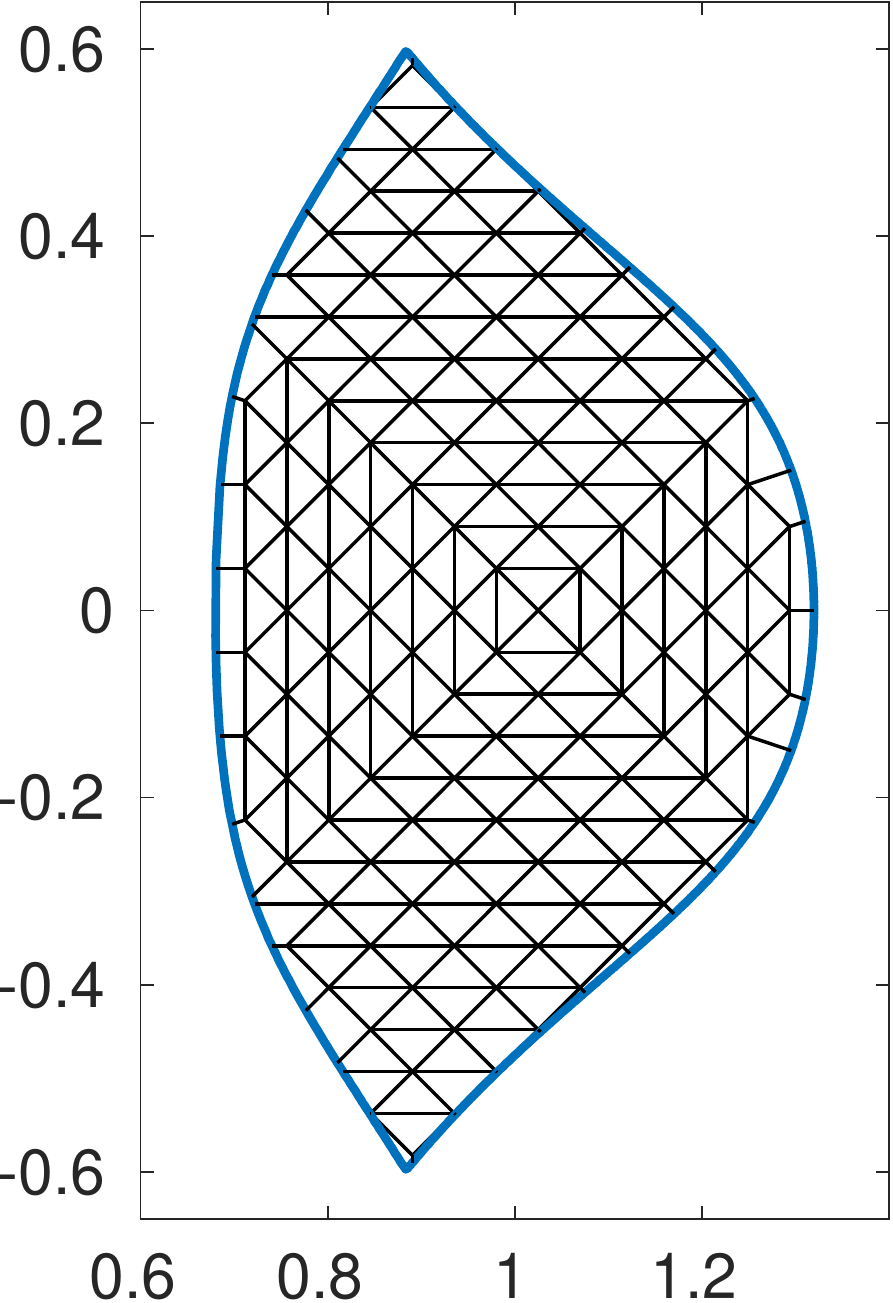}} \\ 
\includegraphics[width=0.26\linewidth]{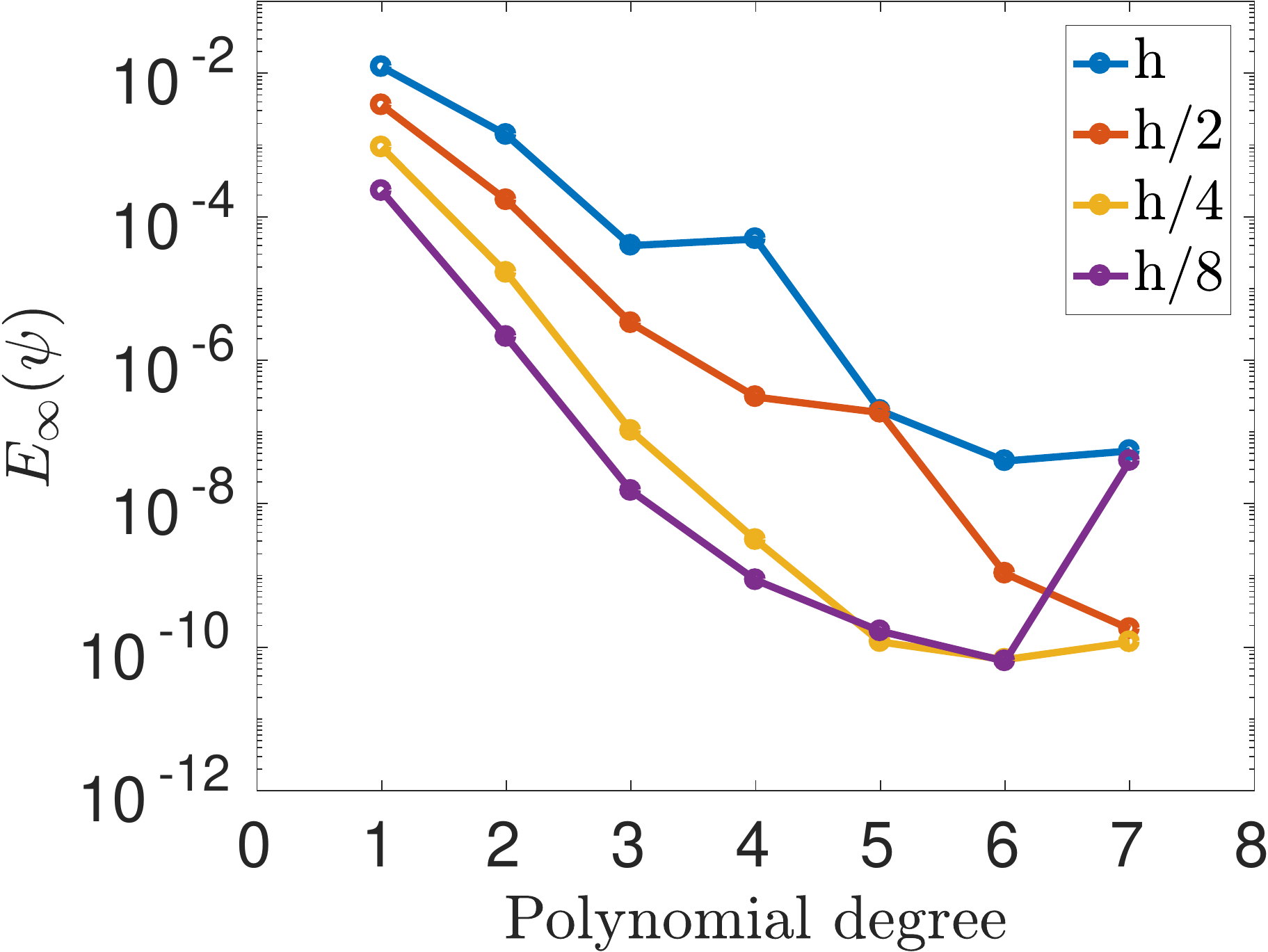} \quad & \includegraphics[width=0.26\linewidth]{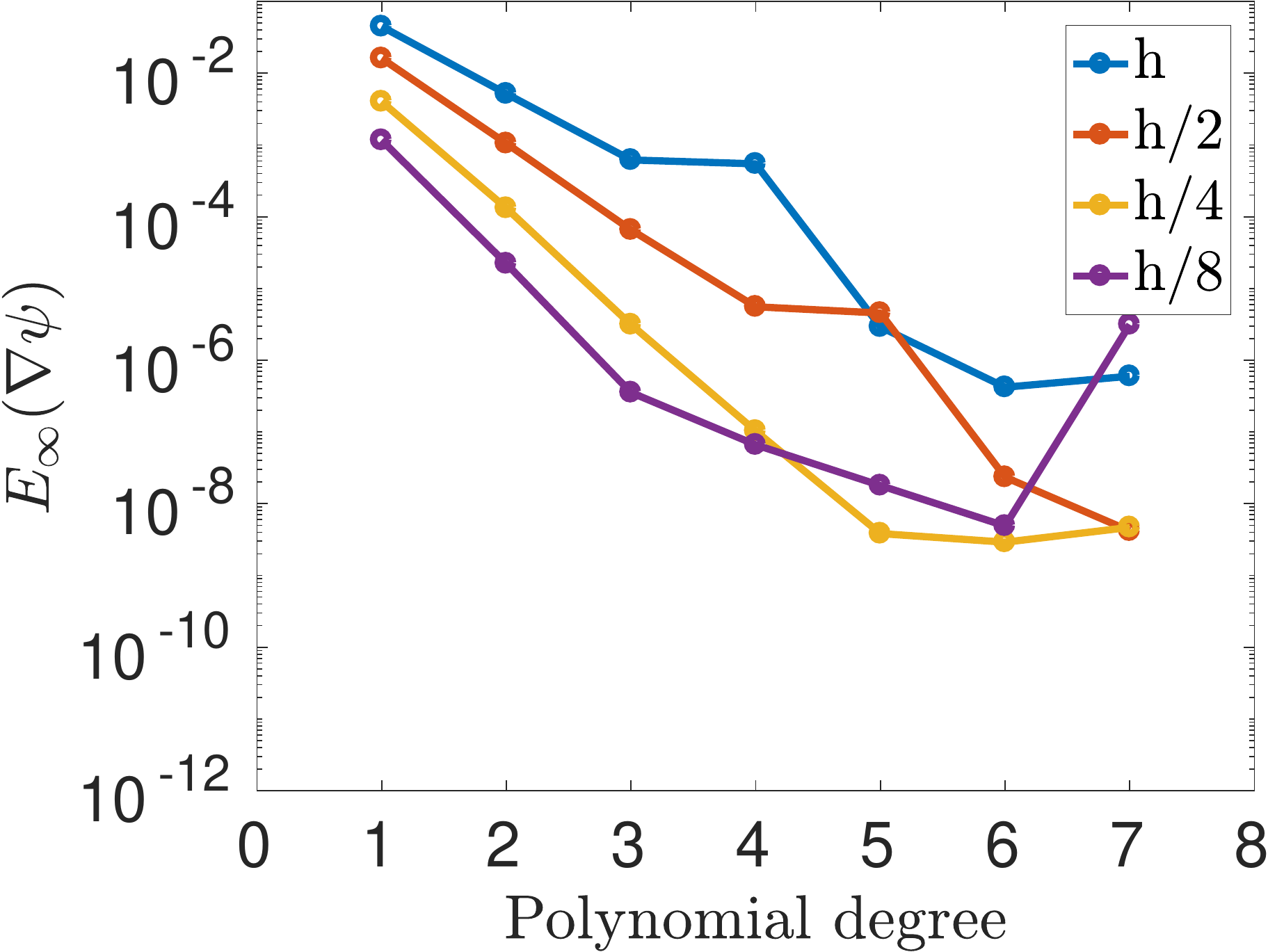} \quad & 
\end{tabular}
\caption{{\footnotesize Convergence plots for Example 5 (ITER with a double null divertor) for successive refinements of the computational grid and increasingly higher polynomial degrees. Left column: $E_2$ (top) and $E_\infty$ (bottom) errors for the poloidal flux $\psi$. Center column: $E_2$ (top) and $E_\infty$ (bottom) errors for $\nabla\psi$. Right column: Confinement region and sample grid corresponding to the second level of refinement (red curve in the convergence plot). The reader is referred to the on-line version of the manuscript for the color scheme.}\label{fig:nonlinear}}
\end{figure}

\begin{figure}
\centering
\begin{tabular}{ccc}
\includegraphics[width=0.26\linewidth]{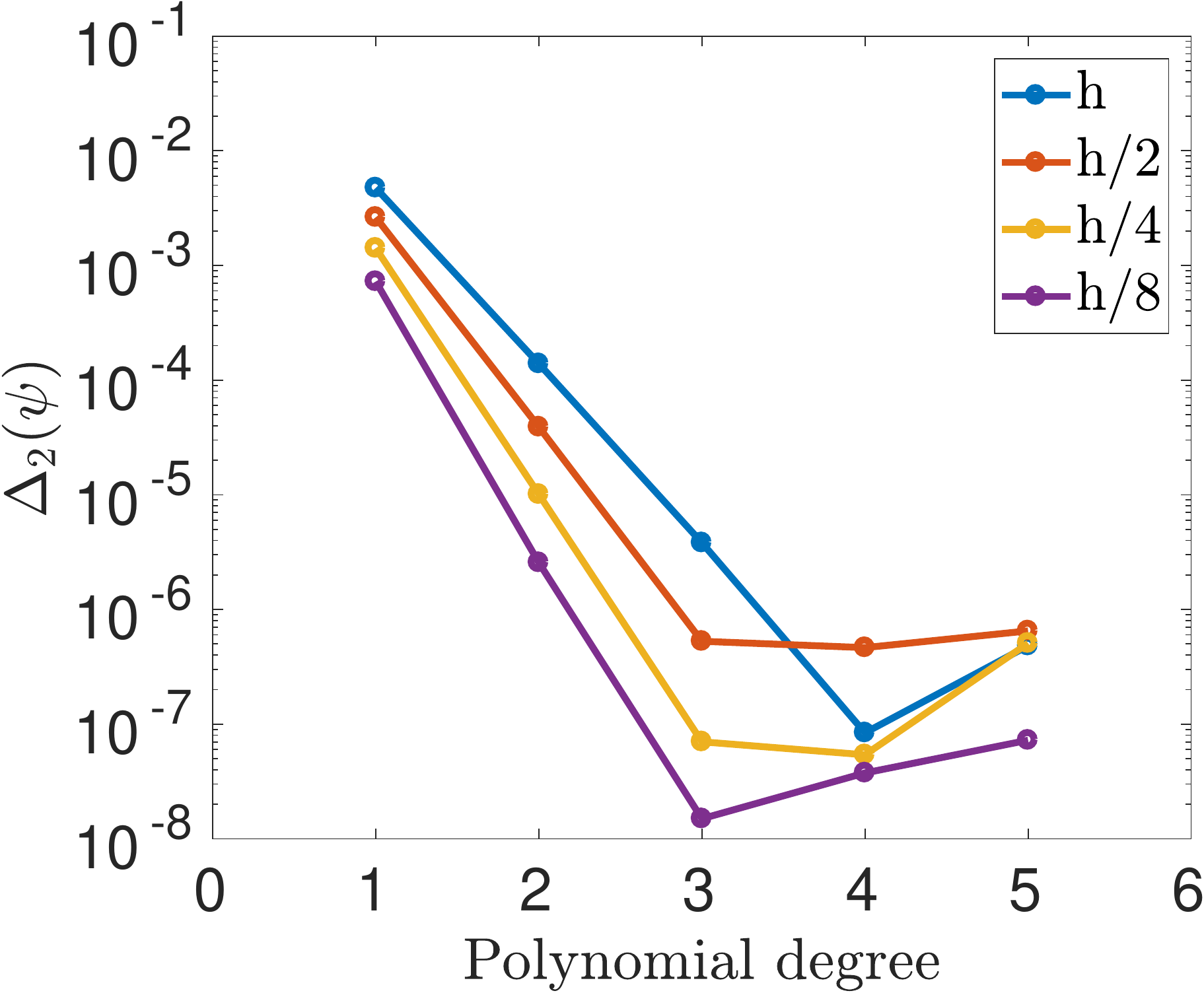} \quad & \includegraphics[width=0.26\linewidth]{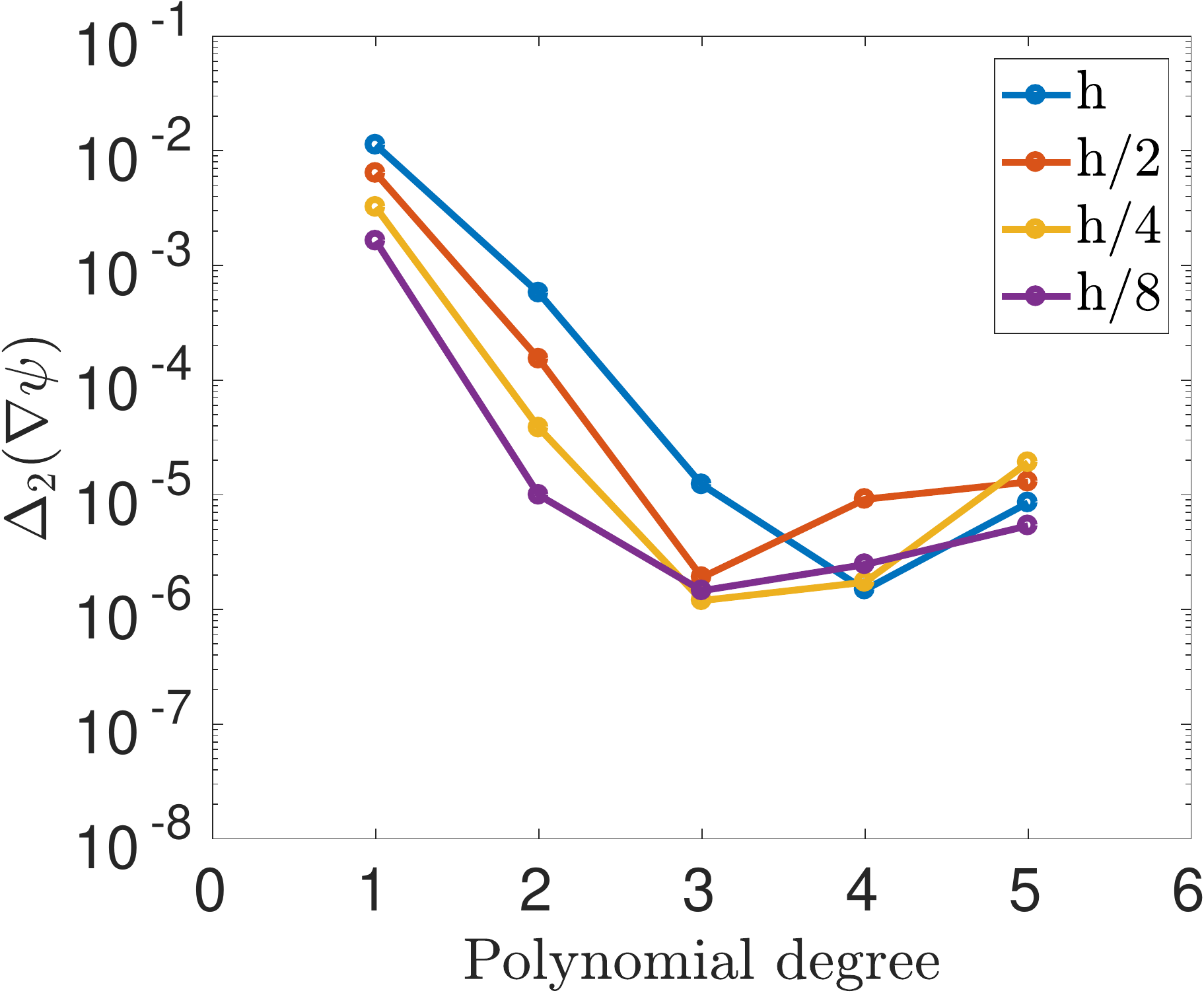} \quad & 
\multirow{5}{*}[\dimexpr 1.2in]{\includegraphics[width=0.3\linewidth]{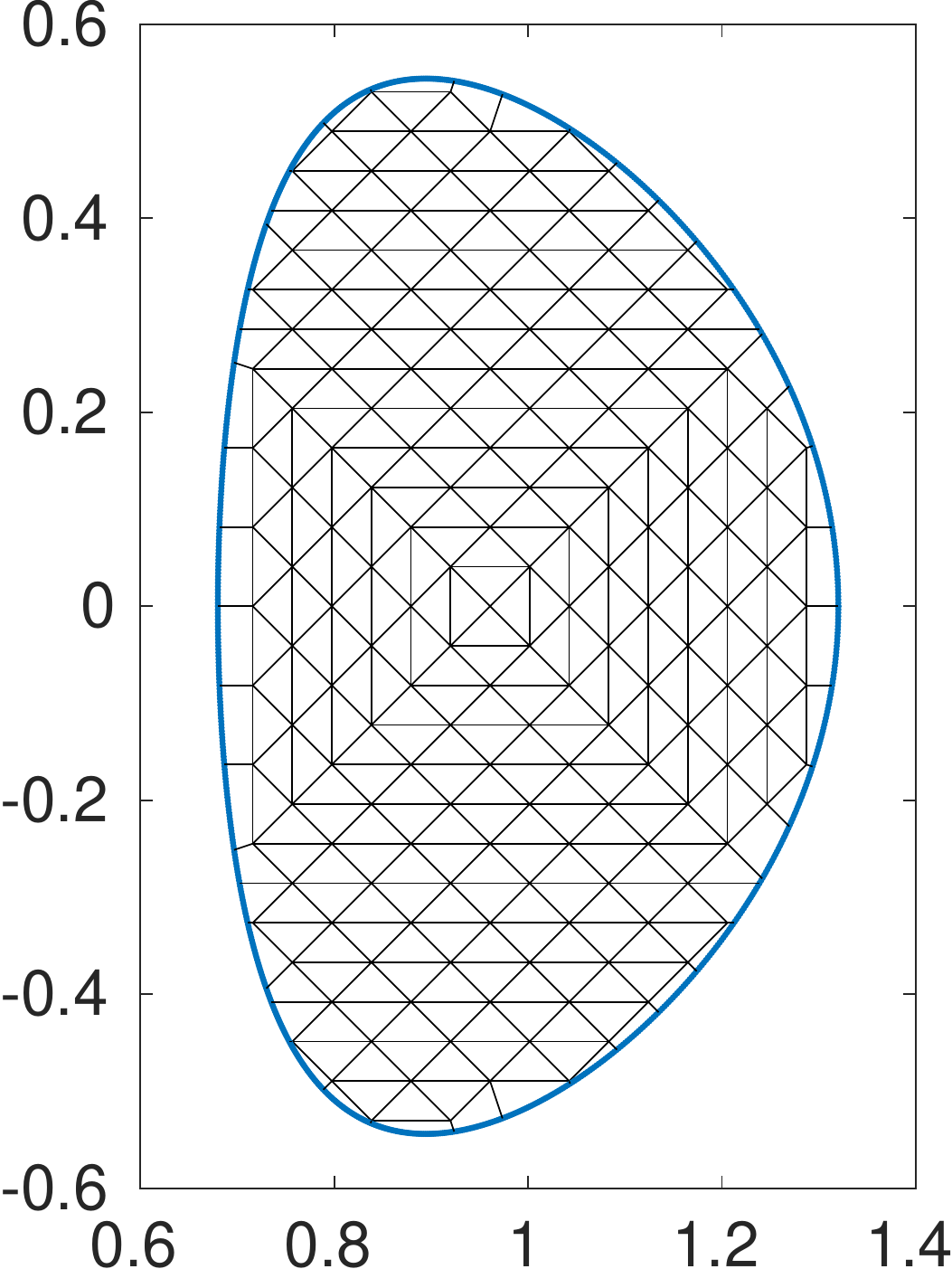}} \\ 
\includegraphics[width=0.26\linewidth]{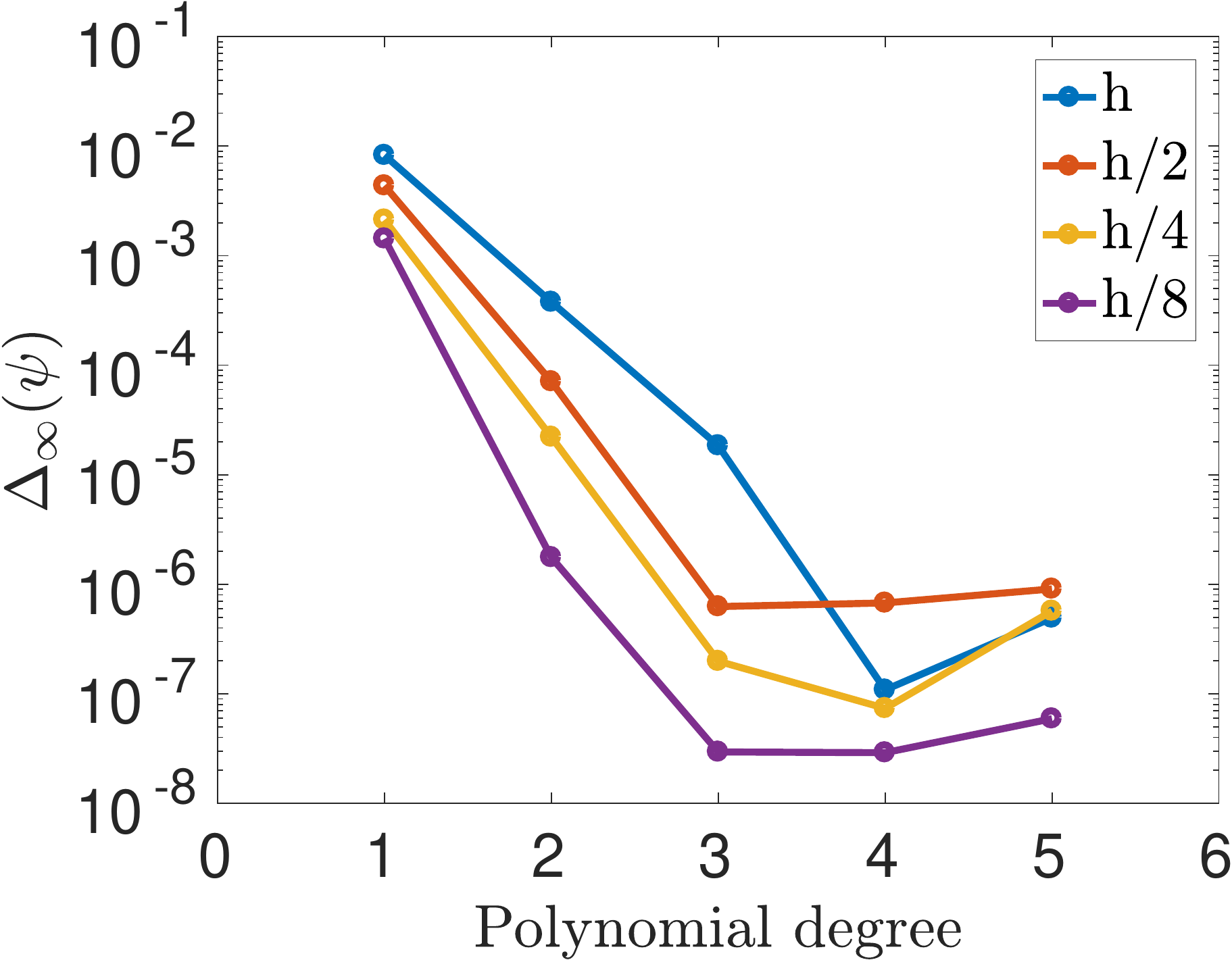} \quad & \includegraphics[width=0.26\linewidth]{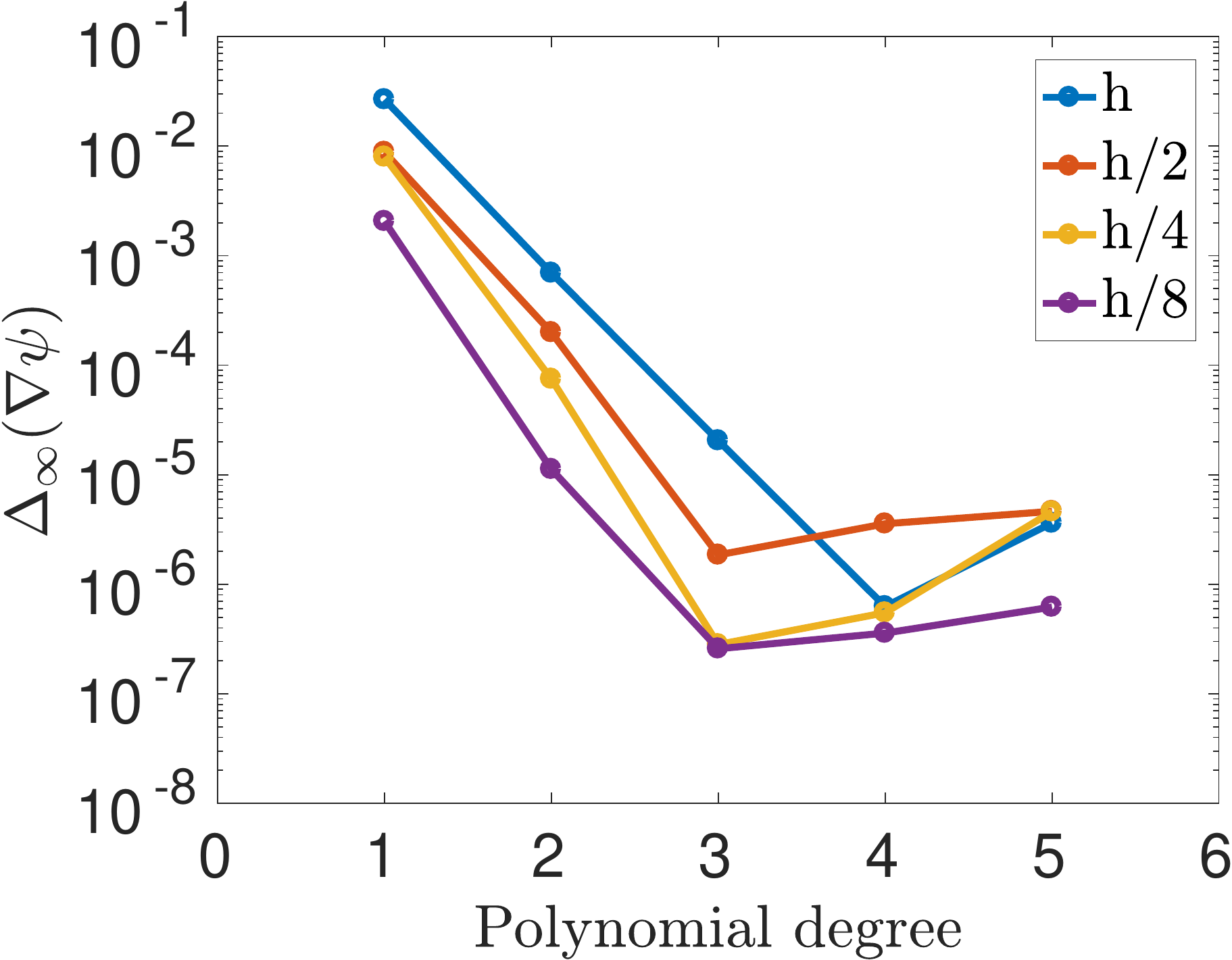} \quad & 
\end{tabular}
\caption{{\footnotesize Convergence estimates for Example 6 for successive refinements of the computational grid and increasingly higher polynomial degrees. The $\Delta_2$ and $\Delta_\infty$ differences between levels of refinement for the poloidal flux $\psi$ (left) and $\nabla\psi$ (center) are shown respectively along the top and bottomm rows. The confinement region and sample grid corresponding to the second level of refinement (red curve in the convergence plot) is shown on the right column. The reader is referred to the on-line version of the manuscript for the color scheme.}\label{fig:nonlinearSmooth}}
\end{figure}

On the convergence plots shown in Figures \ref{fig:frc}, \ref{fig:iter}, \ref{fig:nstx}, \ref{fig:asdexup}, \ref{fig:nonlinear}, and \ref{fig:nonlinearSmooth} the polynomial degree is varied along the horizontal axis, while the convergence curves for meshes with different diameters are superimposed with a different color on the same graph. The reader is referred to the electronic version of the paper for the color reference.

When the mesh is refined dyadically as in our experiments, the estimated convergence rate (e.c.r.) between two successive levels of $h$-refinement can be estimated by
\[
\text{e.c.r} \approx \log{\left(E^k_j(f)/E^{k+1}_j(f)\right)}/\log{(2)},
\]
where $E^k_j(f)$ corresponds to the error associated to the approximation of $f$ in a mesh with diameter $h/2^k$ and the index $j \in \{2,\infty\}$ denotes the mean square and maximum measures. Tables \ref{tab:ecrFRC}, \ref{tab:ecrITER}, \ref{tab:ecrNSTX}, \ref{tab:ecrASDEXUP}, and \ref{tab:ecrNONLINEAR} show the estimated convergence rates for each of the experiments as computed by the above relation.
%
\paragraph{Solov'ev profiles in smooth geometries} We start by testing the performance of the method on a smooth geometry corresponding to a Field Reversed Configuration modeled using the parametrization given in \cite{CeFr:2010}. The parametrization of the plasma-vacuum boundary uses parameter values  $\epsilon=0.99, \delta = 0.7, \kappa = 10$, and $A=0$  and the coarsest mesh for this geometry is characterized by $h = 1.25$. 

For this configuration, Example 1, the exact Solov'ev solution is a linear combination of the functions  $\psi_p$ from equation \eqref{eq:psir} with $A=0$ and $\psi_1$, $\psi_2$, and $\psi_4$ from equation \eqref{eq:solovevSol}, yielding an up-down symmetric configuration which is a bivariate polynomial of fourth degree. This explains the sharp drop of the $L^2$ approximation error when the polynomial degree reaches 4 (Figure \ref{fig:frc}).

Despite the smoothness of the geometry and the relative simplicity of the exact solution,  configurations like this can present challenges to solvers due two factors. The first one, and perhaps the most evident, is the proximity of the plasma boundary to the origin, where the toroidal operator $\Delta^*$ becomes singular. The second one is the high elongation of the confinement region which can lead to overcrowding in the regions of high curvature for methods relying on conformal mapping, or to meshes with elements that are either too large and resolve the geometry poorly, or too many and result in large numbers of unknowns.

As it is shown in Figure \ref{fig:frc}, the algorithm performs remarkably well despite the geometrical challenges mentioned above and the comparatively large mesh diameters.  Moreover, Table \ref{tab:ecrFRC}  shows that the estimated convergence rate is of order at least $k+1$ ($k<4$) for $\psi$ and its gradient, which agrees with the theory in \cite{CoQiSo:2014}. The estimated convergence rate for $k\geq 4$, in this case,  does not provide any useful information because the exact solution is a polynomial of degree four. For smooth geometries and Solov'ev profiles this kind of performance was also observed on other geometries such as a Spheromak and ITER-like and NSTX-like configurations without an x-point; for the sake of conciseness these examples were left out of the paper.
\begin{table}\centering
\scalebox{0.675}{
\begin{tabular}{c|ccc|ccc|ccc|ccc|}
\cline{2-13}
 & \multicolumn{3}{|c|}{$E_2(\psi)$} & \multicolumn{3}{|c|}{$E_2(\nabla\psi)$} & \multicolumn{3}{|c|}{$E_\infty(\psi)$} & \multicolumn{3}{|c|}{$E_\infty(\nabla\psi)$} \\
\hline
\multicolumn{1}{|c|}{Degree} & \multicolumn{1}{|c}{$h\rightarrow h/2$} & \multicolumn{1}{|c}{$h/2\rightarrow h/4$} & \multicolumn{1}{|c|}{$h/4\rightarrow h/8$} & \multicolumn{1}{|c}{$h\rightarrow h/2$} & \multicolumn{1}{|c}{$h/2\rightarrow h/4$} & \multicolumn{1}{|c|}{$h/4\rightarrow h/8$} & \multicolumn{1}{|c}{$h\rightarrow h/2$} & \multicolumn{1}{|c}{$h/2\rightarrow h/4$} & \multicolumn{1}{|c|}{$h/4\rightarrow h/8$} & \multicolumn{1}{|c}{$h\rightarrow h/2$} & \multicolumn{1}{|c}{$h/2\rightarrow h/4$} & \multicolumn{1}{|c|}{$h/4\rightarrow h/8$}\\ \hline
\multicolumn{1}{|c|}{1} & 2.75&2.40&2.13&2.67&2.12&1.98&1.32&2.40&1.65&1.41&1.77&1.69 \\ \hline
\multicolumn{1}{|c|}{2} & 3.44&3.442&3.143&3.159&3.318&3.252&2.888&3.228&2.449&2.741&3.056&1.98 \\ \hline
\multicolumn{1}{|c|}{3} & 6.86&4.99&4.44&6.42&4.31&3.78&4.55&4.46&4.16&5.92&3.38&4.48 \\ \hline
\multicolumn{1}{|c|}{4} & 9.90&0.96&0.68&8.46&-0.20&-0.54&5.96&-0.23&0.12&6.70&-0.87&-0.37 \\ \hline
\multicolumn{1}{|c|}{5} & 13.40&1.01&1.18&11.42&0.02&-0.24&9.74&-0.66&0.15&10.10&-0.70&-0.39 \\ \hline 
\end{tabular}
}
\caption{{\footnotesize Estimated $h$-convergence rates between two successive levels of refinement for Example 1 for polynomial orders ranging between 1 and 5. Beyond this order, round-off error becomes significant as can be seen from the convergence plots. The coarsest level mesh had diameter $h = 1.25$.}\label{tab:ecrFRC} }
\end{table}
%
\paragraph{Solov'ev profiles on geometries with an x-point} The second set of test problems consists of Solov'ev profiles in geometries that present an x-point. The geometries are up-down asymmetric with a downwards oriented x-point and are modeled by the parametrization detailed in \cite{CeFr:2010}. The exact solutions for these configurations involve all the terms in equations \eqref{eq:psir} and \eqref{eq:solovevSol} and are therefore more challenging tests not only from the geometrical point of view. 

Example 2 corresponds to an ITER-like configuration with parameters  $\epsilon=0.32, \delta = 0.33, \kappa = 2$, and $A=-0.115$ the mesh on the computational domain has an initial diameter of $h = 0.175$ that is successively halved. Example 3 corresponds to an NSTX-like configuration modeled using the values $\epsilon=0.78, \delta = 0.335,  \kappa = 1.7$, and $A=-0.115$. In this case the coarsest computational mesh has diameter $h = 0.5$. 

The presence of an x-point in these configurations makes the problem challenging for many available solvers, nevertheless as can be seen in Figure \ref{fig:iter} for the ITER geometry and Figure \ref{fig:nstx} for the NSTX configuration, our algorithm performs satisfactorily both on and off the grid, with the NSTX geometry presenting a bigger challenge due to its larger elongation and proximity to the origin as compared to the ITER configuration.

In addition, for both examples we also observe in Tables \ref{tab:ecrITER} and \ref{tab:ecrNSTX}  optimal estimated convergence rate as predicted by the theory in \cite{CoQiSo:2014}. We point out that when $k=5$ and the meshsize is $h/4$ or smaller, the errors are affected by round-off errors and then computed convergence does not provide any useful information in this case.
\begin{table}\centering
\scalebox{0.675}{
\begin{tabular}{c|ccc|ccc|ccc|ccc|}
\cline{2-13}
 & \multicolumn{3}{|c|}{$E_2(\psi)$} & \multicolumn{3}{|c|}{$E_2(\nabla\psi)$} & \multicolumn{3}{|c|}{$E_\infty(\psi)$} & \multicolumn{3}{|c|}{$E_\infty(\nabla\psi)$} \\
\hline
\multicolumn{1}{|c|}{Degree} & \multicolumn{1}{|c}{$h\rightarrow h/2$} & \multicolumn{1}{|c}{$h/2\rightarrow h/4$} & \multicolumn{1}{|c|}{$h/4\rightarrow h/8$} & \multicolumn{1}{|c}{$h\rightarrow h/2$} & \multicolumn{1}{|c}{$h/2\rightarrow h/4$} & \multicolumn{1}{|c|}{$h/4\rightarrow h/8$} & \multicolumn{1}{|c}{$h\rightarrow h/2$} & \multicolumn{1}{|c}{$h/2\rightarrow h/4$} & \multicolumn{1}{|c|}{$h/4\rightarrow h/8$} & \multicolumn{1}{|c}{$h\rightarrow h/2$} & \multicolumn{1}{|c}{$h/2\rightarrow h/4$} & \multicolumn{1}{|c|}{$h/4\rightarrow h/8$}\\ \hline
\multicolumn{1}{|c|}{1} & 1.91 & 1.96 & 1.98 & 1.98 & 2.30 & 2.06 & 1.91 & 1.18 & 1.81 & 0.99 & 2.03 & 0.56 \\ \hline
\multicolumn{1}{|c|}{2} & 3.16 & 3.07 & 3.04 & 2.88 & 3.74 & 3.22 & 2.28 & 3.86 & 2.16 & 1.44 & 4.12 & 1.93 \\ \hline
\multicolumn{1}{|c|}{3} & 4.28 & 4.56 & 4.49 & 4.19 & 4.68 & 4.37 & 3.64 & 4.75 & 4.17 & 2.83 & 4.70 & 3.24 \\ \hline
\multicolumn{1}{|c|}{4} & 5.95 & 5.85 & 5.69 & 5.27 & 5.83 & 5.28 & 5.57 & 5.75 & 5.34 & 4.01 & 5.83 & 4.49 \\ \hline
\multicolumn{1}{|c|}{5} & 6.09 & 7.12 & 0.62 & 5.80 & 6.98 &-1.00 & 5.92 & 6.98 &-2.39 & 4.78 & 6.31 &-3.73 \\ \hline
\end{tabular}
}
\caption{{\footnotesize Estimated $h$-convergence rates between two successive levels of refinement for Example 2 for polynomial orders ranging between 1 and 5. Beyond this order, round-off error becomes significant as can be seen from the convergence plots. The coarsest level mesh had diameter $h = 0.5$.}\label{tab:ecrITER} }
\end{table}
\begin{table}[h]\centering
\scalebox{0.675}{
\begin{tabular}{c|ccc|ccc|ccc|ccc|}
\cline{2-13}
 & \multicolumn{3}{|c|}{$E_2(\psi)$} & \multicolumn{3}{|c|}{$E_2(\nabla\psi)$} & \multicolumn{3}{|c|}{$E_\infty(\psi)$} & \multicolumn{3}{|c|}{$E_\infty(\nabla\psi)$} \\
\hline
\multicolumn{1}{|c|}{Degree} & \multicolumn{1}{|c}{$h\rightarrow h/2$} & \multicolumn{1}{|c}{$h/2\rightarrow h/4$} & \multicolumn{1}{|c|}{$h/4\rightarrow h/8$} & \multicolumn{1}{|c}{$h\rightarrow h/2$} & \multicolumn{1}{|c}{$h/2\rightarrow h/4$} & \multicolumn{1}{|c|}{$h/4\rightarrow h/8$} & \multicolumn{1}{|c}{$h\rightarrow h/2$} & \multicolumn{1}{|c}{$h/2\rightarrow h/4$} & \multicolumn{1}{|c|}{$h/4\rightarrow h/8$} & \multicolumn{1}{|c}{$h\rightarrow h/2$} & \multicolumn{1}{|c}{$h/2\rightarrow h/4$} & \multicolumn{1}{|c|}{$h/4\rightarrow h/8$}\\ \hline
\multicolumn{1}{|c|}{1} & 1.95 & 1.97 & 1.99 & 2.20 & 2.10 & 2.05 & 0.60 & 2.07 & 3.18 & 2.33 & 1.75 & 2.22 \\ \hline
\multicolumn{1}{|c|}{2} & 3.09 & 3.01 & 2.99 & 3.40 & 3.58 & 3.32 & 3.68 & 2.68 & 2.49 & 3.42 & 2.71 & 2.27 \\ \hline
\multicolumn{1}{|c|}{3} & 4.63 & 4.69 & 4.23 & 4.28 & 4.55 & 4.43 & 4.66 & 4.66 & 1.95 & 3.99 & 3.70 & 2.39 \\ \hline
\multicolumn{1}{|c|}{4} & 4.86 & 5.96 & 5.98 & 5.03 & 5.40 & 5.50 & 4.95 & 5.27 & 5.52 & 5.45 & 3.91 & 4.36 \\ \hline
\multicolumn{1}{|c|}{5} & 5.77 & 5.12 & 4.26 & 5.99 & 5.05 & 5.22 & 6.75 & 2.30 & 4.74 & 7.33 & 2.19 & 3.91 \\ \hline
\end{tabular} 
}
\caption{{\footnotesize Estimated $h$-convergence rates between two successive levels of refinement for Example 3 for polynomial orders ranging between 1 and 5. Beyond this order, round-off error becomes significant as can be seen from the convergence plots. The coarsest level mesh had diameter $h = 0.5$.}\label{tab:ecrNSTX} }
\end{table}
%
\paragraph{Dissimilar source terms on geometries with an x-point} 
The test chosen for Example 4 is based on the model for the ASDEX upgrade experiment discussed in \cite{McCarthy:1999} and described by equation \eqref{eq:dissimilarSol}. With the pressure and current profiles determined by this model, the source term of the Grad-Shafranov equation depends linearly on $\psi$. This kind of source terms can be absorbed directly into a modified bilinear form and dealt with computationally by introducing a mass matrix in the system. However, our goal is to provide a solver that relies only on the discretization of the toroidal operator $\Delta^*$ and allows the user to provide the source terms as problem data. In this framework, a general source term is always dealt with iteratively.

The exact solution associated to this configuration of the ASDEX Upgrade experiment corresponds to equation \eqref{eq:dissimilarSol} with the set of coefficients given by: 
\begin{alignat*}{12}
c_1 = &&\, 0.17795 &\qquad& c_6 =&&\, -0.162   &\qquad& c_{11} =&&\, 1.5820 &\qquad& c_{16} =&&\, -0.4265  \\
c_2 =&&\, -0.03291 &\qquad& c_7 =&&\, 0.3722 &\qquad& c_{12} =&&\,-0.009059 &\qquad& c_{17} =&&\, 0.8057 \\
c_3 =&&\, 1.4934	  &\qquad& c_8 =&&\, 0.07697   &\qquad&  c_{13}=&&\, 2.2388 &\qquad& c_{18} =&&\, -0.004804 \\
c_4 =&&\,-0.4818   &\qquad& c_9 =&&\, 1.2959 &\qquad& c_{14} =&&\,0.4186   &\qquad& T =&&\,  17.8116.\\   
c_5 =&&\, -1.1759  &\qquad& c_{10} =&&\,0.5881  &\qquad& c_{15} =&&\, 1.195 &&&& 
\end{alignat*}
These values give rise to a configuration with an upwards-oriented x-point as shown in Figure \ref{fig:asdexup}. The background mesh at the coarsest level of refinement had diameter $h = 0.275$. As confirmed by the convergence plots on Figure \ref{fig:asdexup}, the iterative process performs well both in the point-wise and mean square senses. Moreover, the behavior of the rate of convergence observed in Table \ref{tab:ecrASDEXUP} is similar to that of previous examples.
\begin{table}[h]\centering
\scalebox{0.675}{
\begin{tabular}{c|ccc|ccc|ccc|ccc|}
\cline{2-13}
 & \multicolumn{3}{|c|}{$E_2(\psi)$} & \multicolumn{3}{|c|}{$E_2(\nabla\psi)$} & \multicolumn{3}{|c|}{$E_\infty(\psi)$} & \multicolumn{3}{|c|}{$E_\infty(\nabla\psi)$} \\
\hline
\multicolumn{1}{|c|}{Degree} & \multicolumn{1}{|c}{$h\rightarrow h/2$} & \multicolumn{1}{|c}{$h/2\rightarrow h/4$} & \multicolumn{1}{|c|}{$h/4\rightarrow h/8$} & \multicolumn{1}{|c}{$h\rightarrow h/2$} & \multicolumn{1}{|c}{$h/2\rightarrow h/4$} & \multicolumn{1}{|c|}{$h/4\rightarrow h/8$} & \multicolumn{1}{|c}{$h\rightarrow h/2$} & \multicolumn{1}{|c}{$h/2\rightarrow h/4$} & \multicolumn{1}{|c|}{$h/4\rightarrow h/8$} & \multicolumn{1}{|c}{$h\rightarrow h/2$} & \multicolumn{1}{|c}{$h/2\rightarrow h/4$} & \multicolumn{1}{|c|}{$h/4\rightarrow h/8$}\\ \hline
\multicolumn{1}{|c|}{1} & 3.14 & 2.62 & 2.40 & 3.14 & 2.60 & 2.55 & 2.99 & 2.43 & 2.09 & 2.67 & 2.47 & 1.56 \\ \hline
\multicolumn{1}{|c|}{2} & 4.97 & 3.53 & 3.18 & 4.15 & 3.53 & 3.70 & 4.47 & 2.60 & 2.98 & 3.26 & 3.05 & 3.08 \\ \hline
\multicolumn{1}{|c|}{3} & 5.42 & 4.91 & 5.21 & 5.11 & 4.57 & 4.73 & 5.32 & 4.83 & 4.96 & 4.65 & 4.46 & 3.95 \\ \hline
\multicolumn{1}{|c|}{4} & 7.32 & 5.77 & 7.47 & 6.06 & 5.42 & 5.90 & 7.21 & 5.73 & 6.22 & 5.86 & 4.18 & 5.10 \\ \hline
\multicolumn{1}{|c|}{5} & 7.03 & 7.28 & 3.46 & 5.96 & 6.94 & 4.26 & 7.03 & 7.27 & 3.54 & 4.61 & 6.97 & 2.28 \\ \hline
\end{tabular} 
}
\caption{{\footnotesize Estimated $h$-convergence rates between two successive levels of refinement for Example 4 for polynomial orders ranging between 1 and 5. Beyond this order, round-off error becomes significant as can be seen from the convergence plots. The coarsest level mesh had diameter $h = 0.275$.} \label{tab:ecrASDEXUP} }
\end{table}
%
\paragraph{Non-linear source terms} Example 5 uses an up-down symmetric geometry with two x-points corresponding to a double-null divertor. The parameters, $\epsilon=0.32, \delta=0.33, \kappa=1.7$, and $ A=0$, as given in \cite{CeFr:2010}, are those of an  ITER-like equilibrium. 

Analytic solutions for pressure and current profiles that result in non linear source terms as a function of $\psi$ are not available. Therefore to test this example we resorted to a manufactured solution. We opted for including a linear, a quadratic and an exponential term as functions of $\psi$ in the source, which was then complemented so that the function
\[
\psi = \sin{\big(\kappa_r(r+r_0)\big)}\cos{\big(\kappa_z z\big)}, \qquad \Big(r_0 = -0.5, \; \kappa_r = 1.15\pi, \; \kappa_z = 1.15 \Big)
\]
satisfies Equation \eqref{eq:BVP2D}. The effective source term was
\[
F(r,z,\psi) := (\kappa_r^2+\kappa_z^2)\psi+\frac{\kappa_r}{r}\cos{\big(\kappa_r(r+r_0)\big)}\cos{\big(\kappa_z z\big)}+ r\,\Big(\sin^2{\big(\kappa_r(r+r_0)\big)}\cos^2{\big(\kappa_z z\big)}-\psi^2 + e^{-\sin{\big(\kappa_r(r+r_0)\big)}\cos{\big(\kappa_z z\big)}}-e^{-\psi}\Big), 
\]
and non-homogeneous Dirichlet boundary conditions were imposed correspondingly.

With the parameters chosen as above, the imposed solution has similar qualitative behavior to what can be expected from a flux function, with values close to zero near the boundary and a critical point on the interior. Due to the nature of the source term, the problem requires the iterative treatment described in Section \ref{sec:iterative}.

As in the previous examples, both $h$ and $p$ refinements were considered with satisfactory results, as presented in Figure \ref{fig:nonlinear} for increasing polynomial degree. The estimated convergence rates with respect to successive refinements of the spatial grid can be seen in Table \ref{tab:ecrNONLINEAR}, where we observe an estimated convergence rate of order at least $k+1$ when $k\leq 4$. For $k=4$, this rate is $1.68$ from the first to the second mesh but it becomes $10.04$ from the second to the third mesh. In other words, a least squares fitting of the estimated convergence rate considering the three first meshes shows a convergence rate of order $k+1=6$. We emphasize that there are no theoretical estimates on the rate of convergence when the source term is non-linear as it is in this example. We are currently working on the error analysis to cover this case.
\begin{table}[h]\centering
\scalebox{0.675}{
\begin{tabular}{c|ccc|ccc|ccc|ccc|}
\cline{2-13}
 & \multicolumn{3}{|c|}{$E_2(\psi)$} & \multicolumn{3}{|c|}{$E_2(\nabla\psi)$} & \multicolumn{3}{|c|}{$E_\infty(\psi)$} & \multicolumn{3}{|c|}{$E_\infty(\nabla\psi)$} \\
\hline
\multicolumn{1}{|c|}{Degree} & \multicolumn{1}{|c}{$h\rightarrow h/2$} & \multicolumn{1}{|c}{$h/2\rightarrow h/4$} & \multicolumn{1}{|c|}{$h/4\rightarrow h/8$} & \multicolumn{1}{|c}{$h\rightarrow h/2$} & \multicolumn{1}{|c}{$h/2\rightarrow h/4$} & \multicolumn{1}{|c|}{$h/4\rightarrow h/8$} & \multicolumn{1}{|c}{$h\rightarrow h/2$} & \multicolumn{1}{|c}{$h/2\rightarrow h/4$} & \multicolumn{1}{|c|}{$h/4\rightarrow h/8$} & \multicolumn{1}{|c}{$h\rightarrow h/2$} & \multicolumn{1}{|c}{$h/2\rightarrow h/4$} & \multicolumn{1}{|c|}{$h/4\rightarrow h/8$}\\ \hline
\multicolumn{1}{|c|}{1} & 1.86 & 2.03 & 2.01 & 1.96 & 2.11 & 1.98 & 1.77 & 1.95 & 2.02 & 1.48 & 2.00 & 1.77 \\ \hline
\multicolumn{1}{|c|}{2} & 3.41 & 3.27 & 3.01 & 2.48 & 3.62 & 3.12 & 3.02 & 3.36 & 2.97 & 2.30 & 3.00 & 2.57 \\ \hline
\multicolumn{1}{|c|}{3} & 3.95 & 4.41 & 3.82 & 4.10 & 4.95 & 3.55 & 3.59 & 4.99 & 2.78 & 3.24 & 4.39 & 3.15 \\ \hline
\multicolumn{1}{|c|}{4} & 6.50 & 6.88 & 3.74 & 6.73 & 6.20 & 2.40 & 7.31 & 6.62 & 1.86 & 6.62 & 5.75 & 0.64\\ \hline
\multicolumn{1}{|c|}{5} & 1.68 &10.04 & 1.95 & 0.52 &10.09 & 0.32 & 0.08 &10.62 &-0.50 &-0.64 &10.21 &-2.23 \\ \hline
\end{tabular}
}
\caption{{\footnotesize Estimated $h$-convergence rates between two successive levels of refinement for Example 5 for polynomial orders ranging between 1 and 5. Beyond this order, round-off error becomes significant as can be seen from the convergence plots. The coarsest level mesh had diameter $h = 0.1792$.} \label{tab:ecrNONLINEAR} }
\end{table}
To close this section, the final Example 6 is set on a smooth D-shaped geometry. The $(r,z)$ coordinates of the boundary are given by a Miller parametrization \cite{Miller:1998} in terms of $t\in[0,2\pi)$ of the form
\[
r(t) = 1 + \epsilon\cos{\big(t+\arcsin{(\delta\sin{(t)})}\big)}, \qquad z(t) = \epsilon\kappa\sin{(t)}, 
\]
where $\epsilon=0.32$, $\delta=0.33$, and $\kappa = 1.7$, corresponding to an ITER-like configuration. In this instance the source term has a polynomial non-linearity corresponding to
\[
p = \frac{\psi}{2}\left(1+\frac{2\psi^2}{3}-\frac{\psi^5}{5}\right) \quad \text{ and }\quad g=0,
\]
which yield a source of the form
\[
F(r,z,\psi) = r^2\left(1-\frac{(1-\psi^2)^2}{2}\right).
\]
For this final example we do not impose a manufactured solution; instead we set homogeneous boundary conditions and estimate the convergence by measuring the maximum difference between two consecutive levels of refinement
\[
\Delta_j^k(f) := \|f_h^{k}-f_h^{k-1}\|_j.
\]
Here as before, the super index refers to the $k$-th level of discretization, while the lower index $j\in\{2,\infty\}$ refers to the norm being used. In this example, the background mesh had a parameter $h=0.1632$ at the coarsest level of refinement. With respect to the approximate convergence plots shown in Figure \ref{fig:nonlinearSmooth}, we remark that even if the relative difference between two successive approximations is a pessimistic estimate of the true error, the method still performs satisfactorily.
%
\section{Discussion}\label{sec:conclusions}
We have presented a high order accurate solver for the Grad-Shafranov equation based on the Hybridizable Discontinuous Galerkin method. As the numerical examples show, the method is very robust with respect to the geometrical properties of the confinement region and provides competitive convergence rates for the poloidal flux function and, more importantly, for the magnetic field even for relatively coarse grids. As the numerical experiments confirm, the proposed Anderson-accelerated iterative strategy satisfactorily accommodates for source terms with linear or non-linear dependence in $\psi$.

The use of HDG provides a framework with great potential for parallel computations, due to the fact that, once the global hybrid unknown has been determined, the local problems decouple and can be solved independently from each other, and that the associated global system has reduced bandwidth. This feature enhances the speed of the calculations specially in fine grids with a large number of elements. The transferring path method used to enforce Dirichlet boundary conditions on a polygonal subdomain gives the algorithm great geometric flexibility and avoids the need for constant re-meshing if the geometry has to be adjusted, as it is often the case for real time monitoring or free boundary applications. Moreover, the technique can be applied to both fitted and unfitted meshes.

On the downside, the extrapolation strategy used to define the approximation in the extension $\Omega^h_{ext}$ may fail to resolve possible boundary layers close to the separatrix. In order to address this possibility, the authors are currently working on an adaptive mesh refinement strategy that when combined with the transferring technique can resolve the fine-scale structure up to a prescribed tolerance without increasing the computational cost excessively.

Other areas of improvement include the implementation of a combined two-grid fixed-point + Newton step strategy, where accelerated fixed point iterations are carried out on the coarse grid to generate a good initial guess which is used as the starting point for a Newton iteration on the fine grid. This technique promises to dramatically decrease the number of iterations required on the fine grid, speeding up the computation even further. All these enhancements are the subject of ongoing work and will be implemented in a subsequent stage in order to enable the treatment of the free boundary problem and other situations in which other quantities of physical relevance are provided as input in combination with the pressure profile instead of the function $g(\psi)$ appearing in the current application. 
%
\section{Acknowledgements}
The authors are deeply thankful to Antoine Cerfon for his numerous suggestions and remarks that contributed to greatly improve the quality of the work and to point out several possible paths for future and current improvements. Tonatiuh S\'anchez-Vizuet was partially supported by the US Department of Energy. Grant No. DE-FG02-86ER53233. Manuel Solano was partially supported by CONICYT--Chile through FONDECYT project No. 1160320, by BASAL project CMM, Universidad de Chile and by Centro de Investigaci\'on en Ingenier\'ia Matem\'atica (CI$^2$MA), Universidad de Concepci\'on.
%
\bibliographystyle{elsarticle-num}
\bibliography{ReferencesGSHDG}
%
\end{document}